\documentstyle[preprint]{aastex}
\def\like{{\cal L}}

\newbox\grsign \setbox\grsign=\hbox{$>$} \newdimen\grdimen \grdimen=\ht\grsign
\newbox\simlessbox \newbox\simgreatbox \newbox\simpropbox
\setbox\simgreatbox=\hbox{\raise.5ex\hbox{$>$}\llap
     {\lower.5ex\hbox{$\sim$}}}\ht1=\grdimen\dp1=0pt
\setbox\simlessbox=\hbox{\raise.5ex\hbox{$<$}\llap
     {\lower.5ex\hbox{$\sim$}}}\ht2=\grdimen\dp2=0pt
\def\simgreat{\mathrel{\copy\simgreatbox}}
\def\simless{\mathrel{\copy\simlessbox}}

\begin{document}

\title{Resonant Cyclotron Radiation Transfer Model Fits to \\
Spectra from Gamma-Ray Burst GRB870303}

\author{P.~E.~Freeman$^{1,7}$, D.~Q.~Lamb$^{1}$, J.~C.~L.~Wang$^{2,8}$, \\
I.~Wasserman$^{3}$, T.~J.~Loredo$^{3}$, E.~E.~Fenimore$^4$, T.~Murakami$^5$,
A.~Yoshida$^6$}
\altaffiltext{1}{Dept. of Astronomy and Astrophysics, University
of Chicago, Chicago, IL 60637}
\altaffiltext{2}{Dept. of Astronomy, University of Maryland, College Park, MD 20742}
\altaffiltext{3}{Dept. of Astronomy, Space Sciences Building, Cornell University
, Ithaca, NY 14853}
\altaffiltext{4}{Mail Stop D436, Los Alamos National Laboratory, Los Alamos, NM 87545}
\altaffiltext{5}{Institute of Space and Astronautical Science, 1-1, Yoshinodai 3-chome, Sagamihara, Kanagawa 229, Japan}
\altaffiltext{6}{Institute of Physical and Chemical Research, 2-1, Hirosawa, Wako, Saitama 351-01, Japan}
\altaffiltext{7}{Presented as a thesis to the Department of Astronomy and Astrophysics, The University of Chicago, in partial fulfillment of the requirements for the Ph.D. degree.}
\altaffiltext{8}{Deceased 8 August 1998.  His friends and colleagues lament his passing.}

\begin{abstract}
We demonstrate that models of resonant cyclotron radiation 
transfer in a strong field (i.e.~cyclotron scattering)
can account for spectral lines seen at two epochs,
denoted S1 and S2, in the {\it Ginga} data for GRB870303.
S1, which extends 4~s, exhibits one line at $\approx$
20 keV, while S2, which extends 9~s, exhibits harmonically spaced
lines at $\approx$ 20 and 40 keV.  The midpoints of S1 and
S2 are separated by 22.5~s. 
Using a generalized version of the Monte Carlo code
of Wang et al.~(1988,1989b), we model line formation
by injecting continuum photons into a static plane-parallel slab of electrons
threaded by a strong neutron star magnetic field ($\sim$ 10$^{12}$ G)
which may be oriented at an arbitrary angle relative to the slab normal.
We examine two source geometries, which we denote
``1-0'' and ``1-1,'' with
the numbers representing the 
relative electron column densities above and below the
continuum photon source plane.
The 1-0 geometry may represent, e.g., a line-formation region levitating
above the surface of the neutron star, or possibly a plasma-filled flux tube
illuminated from below.  The 1-1 geometry, on the other hand, corresponds
to line formation in a semi-infinite atmosphere at the surface of a 
neutron star.  We apply rigorous statistical inference to compare
azimuthally symmetric models, i.e. models in which the magnetic field
is parallel to the slab normal, with models
having more general magnetic field orientations.  If the bursting source
has a simple dipole field, these two model classes represent
line formation at the magnetic pole, or elsewhere on the stellar surface.
We find that the data of S1 and S2, considered individually, are
consistent with both
geometries, and with all magnetic field orientations,
with the exception that the
S1 data clearly favors line formation 
away from a polar cap in the 1-1 geometry,
with the best-fit model placing the line-forming region 
at the magnetic equator.  Within both geometries, fits to the combined
(S1+S2) data marginally favor models which feature equatorial line formation,
and in which the observer's orientation with respect to the slab changes
between the two epochs.  We interpret this change as being due to neutron star
rotation, and we place limits on the rotation period.
\end{abstract}


\section{Introduction}

The low-energy spectral lines detected between $\approx$ 20-50 keV 
in the spectra of some gamma-ray bursts (GRBs) provide
the most powerful means by which an analyst
can probe the burst source environment.
This is because the study of lines, unlike the study of continuum
spectra, can yield precise information about
the temperature and column depth of the line-formation
region, the velocity and orientation of that region, and
the strength and orientation of any magnetic field in that region.

Candidate lines in GRB spectra were first reported by
Mazets et al.~(1980, 1981), who detected single absorption-like dips and 
troughs in the spectra of $\approx$ 15\% of bursts 
observed by the Konus detectors of {\it Venera 11} and {\it 12}.
Hueter (1987) reported similar absorption-like features
in the spectra of two bursts observed by the {\it HEAO-1} A4 detector,
and the {\it Ginga} Gamma-Ray Burst Detector (GBD) observed three 
bursts$-$GRB870303, GRB880205, and GRB890929$-$whose spectra
exhibited harmonically spaced absorption-like line candidates
with centroid energies between $\approx$ 20-50 keV
(Murakami et al.~1988; Fenimore et al.~1988; Yoshida et al.~1991; 
Graziani et al.~1992, 1993; Freeman et al.~1999a, hereafter Paper I).
A spectrum from an earlier epoch of GRB870303, denoted S1, was 
subsequently found to exhibit a single absorption-like line candidate
at $\approx$ 20 keV (Graziani et al.~1992, who used S2 to denote the other
line candidate spectrum of GRB870303; Graziani et al.~1993, Paper I).
Fits with phenomenological continuum-plus-line(s) 
models demonstrate that the {\it Ginga}
line candidates have statistical significance $\sim$ 10$^{-3}$-10$^{-8}$.
The enhanced low-energy coverage of the {\it Ginga} GBD
($E_{\rm low} \approx$ 1.5 keV, compared with
$E_{\rm low} \simgreat$ 20 keV for Konus and {\it HEAO-1} A4),
allowed analysts to
rule out the possibility that these line candidates could be explained
by a sudden change in continuum shape.

Mazets et al.~posited that the Konus line candidates
were formed by either cyclotron absorption or cyclotron scattering in the 
strong magnetic fields of galactic neutron stars ($\sim$ 10$^{12}$ G).
The observation of harmonically spaced line candidates 
gave further credence to this hypothesis,
because the quantization of an electron's energy
perpendicular to a strong field
can lead to the formation of evenly spaced lines with 
energies $E_n \approx$ 11.6$nB_{12}$ keV.  Rigorous proof of the
viability of this hypothesis came when 
a number of analysts used theoretical models of cyclotron scattering
(and not cyclotron absorption) to produce
emergent spectra from line-forming regions, which were found
to compare favorably with the {\it Ginga} data 
(Wang et al.~1989a; Alexander {\&} M{\'e}sz{\'a}ros 1989;
Nishimura et al.~1992; Wang, Wasserman, \& Lamb 1993).
For instance, Wang et al.~(1989a) used a Monte Carlo radiative transfer code 
(Wang, Wasserman, \& Salpeter 1988, 1989b; Lamb et al.~1989) 
to demonstrate that the harmonic line candidates in the GRB880205 spectrum
could be produced in a line-forming layer with magnetic
field 1.7 $\times$ 10$^{12}$ G and column density 1.2 $\times$
10$^{21}$ cm$^{-2}$.  Freeman et al.~(1992) later demonstrated that 
deepening the scattering region behind the continuum source led to
a substantially better fit to these same data.

The cyclotron scattering interpretation for line candidates
went largely unchallenged when it was consistent with the prevailing view
that GRB sources were neutron stars
residing in a thick disk in the Milky Way 
(with scale height $\approx$ 2 kpc; see, e.g.,
Higdon \& Lingenfelter 1990, Harding 1991).  However, recent
developments, while not challenging the
cyclotron scattering picture {\it per se},
have led many to question whether the reported line candidates 
actually exist at all.
First, the discovery of optical transients (OTs) associated
with GRBs (e.g.~van Paradijs et al.~1997 and references therein), and
the apparent determination of redshifts for five of them$-$GRB970508
(Metzger et al.~1997), GRB971214 (Kulkarni et al.~1998),
GRB980613 (Djorgovski et al.~1999), GRB980703 (Djorgovski et al.~1998),
and GRB990123 (Kelson et al.~1999)$-$have indicated that some
(if not all) GRBs occur at cosmological distances.
Second, there have been no reports of 
definitive detections of line candidates in spectra
of the Burst and Transient Source Experiment Spectroscopy Detectors
(BATSE SDs) on the {\it Compton Gamma-Ray Observatory}
(Palmer et al.~1994, Band et al.~1996, Briggs et al.~1996, 1998).
These developments, combined with the
apparent difficulty of forming lines in the circumstellar 
environments of cosmological burst sources
(cf.~Stanek, Paczy{\'n}ski, \& Goodman 1993 and
Ulmer \& Goodman 1995, who attempt to account for lines by invoking
gravitational femtolensing), have led many
to adopt the viewpoint that {\it all} bursts
are cosmological and that the reported line candidates, for whatever 
reason, are not real.

This viewpoint may be intuitively reasonable on its surface.
However, while we do not seek to disprove the posited cosmological
origin of the GRBs listed above,
we feel that there are several reasons why we should continue to
test the cyclotron scattering model.  While none are completely 
compelling by themselves, they cumulatively indicate that the complete
solution of the GRB mystery may have not yet been provided by the
discovery of GRB redshifts.

The first reasons are the suggestive pieces of direct observational evidence 
that there are two or more classes of classical
gamma-ray bursts.  These pieces of evidence range from the well-established to
the speculative.  Well-established is the identification of two classes
with differing light-curve morphologies (Kouveliotou et al.~1993, using
burst duration; Lamb, Graziani, \& Smith 1993, using light curve variability).
Further studies lend credence to the reality of separate classes
(Kouveliotou et al.~1996; Katz \& Canel 1996),
but it is not yet proved that the correlation of burst
properties cannot be explained using a single underlying mechanism.

Less well-established, but still highly suggestive, evidence is
provided by burst repetition (e.g.~Quashnock \& Lamb 1993, 
Wang \& Lingenfelter 1995).  Unlike classes based on light-curve morphology,
repetition indicates directly that there must be a galactic component to the
overall GRB source population, since it is generally considered impossible for 
cosmological bursters caused by one-time events such as
neutron star-neutron star or neutron star-black hole mergers to repeat.
Evidence for repetition is absent from the Third BATSE
(3B) catalog (Meegan et al.~1996, Tegmark et al.~1996),
but Graziani \& Lamb (1996) have questioned how the
systematic errors of this catalog are computed.  (For details on 
the BATSE burst location algorithm, see Pendleton et al.~1999).
Also, in October 1996, after the publication of the 3B catalog,
BATSE observed four bursts in less than two days which
all came from directions consistent with a single source.
Graziani, Lamb, \& Quashnock (1998) perform simulations which indicate that
the probability of such a spatial and temporal coincidence of four 
distinct bursts from four different sources is 3.1 $\times$ 10$^{-5}$.
The probability is increased to 1.6 $\times$ 10$^{-3}$ if only three bursts
occurred, but one of the three would have to be the longest ever observed
by BATSE.

The most speculative piece of evidence
is the apparent lack of consistency that recently localized bursts
show when examined in other wavebands.
For instance, since the beginning of 1997, there
have been eleven GRBs observed with
the {\it BeppoSAX} WFC and/or the {\it RXTE} ASM detectors
whose positions have been refined through the use of
the Interplanetary Network (IPN).\footnote{
This information was compiled using
{\tt http://www.obs.aip.de/$\sim$jcg/grbgen.html}, which contains 
complete references, including texts of circulars,
for each burst.}
Of these eleven bursts, X-ray afterglows, OTs, and 
radio afterglows were detected for eight, six, and three of them 
respectively.  (Two have measured redshifts and are thus
conclusively cosmological$-$GRB971214 and GRB990123.)
The lack of counterparts may be, of course, 
more simply explained by invoking, e.g., selection effects,
rather than by invoking a separate class of bursters whose broad-band behavior
is such that they do not show optically transient emission.  Also,
if there is a separate class of bursters which are
defined by counterpart behavior, it would not necessarily
have to be associated with the Milky Way.

Another reason for us to continue to test the cyclotron scattering
model is the result of the exacting statistical analysis of the 3B catalog
performed by Loredo \& Wasserman (1995, 1998a,b).
They determine that the 3B catalog is consistent with
the hypothesis that there is a component of the GRB source population
residing within the galaxy, with that component comprised of either
dim local halo sources at distances $\simless$ 1 kpc
(accounting for up to $\approx$ 60\% of all bursts), or
luminous halo sources at distances $\simgreat$ 50 kpc
(accounting for up to $\approx$ 10\% of all bursts).  Studies show that
neutron stars which receive high initial kick velocities when they
are formed can populate the halo
(e.g.~Li \& Dermer 1992; Podziadlowski, Rees, \& 
Ruderman 1995; Bulik, Lamb, \& Coppi 1998).
Such ``hybrid" models with both galactic and cosmological bursters
fit to the 3B data better than models with
only cosmological bursters, in part because the existence of a highly
isotropic, extragalactic burst component weakens greatly the isotropy
constraints on any anisotropic Galactic component.
However, the fit is not so much better as to rule out models containing
only cosmological bursters.

A final reason to continue to test the cyclotron scattering
model is the fact that the BATSE SDs {\it may} 
lack the low-energy spectral sensitivity that is necessary for
low-energy lines to be easily detected.  The gain settings on the
individual SDs differ; those with the highest gain settings can,
in principle, observe GRBs at energies $\simgreat$ 10 keV.
An electronic artifact discovered after launch
affects energy calibration such that spectra are distorted in the first
$\approx$ 10 channels above the low-energy cutoff
(the so-called ``SLED'' effect; see Band et al.~1992).
Despite this, studies using simulated {\it Ginga} line candidate spectra
indicated that the SDs were still capable of detecting low-energy
line candidates (Band et al.~1995), and
no line candidates were definitively detected during initial
visual searches of those spectra with the largest signal-to-noise
ratios (Palmer et al.~1994, Band et al.~1996).
An automated line candidate search algorithm designed by the BATSE SD team
(Briggs et al.~1996) was then
applied to spectra in 117 bright bursts for which there is at least one
spectrum with signal-to-noise $>$ 5 at $\approx$ 40 keV (Briggs et al.~1998).
This automated search, which is considerably more sensitive than
a visual search, yielded 12 candidate spectral line candidates for
which the change in $\chi^2$ between the continuum and continuum-plus-line fits
is $> 20$ (significance $< 5 \times 10^{-5}$).
Perhaps problematically, all the line candidates except one are 
emission-like lines observed at $\approx$ 40 keV, with the one exception being
an absorption-like line candidate observed at $\approx$ 60 keV.
Briggs et al.~(1998) cannot rule out the possibility that sharp breaks
in the continua, instead of lines, cause the observed features,
but such breaks would be inconsistent with low-energy
{\it Ginga} GRB data.  Briggs et al.~estimate
the ensemble chance probability of the most-significant feature
as $\simless$ 10$^{-3}$,
and state that few of these features, if any, result from statistical
fluctuations.  However, these lines
should not be considered definitively detected, as
the contemporaneous data from other SDs is still being examined
(Briggs et al.~1999).

We feel that the aforementioned reasons give us
ample justification to test the cyclotron scattering model
further.  In this paper, we use it to examine the lines\footnote{
The rigorous statistical analysis carried out in Paper I on the
data of GRB870303 S1 and S2 demonstrates that when the data
are considered jointly, the significance of the continuum-plus-lines
model is $\sim$ 10$^{-8}$.
Thus we feel that we may drop the 
word ``candidates" when referring to these lines throughout
the remainder of this paper.}
exhibited by GRB870303 S1 and S2.
Successful fits of the cyclotron scattering model to these data,
when paired with the successful fits of cyclotron scattering models to
the data of GRB880205 by Wang et al.~(1989a) and Freeman et al.~(1992),
would further strengthen the evidence supporting our contention that
that some (though not all) GRBs are galactic in origin.

In {\S}\ref{sect:geom}, we describe the spatial geometry of
the line-formation region.
We assume it to be a static plane-parallel slab of electrons
threaded by a strong magnetic field, which can be
oriented at an arbitrary angle relative to the slab normal
(unlike in Wang et al.~1989a, where the field was
parallel to the slab normal).
The assumption of a static slab is strictly valid only
for bursters located within $\sim$ 100 pc, {\it if} line formation
occurs at the magnetic pole, since otherwise the inferred burst
luminosity would exceed the Eddington limit and radiation forces 
would blow the line-forming layer off in a wind (see
Isenberg, Lamb, \& Wang 1998a, who modify the code to examine the theoretical 
ramifications of a wind).
As noted above, such distances are consistent with current observational 
limits (Loredo \& Wasserman 1998b). 
If line formation occurs away from the magnetic pole, then the assumption
of a static slab may be valid even for luminous halo bursters at
distances $\sim$ 100 kpc
because of the confinement provided by closed magnetic field lines
(e.g.~Zheleznyakov and Serber 1994, 1995).
We examine two slab geometries, which we denote
``1-0'' and the ``1-1'', where
the numbers represent relative electron column densities above and below the
continuum photon source plane.
A slab illuminated from below represents a line-forming region in the
magnetosphere of a neutron star,
while a source plane embedded within a slab
corresponds to line formation within a semi-infinite neutron star
atmosphere.

In {\S}\ref{sect:phys}, we summarize the physics of radiation transfer
in strong magnetic fields that is incorporated into the Monte Carlo code
we use to generate spectra
(Wang et al.~1988, 1989b; Lamb et al.~1989; Lamb, Wang, \& Wasserman~1990;
and references therein),
and in {\S}\ref{sect:samp} we provide examples of these spectra.
As noted above, resonant cyclotron scattering, and not cyclotron absorption, 
describes the
peculiar properties of the {\it Ginga} lines$-$the 
comparable strengths of first and second harmonics,
the absence of third and higher harmonics,
and the narrowness of the lines.
The appearance of the lines is greatly affected
by Raman scattering, i.e.~resonant cyclotron scattering in which the
electron is excited from the ground state directly 
to the second or higher Landau level, but then deexcites back to the
ground state indirectly via intermediate energy levels.
In Raman scattering, the original photon is destroyed, and
two or more photons are created (or {\it spawned}) as the
electron deexcites.  Spawned photons with energies near that 
of the first harmonic line alter its profile and reduce its 
equivalent width to a value similar to that of the second harmonic line.
Because the majority of photons which undergo scattering at the
second and higher harmonics are destroyed, the lines have an approximately
absorption-like profile.  The third harmonic is not seen because its
optical depth is small compared to that of the second harmonic
($\tau_3 \approx$ 0.05$\tau_2$).
Narrow lines result from the fact that the line-forming region is 
optically thin to all continuum photons except for those 
with energies equal to the first and second harmonic energies.
Thus scattered photons at $\approx$ 20 keV, and not continuum
photons $\simgreat$ 1 MeV, dictate the temperature of the 
line-forming region, and the line width.

In {\S}\ref{sect:stat} we summarize the statistical concepts which we use
in this paper.  These concepts are described in more detail in
Paper I and Freeman et al.~(1998b; hereafter Paper III).  In those
works, we present general, rigorous methodologies that address the problem
of establishing the existence of a line in a spectrum, that are based
upon both the so-called ``frequentist," and Bayesian, paradigms of 
statistical inference.  In this work, instead of establishing the
existence of lines, our statistical goal is to compare azimuthally symmetric 
models of line formation, i.e.~models in which the magnetic field
is parallel to the slab normal, with models having
arbitrary magnetic field orientations.
If the GRB source is a neutron star with a simple dipole field, these two
classes may be interpreted as representing
line formation at the magnetic pole, or elsewhere.
We use a rigorous method of model comparison (described
in {\S}\ref{sect:modcomp}) to select between
polar cap and more general models, demanding, e.g., that the increase in the
quality of the fit of general models be sufficiently great to justify
considering of the ramifications of line formation away from the pole.
Once best-fit models are selected, 
we use a rigorous method of parameter estimation (described
in {\S}\ref{sect:parest})
to place limits on each of the model parameters.

We apply the cyclotron scattering model to the data of GRB870303 S1 and S2
in {\S}\ref{sect:appl}.  S1, showing 4 s of data, exhibits a saturated line at 
$\approx$ 20 keV, while S2, showing 9 s of data, exhibits two
harmonically spaced lines at $\approx$ 20 and 40 keV.  
The midpoints of the S1 and S2 time intervals are 22.5 s apart.
In Paper I, we establish the evidence for the GRB870303 lines 
using simple phenomenological models.
In that work, we establish 
the frequentist significance of, and Bayesian odds favoring, 
the S1 line to be $3.6 \times 10^{-5}$ and 114:1, respectively;
for S2, the respective figures are $1.7 \times 10^{-4}$ and 7:1,
while for the combined (S1+S2) data,
they are $4.2 \times 10^{-8}$ and 40,300:1.
In joint fits to the combined (S1+S2) data in this work, we find that
the best-fit values of $\mu$ and/or $\phi$
change during the 22.5 s between S1 and S2 for both the 1-0 and 1-1
geometries, meaning that the orientation of the observer changes with
time.  We interpret the change in orientation by invoking neutron star 
rotation, and in {\S}\ref{sect:rot} (and the Appendix) we describe how
we place limits on the rotation period.

In {\S}\ref{sect:disc}, we discuss our results.

\section{Cyclotron Scattering in Strong Magnetic Fields}

\label{sect:reso}

In this paper, we assess the hypothesis that the lines 
exhibited at $\approx$ 20 keV in GRB870303 S1 and
$\approx$ 20 and 40 keV in GRB870303 S2 were
formed within a strong magnetic field ($B \sim$ 10$^{12}$ G).
We make the further assumption that the lines were formed on
the surface of, or near, a neutron star, the only
astronomical object where such strong field strengths have been observed.
We assess the hypothesis by generating spectra with a Monte
Carlo code that numerically treats radiation transfer in
strong fields (Wang et al.~1988),
and fitting these spectra to the observed data.
In this section, we describe the spatial geometry
of the line-forming region, and then summarize the physics of
radiation transfer included in the Monte Carlo code.
We then provide examples of spectra generated with the code,
so as to build the reader's intuition before we discuss the
results of model fitting in {\S}\ref{sect:appl}.

\subsection{Spatial Geometry of the Line-Forming Region}

\label{sect:geom}

We assume that line formation occurs in a plane-parallel slab of
electrons, which has infinite horizontal extent and height similar
to a neutron star atmospheric scale height ($h \sim $10$^{-5}R_{NS}$).
This slab contains an electron-proton plasma,
threaded with a magnetic field ${\vec B}$ which is
oriented at angle $\Psi$ with respect to the slab normal ${\hat n}$
(Figure \ref{fig:coord}).\footnote{
Wang et al.~(1989a) consider only $\Psi$ = 0 in their fits to the {\it Ginga}
data of GRB880205.}
We neglect variations in the magnitude and direction of ${\vec B}$
within the slab.

Continuum photons are injected into the slab at a source plane,
travel through it, and emerge from one of its faces.  
We refer to photons that emerge from the top of the slab as ``transmitted"
and those that emerge from the bottom as ``reflected."  Transmitted photons
reach the observer, whose orientation relative to the slab is
given by $\theta$, the polar angle from the slab normal
(or equivalently, $\mu = \cos{\theta}$), and
$\phi$, the azimuthal angle from the projection of ${\vec B}$ onto 
the slab (Figure \ref{fig:coord}).

The location of the source plane 
with respect to the slab determines the geometry of the system.
We apply the nomenclature ``1-$x$" to denote geometries;
1 and $x$ (a number)
represent the relative values, above and below the source plane, of the
electron column density $N_{\rm e}$ between the photon source plane and the
observer.
We examine two geometries in this work (Figure \ref{fig:geom}).
The 1-0 geometry is a slab illuminated from below; this geometry is used
by Wang et al.~(1989a).
This geometry represents a line-formation region
physically separated from an isotropically-emitting source of continuum 
photons, e.g.~an illuminated flux tube.  
We assume that the reflected photons return to the neutron star surface,
where they are thermalized, i.e.~absorbed by non-resonant inverse
magnetic bremsstrahlung.
We use a 1-1 geometry to model line formation in an isothermal
semi-infinite neutron star atmosphere.  Slater et al.~(1982) 
and Wang et al.~(1988) determine that the mean number of scattering events 
between the
source and the top edge that a resonant photon experiences prior to its
escape approaches a limiting value as the atmosphere becomes semi-infinite
(i.e.~a 1-$\infty$ geometry).  
The number of scattering events experienced in the 1-1 geometry is 
within $\approx$ 10\% of its limiting value, while the number experienced
in the 1-4 geometry (used, e.g., in Freeman et al.~1992) matches the 
limiting value (Isenberg, Lamb, \& Wang 1998b).  
We do not use the 1-4 geometry, despite its
greater accuracy, because the overall number of scattering events
(both resonant and non-resonant)
is approximately three times that using the 1-1 geometry;
the computer time required for the 1-4 simulation increases proportionally.
Also, in the 1-1 geometry, the reflection symmetry of the line-forming
region allows us to limit injection of photons to a hemisphere facing the
observer.  The sum of the resulting transmitted and reflected spectra is
equivalent to the spectrum which would result 
from spherically isotropic photon injection.
By using hemispherical input, we reduce the amount of computer time per
simulation by a factor of two.

\subsection{The Physics of Resonant Cyclotron Scattering}

\label{sect:phys}

Here, we summarize the physics of strong field
radiation transfer that is incorporated
into the Monte Carlo code.  Unless we specify otherwise, the reader may
find more details on any of the elements of the code described below
in Wang et al.~(1988).

We assume that the line-forming plasma is cold: $kT_{e,\parallel} \ll 
E_B = \frac{{\hbar}e}{{\gamma}m_ec}B \approx 
\frac{{\hbar}e}{m_ec}B$, where $\gamma$ is the Lorentz factor ($\approx$ 1
for the electron energies assumed in this work), and the symbol
$\parallel$ indicates that it is only the energies
parallel to ${\vec B}$ that follow the (continuous)
Maxwell-Boltzmann distribution.
The allowed electron energies perpendicular to the field are the
Landau levels.  The spacing of these levels is assumed to be 
much larger than typical electron energies, so that
they are not collisionally populated,
and the photon densities are assumed to be small,
so that they are also not radiatively populated.
Thus, in each scattering event,
the initial and final electron energy state is the Landau ground state
($n$ = 0).

Electrons which are excited to states $n \geq$ 2 
may either deexcite directly to the ground
state (in which case the photon has undergone
{\it resonant} scattering), or it may reach the ground state
via intermediate excitation levels ({\it Raman} scattering).  
Raman scattering is dominant in weak fields ($\frac{B}{B_c} \ll$ 1,
where B$_c$ is the critical field strength, 4.4$\times$10$^{13}$ G,
at which the first harmonic energy matches the electron mass energy
511 keV).
For instance, for the magnetic field strengths considered in this
paper, electrons excited to $n$ = 2 have probability
$\approx 1-\frac{E_B}{m_ec^2}$ = 1$-\frac{B}{B_c} \approx$ 0.95
of deexciting to the first excited state (Daugherty \& Ventura 1977).
When this occurs, the incident photon with energy $E \approx 2E_B$
is destroyed, and two photons with energies $E \approx E_B$ are 
{\it spawned}.
The second and higher harmonics thus have an absorption-like line profile
in weak fields, while the spawned photons act to
reduce the equivalent width of the first harmonic line from the
value it would have had if Raman scattering did not occur.
Because the equivalent widths of the second and higher harmonic lines 
are $\propto (\frac{B}{B_C})^{n-2}$, the code does not
treat excitation to states $n \geq$ 4.

The code uses scattering cross-sections provided by
Herold (1979) and Daugherty \& Ventura (1977).
Herold states how to formulate
exact relativistic scattering cross-sections, and provides 
leading-order expressions for the case where the initial
and final electron states are the Landau ground state,
valid in the limit
\begin{equation}
\left(\frac{E}{m_e c^2}\right)^2 \frac{B_c}{B} \ll 1 .
\end{equation}
The code uses these
expressions, which are valid in the energy and magnetic field strengths
regimes which are of interest in this work.  However, these
expressions make no provision for the natural line width of the 
resonances, and as a consequence the scattering amplitudes diverge
at the resonances.  To correct the divergence, the natural line-width,
${\Gamma}_n = n\frac{4{\alpha}E_B^2}{3m_ec^2}$, is added 
(Wasserman \& Salpeter 1980, Harding \& Daugherty 1991,
Graziani 1993; $\alpha$ is the fine-structure constant, and the
line-width is given in units of energy).
The code also uses
second and third harmonic resonant scattering cross-sections
derived in the non-relativistic limit by Daugherty \& Ventura,
who use Fermi's Golden Rule.
The first harmonic cross-section 
given by Daugherty \& Ventura does not as accurately portray 
the line wing profile, hence the Herold cross-section, with 
finite line-width added, is used instead.

A problem with this ``hybrid" procedure which mixes exact relativistic
resonant cross-sections and non-relativistic higher harmonic
cross-sections is that in the
treatment of electron-photon scattering, there is the implicit approximation
\begin{equation}
\left\vert \sum_i a_i \right\vert^2 \approx \left\vert a_{0
\rightarrow 0} + a_{0 \rightarrow 1 \rightarrow 0} \right\vert^2 +
\sum_{i\neq 0 \rightarrow 0, \atop 0 \rightarrow 1 \rightarrow 0} \left\vert a_i
\right\vert^2\ ,
\label{BasicAssump}
\end{equation}
where $a_i$ is the matrix element for the $i^{th}$ scattering channel.
This approximation is exact for scattering at the first harmonic both in the
line core (within a few Doppler widths of line center) and wings (far from
the line center).
For higher harmonic scattering, this
approximation is only valid near the line center.
If there is significant scattering at energies between the first and second
harmonics, then the matrix cross-term which is missing from this expression
may be important.
Consequently, we limit use of the code to modeling slabs which are not
optically thick at the first harmonic line wings
(see below; also Wasserman \& Salpeter; Lamb et al.~1989):
\begin{equation}
a_1\tau_1~=~{{\Gamma_1}\over{2E_B(\frac{v_{\rm th}}{c})}}~\tau_1~\approx~0.34\left({{kT_{e,\parallel}}\over{\rm keV}}\right)^{-1}\left({{N_e}\over{\rm 10^{21} cm^{-2}}}\right)~\simless~1,
\end{equation}
where $a_1$ is the ratio of the
natural line width to the Doppler energy width for the first harmonic,
$v_{\rm th}$ is the thermal electron velocity along the magnetic
field lines ($= \sqrt{\frac{2kT_{e,\parallel}}{m_e}}$), 
$N_e$ is the column density of electrons in the slab,
and $\tau_1$ is the polarization-, angle-, and frequency-averaged optical depth
in the first harmonic.
For relevant temperatures ($kT_{e,\parallel} \sim$ 5 keV, see below),
$N_e \simless$ 10$^{22}$ cm$^{-2}$.

(If $N_e \simless$ 10$^{22}$ cm$^{-2}$, the line-forming slab
may be moderately optically thick to photons at the harmonic energies but
will be optically thin to continuum photons.  
Thus scattered photons at $\approx$ 20 keV, and not continuum
photons $\simgreat$ 1 MeV, dictate the temperature of the
line-forming region.  Lamb et al.~1990 derive what they dub the
Compton temperature, $T_C$ [which equals $T_{e,\parallel}$
in this work, but not generally], for this
density regime, and find that
[1] slabs reach this temperature 
on timescales $t \sim 10^{-8}$ s, and [2] $kT_C$ is a fraction
[$\approx$ 25\%] of the first harmonic energy.
Thus the lines are narrow.  This result demonstrates the
appropriateness of the originally stated assumption 
that lines are formed in a cold plasma.)

The resonant cross-sections are averaged over the initial polarization 
states and summed over the final states.  Polarization-averaged
cross sections are appropriate for first harmonic scattering in optically
thick media when the vacuum contribution to the dielectric tensor 
dominates the plasma contribution, a condition we may state as
$\left(\frac{n_e}{10^{22} {\rm cm}^{-3}}\right)\left(\frac{B}{10^{12} 
{\rm G}}\right)^{-4} \simless$ 1,
where $n_e$ is the electron number density.
This condition holds for the
physical conditions which we explore here, and is, in fact, less limiting
than the requirement that line-forming regions be optically thin in the
line wings.

The line profile at each harmonic is created by making a Lorentz transformation
to the lab frame, and averaging over $f(p)$, the one-dimensional electron
momentum distribution (along ${\vec B}$).
The total scattering profile is assumed to be
the sum of the profiles for the individual Landau levels (including the
continuum contribution).  Profiles are averaged over azimuthal angle,
$\phi$, despite the fact that for $\Psi \neq$ 0, azimuthal 
symmetry is broken.  The $\phi$-dependent part of the scattering cross 
section is only significant in the line wings and continuum, and in the
present work we considerly only line-forming regions which are not 
optically thick in the line wings.

While we have used the terminology ``line core" and ``line wings,"
whose meanings are intuitively well known, we can define them mathematically
as fulfilling the conditions
$\vert\frac{x_n}{\mu_B}\vert \ll 1$ (core) and
$\vert{x_n}\vert \gg 1$ (wings),
where $\mu_B$ is the cosine of the angle between ${\vec B}$ and the
direction of propagation of the photon, and
$x_n \equiv \frac{E - E_n}{E_n(\frac{v_{\rm th}}{c})}$ 
is a dimensionless energy shift
(with the denominator being the Doppler width associated with the 
$n^{\hbox{th}}$ harmonic energy $E_n = nE_B$;
Wasserman \& Salpeter).   In the line core, the thermal
electron distribution dominates the profile so that it is $\propto
\exp\left[-\left(\frac{x_n}{\mu_B}\right)^2\right]$,
while in the wings, the tail of the
Lorentzian distribution dominates so that the profile is 
$\propto a_n x_n^{-2}$,
where $a_n$ is again the ratio of the natural line width
to the Doppler energy width.
If $\mu_B$ = 0 the line wings extend to $x(\mu_B)$ = 0, i.e.~the 
line profile is purely Lorentzian for all $x$.  This is because the
Doppler effect vanishes, to first order in $\frac{v}{c}$, for photons
propagating perpendicularly to ${\vec B}$.
Wasserman \& Salpeter
showed that for the first harmonic, the core-wing
boundary appears at $\frac{x}{\mu_B}
\approx 2.62 - 0.19 \log(\frac{100a}{\vert\mu_B\vert})$.
We refer to the wings at energies below and above the line center as the
red and blue wings respectively.

Lamb et al.~(1989) showed that relativistic kinematics has a significant
effect on the shape of the absorption profile, even in the limits
$E,kT_{e,\parallel} \ll m_e c^2$.  For zero natural line width,
relativistic kinematics prohibits
scattering at the $n^{th}$ harmonic above a cutoff energy
$E_c=\frac{(\sqrt{1+2nb}-1)m_ec^2}{\sqrt{1-\mu_B^2}}$, where
$b \equiv \frac{B}{B_c}$ (Daugherty \& Ventura
1978; Harding \& Daugherty 1991; see the Appendix of Wang et al.~1993
for a physical derivation).
Wasserman \& Salpeter show that for
the physical conditions where electron recoil is important, photons will
escape more readily in the red wing than in the blue.  
This result motivated Lamb et al.~(1989), Wang et al.~(1989b), and
Wang et al.~(1993) to make the simplistic assumption that the
resonant scattering profile
was zero for $E > E_c$, i.e.~to ignore the effects of the finite natural
line width.  For $E \leq E_c$, the effects of finite line width
were included.
The simplification led to the appearance of spikes in some cyclotron
scattering spectra at energies just above that of the first harmonic
(e.g.~as $\mu \rightarrow$ 0 for $\Psi$ = 0).
This spike contained both photons which were scattered
at the first harmonic, and spawned photons resulting from
Raman scattering at the second and higher harmonics, which
then immediately escaped from the slab.
The code used in this work does include
the effect of finite natural line width
for $E > E_c$ so that the resonant scattering profile is now
small but
finite above $E_c$.  With this enhancement, the spikes are smeared
by scattering and no longer appear.
We stress that even though the scattering profile is finite above the
cutoff energy when the effects of natural line width are properly
treated, the profile still falls off sharply above
$E_c$, leading to a strongly asymmetric line shape in some spectra.

\subsection{Examples of Monte Carlo Spectra}

\label{sect:samp}

To help build the reader's intuition
before describing the results of spectral fits to the data of 
GRB870303 S1 and S2, we present examples of spectra produced
with the Monte Carlo code.
In Figures \ref{fig:10samp} and \ref{fig:11samp}, we
show output counts spectra produced with $\Psi$ = 0, 
for the 1-0 and 1-1 geometries respectively.
In Figure \ref{fig:eqsamp} we show spectra produced with
$\Psi = \frac{\pi}{2}$, for the 1-0 geometry (the 1-1 spectra are 
similar and are not shown).
Each spectrum is produced assuming $B_{12}$ = 1.7 and $N_{\rm e,21}$ = 0.6.
The relative strengths of the harmonics in each figure may be understood
as arising from the 
interplay of line-of-sight column density (i.e.~whether we observe
line formation from above or the side) and the angular dependence of the
resonant scattering cross-section of the $N^{th}$ harmonic:
\begin{equation}
\sigma_{N}~\propto~(1~+~\mu_B^2)~(1~-~\mu_B^2)^{N-1} .
\end{equation}
If $\mu_B$ = 0, then the photon travels perpendicularly to ${\vec B}$,
and $\sigma_{N}$ is non-zero for all $N$, and a maximum for all $N >$ 1.
If $\mu_B$ = 1, then the photon travels along ${\vec B}$,
and $\sigma_{N}$ equals zero for all $N$,
except $N$ = 1, for which case the cross-section is maximized.
Radiative decay to
the Landau ground state is an electric dipole transition, 
and the probability of photon emission in a particular direction is
$\propto 1+\mu_B^2$ (Daugherty \& Ventura 1977); the emitted photon
is thus most likely to be emitted in the direction of ${\vec B}$.

If we compare the line profiles in Figures \ref{fig:10samp}-\ref{fig:eqsamp},
we see that they are broadened if the observer is oriented along ${\vec B}$.
This is because scattering will only occur if the energy of the photon,
as calculated in the electron rest frame, matches the resonant energy, i.e.
if $E_{\gamma}(1-\beta\mu_B)$ = $E_{\rm c}$, where $\beta$ = $\frac{v_e}{c}$, 
$E_{\gamma}$ is the photon lab frame energy,
and $E_{\rm c}$ is the line centroid energy for the first harmonic.
Hence if $\mu_B$ = 0, only those photons with {\it exactly} 
the resonant energy can scatter; if $\mu_B$ = 1, then
there is a range of $E_{\gamma}$ such that scattering is possible, so that
the scattering profile is broadened.

If we compare Figures \ref{fig:10samp}a 
and \ref{fig:11samp}a, we see distinctive ``shoulders" arising in
the red and blue wings of the first harmonic in
the 1-1 geometry.
They appear most prominently when the observer is oriented along
the ${\hat n}$ ($\mu \rightarrow$ 1), regardless of the angle $\Psi$
or the geometry (Isenberg et al.~1998b).
Line shoulders are a predicted element of radiation transfer in strong
fields (Wasserman \& Salpeter), and 
many authors discuss their appearance
(Alexander \& M\'esz\'aros 1989, Nishimura
\& Ebisuzaki 1992, Araya \& Harding 1996, Isenberg et al.~1998b; 
cf.~Nishimura 1994, 
whose results contradict those of Alexander \& M\'esz\'aros 
and Isenberg et al.).
If $a_1$, the ratio of the natural line width to the Doppler energy width for
the first harmonic, $\rightarrow 0$,
then the energy of a scattered photon must lie within the regime
of the optically thick line core.  That photon will thus only escape after
$\sim \tau_1^2$ scatters.  If $a_1 \neq$ 0, then with each scatter
there is a probability $\sim a_1$ that a photon
will be redistributed into the optically thin line wings,
so that after $\sim a_1^{-1}$ scatters, that photon will escape.
Photons in the first harmonic core will, on average, scatter at least this
many times for the 1-1 geometry, but not for the 1-0 geometry, 
for the values of $N_e$ considered in this paper.
This is due to the fact that in the 1-0 geometry, photons which scatter
in the line core may cross the source plane, whereupon they are lost from
the calculation.
The size of the shoulders will change depending upon the importance of
Raman scattering in the line-forming region, since this dictates the number of
photons spawned within the first harmonic line core.
Their relative size in the red and blue wings is determined by the
relationship between $T_{e,\parallel}$ and $T_C$;
because we assume these temperatures to be equal,
the equivalent widths of the two shoulders will be equal.

\section{Statistical Methodology}

\label{sect:stat}

In this work, we examine two general classes of models:
the azimuthally symmetric model class, which has
free parameters $B$, $N_e$, and $\mu$ ($\Psi$ = 0 and $\phi$
is undefined); and a general model class which includes $\Psi$ and $\phi$
as additional free parameters.
If the GRB source is a neutron star with a simple dipole field, these two
classes may be interpreted as representing 
line formation at the magnetic pole, or elsewhere.
Hence we will refer to the azimuthally symmetric models as
``polar cap'' models throughout the remainder of this work.
Because many physical processes are known to occur at the magnetic pole
(e.g.~accretion of gas from a companion star onto a neutron star), 
the association of simpler models with the magnetic pole has intuitive appeal.

In this section, we summarize the methods of
statistical inference which we use to fit models from each class to the 
data, to compare the best-fitting models from each class, and to place
limits on values of the free parameters of the models which fit best
overall.  The reader will find fuller descriptions of these methods 
in Papers I and III, and references therein.

\subsection{Model Fitting}

\label{sect:modfit}

We create Monte Carlo spectra at discrete
points on a grid of values for the model parameters $B_{\rm 12}$ and
$N_{\rm e,21}$, and, if applicable, $\Psi$.
We show these values in Table \ref{tab:param}.\footnote{
We can analytically shift the line profiles in
$E$ (or equivalently $B_{\rm 12}$) by up to $\pm$ 10\%, 
with no loss in accuracy, during fits.
}
For each given set of parameter values,
we determine the Compton temperature $T_C$ ($= T_{e,\parallel}$)
by carrying out a series of Monte Carlo calculations for several different
temperatures and determining at which temperature heating
and cooling balance within the slab 
(see, e.g., Freeman et al.~1992, Isenberg et al.~1998b).
Hence, $T_C$ is {\it not} a free parameter of the fit.

Photons emerging from the line-forming region are binned.
We generally use eight bins spanning $\mu$ = [0,1] (where
$\mu = \cos{\theta}$; an exception made in fits to GRB870303 S1).
As for $\phi$, if $\Psi \neq$ 0 and $\Psi \neq \frac{\pi}{2}$,
we may take advantage of
symmetry across the $({\vec B},{\hat n})$-plane to use eight bins spanning
$\phi$ = [0,$\pi$]. 
If $\Psi$ = $\frac{\pi}{2}$, further symmetry across the
plane perpendicular to the $({\vec B},{\hat n})$-plane allows us to reduce 
the number of bins to four, spanning $\phi$ = [0,$\frac{\pi}{2}$].

We find that a spectrum is sufficiently accurate,
i.e.~the counts standard deviation in each output bin is sufficiently low,
if $\sim$ 10$^6$ photons are injected into the line-forming region.
We also find that the number of photons needed to accurately portray the
low-energy line profiles is roughly an order of magnitude
fewer than the number needed to accurately
portray the high-energy continuum ($E \simgreat$ 100 keV).
We cannot ignore the high-energy continuum, because the {\it Ginga} GBD
recorded the energy lost by an impinging photon
(e.g.~a 500 keV photon may have lost only 20 keV while passing through
the GBD, leading to the recording of a count within the energy regime
of interest).
Thus we run the code with continuum
photons sampled at energies $\simless$ 100 keV, and attach to the
resulting spectrum a separately created high-energy continuum
($E \approx$ 100-1479 keV).
Each spectrum is properly weighted so that overall spectrum 
is smoothly continuous at the matching point.

For a given set of input parameters, many spectra
will be generated, one for each binned value of ($\mu$,$\phi$).
In principle, we would assess the goodness-of-fit of each generated spectrum
using the Poisson likelihood function ${\cal L}$, and select that spectrum
for which ${\cal L}$ is maximized.  However, for reasons given below, 
we fall back upon the understanding that
in the limit of a large number of counts $n$ in a bin, we can use
Pearson's $\chi^2$ statistic, an approximation of $L = {\log}\like$:
\begin{equation}
s^2~=~\sum_{i=1}^{N} \frac{(m_i - n_i)^2}{\sigma_i^2} \,.
\end{equation}
The fit extends over $N$ data bins, and
$m_i$ and $n_i$ are the model amplitude and data in bin $i$, respectively.
The best-fit parameters are those for which $s^2$ is 
minimized.\footnote{
We follow the notation of Lampton, Margon, \& Bowyer (1976), reserving
the symbol $\chi^2$ for a statistic which is explicitly sampled from the
$\chi^2$ distribution.
}
We set $\sigma_i^2$ = $m_i$ (``model variance"), and
hereafter denote the fitting statistic as $s_{\rm m}^2$.

\subsection{Model Comparison}

\label{sect:modcomp}

In Paper I, we describe both frequentist and Bayesian methods of 
model comparison, and apply both methods to determine
the frequentist significance of, and the Bayesian odds favoring, a
spectral model with exponentiated Gaussian absorption lines.  The 
calculation of the Bayesian odds generally requires the analyst
to numerically
integrate the Poisson likelihood function ${\cal L}$ over
parameter space.  This can be computationally intensive when the number
of parameters is large.  However, we can determine
the odds analytically if the shape of the likelihood surface in
parameter space is similar to that of a multi-dimensional Gaussian
(the Laplace approximation).
In Paper I, we were able to reparametrize the exponentiated Gaussian
line model so that we could use the Laplace approximation.
Unfortunately, the function
${\cal L}(B,\mu,N_e,\Psi,\phi)$ does not have such the required 
Gaussian shape, nor can we reparametrize the model so that ${\cal L}$
will have this shape.  Thus in this work
we use only the frequentist method of model comparison.

The frequentist comparison of two models, the null hypothesis $H_0$,
and the alternative hypothesis $H_1$, is carried out by constructing a test
statistic $T$, which is usually a function of the goodness-of-fit statistics 
for both models.  There are two probability distribution functions (PDFs) which
indicate the a priori probability that we would observe the value $T$,
which are
computed assuming the truth of $H_0$, and $H_1$, respectively.  The
test significance $\alpha$ is calculated by computing the tail
integral of the $H_0$ PDF from $T$ to infinity.  The resulting number
represents the probability of selecting $H_1$
when in fact $H_0$ is correct; if $\alpha$ is
sufficiently small, we reject $H_0$ in favor of $H_1$.  A common threshold
for rejecting the null hypothesis is $\alpha \leq$ 0.05; in this work,
we use the more conservative threshold $\alpha \leq$ 0.01.
The selection of the threshold value is subjective;
we use a more conservative value than is used normally because, e.g.,
we would demand that any increase in the quality of fit provided
by general models be sufficiently great to justify considering
the ramifications of line formation taking place outside the polar cap.

The preferred model comparison statistic $T$ would be
the ratio of the maximum Poisson likelihoods of $H_1$ and $H_0$,
but then simulations are required to estimate $\alpha$.
We can make an analytic estimate of $\alpha$ that is approximately correct
using the $\chi^2$ Maximum Likelihood Ratio ($\chi^2$~MLR) test
(Eadie et al.~1971, pp.~230-232).  The test statistic is
$T = {\Delta}s_{\rm m}^2 = s_{\rm m}^2(H_0) - s_{\rm m}^2(H_1)$.
The PDF $p({\Delta}s_{\rm m}^2 \vert H_0)$ is then assumed to be
the $\chi^2$ distribution for ${\Delta}P = P_1 - P_0$ degrees of freedom,
where $P_0$ and $P_1$ are the number of free parameters in models
$H_0$ and $H_1$ respectively.
In order to use this test, the simpler model $H_0$
must be nested within the more complicated model $H_1$, i.e.~we
arrive at $H_0$ after setting the extra ${\Delta}P$ 
parameters of $H_1$ to default values (often zero).

For a single dataset, we directly compare fits of the 
three-parameter polar cap model, and five-parameter general model,
to the data.
When jointly fitting two (or more) datasets, the process of model selection
becomes more complicated, as there is more than one model per class.  
For instance, for fits to S1 and S2, there are eight polar cap models,
with the simplest specifying that $(B_{\rm S1},N_{e,{\rm S1}},\mu_{\rm S1}) = 
(B_{\rm S2},N_{e,{\rm S2}},\mu_{\rm S2})$, and the next three simplest
specifying that one of the parameters changes between the epochs of
S1 and S2, etc.  (Analogously, there are 32 models in the general class.)
We illustrate how we select models within a model class 
in Figure \ref{fig:flow}.  We begin by comparing the simplest model ($M_1$) 
with all models that have a greater number of free parameters
($M_2$-$M_{\rm N}$ in Figure \ref{fig:flow}),
computing the significance of the additional parameters
of each alternative model 
($\alpha_{\rm 1,2}$-$\alpha_{\rm 1,N}$).  We then adopt the simplest
alternative model for which
$\alpha_{\rm \chi^2MLR} \leq$ 0.01 as our new null hypothesis $M_{\rm i}$.
(If two or more models with the same number of free parameters fulfills
this criterion, we would select the one with the smallest $s_{\rm m}^2$, or
equivalently, the smallest $\alpha_{\rm \chi^2MLR}$.)
We repeat the comparison process, until either
no alternative model is adopted, or the most complex model is selected.

\subsection{Parameter Estimation}

\label{sect:parest}

In Paper I, we describe both frequentist and Bayesian methods of 
parameter estimation, and apply both methods to determine
the confidence (frequentist) and credible (Bayesian) intervals on
the free parameters of the best-fit 
exponentiated Gaussian absorption line models.
If the surface defined by $s_{\rm m}^2$ in parameter space has
paraboloidal shape, then the $n\sigma$ confidence interval for an 
individual parameter is given by those values of that parameter for
which $s_{\rm m}^2$ = $s_{\rm m,min}^2 + n^2$, where the values
of all other free parameters are allowed to vary to new
best-fit values.  Otherwise, simulations
are needed to determine the appropriate confidence intervals
(Eadie et al., pp.~190-201).
Since the $s_{\rm m}^2$ surfaces for fits in this work do not
have the required paraboloidal shape, we limit ourselves to computing
Bayesian credible intervals.

We may determine a credible interval for a particular
parameter $x$ of model $M$, without reference to the other,
``uninteresting," parameters, collectively denoted $x'$, by 
marginalizing the Bayesian posterior function $p(x,x' \vert D,I)$
over the space of parameters $x'$:
\begin{equation}
p(x \vert D,I) = \int dx' p(x,x' \vert D,I) \propto \int dx' p(x,x' \vert I) p(D \vert x,x',I) \,.
\end{equation}
$D$ and $I$ represent the data and background information about the
experiment (such as the detector bandpass) respectively, while 
$p(x,x' \vert I)$ is the prior probability (a quantitative statement
of our state of knowledge about the relative probability of each
possible value of $x$ and $x'$ before the data $D$ are examined).
$p(D \vert x,x',I)$ is simply the likelihood, ${\cal L}(x,x')$,
which is assumed to be $\propto \exp(-s_{\rm m}^2/2)$.  If the prior
is assumed to be uniform, then $p(x \vert D,I)$ is simply proportional
to ${\cal L}(x,x')$.  The credible interval is then the range $[x_1,x_2]$
such that
\begin{equation}
z~=~\frac{\int_{x_1}^{x_2} dx~p(x \vert D,I)}{\int_{{\rm all}~x} dx~p(x \vert D,I)} \,,
\end{equation}
where $z$ is the desired probability content (e.g.~0.683 
for 1$\sigma$ bounds),
and $p(x_1 \vert D,I) = p(x_2 \vert D,I)$.  (Note that 
if the likelihood surface 
is not well-behaved and exhibits a number of modes, then the
credible ``interval" may actually consist of a number of intervals.)

We note two items which affect the interpretation of these regions.
First, 
our use of the Monte Carlo code is limited to $N_{\rm e,21} \simless$ 10.
The upper limit can potentially cut off highly probable
regions of parameter space, affecting the computation of credible intervals.
Second, it is not computationally feasible to create models using all $\Psi$
values when creating credible intervals 
for the parameters of the general model.
Thus the quoted marginal distribution for each model parameter other
than $\Psi$ is of the form $p(x,\Psi = \Psi_{\rm best~fit} \vert D,I)$, 
and we do not compute the marginal distribution for $\Psi$ itself.

\section{Application of the Cyclotron Scattering Model to GRB870303}

\label{sect:appl}

\subsection{The Data}

\label{sect:data}

The Los Alamos/ISAS Gamma-Ray Burst Detector (GBD; Murakami et al.~1989) 
on {\it Ginga} detected GRB870303 at 16:23 UT on 3 March 1987.
Figure \ref{fig:time} shows burst-mode time history data for the GBD
Proportional Counter (PC), which covered 1.4-23.0 keV, and 
the GBD Scintillation Counter (SC), which covered 16.1-335 keV.  
The GBD continuously recorded burst-mode data at 0.5-second
intervals.  These data were not stored in memory until a burst was detected,
at which time the data from 16 seconds
prior to the burst trigger until 48 seconds after the burst trigger
were stored.  The background-subtracted GRB870303 spectra 
(shown in Figure \ref{fig:data}) exhibit lines in two time intervals,
denoted Spectra 1 and 2 (S1 and S2) by Graziani et al.~(1992).
S1 is a 4 s spectrum with photon fluence 1.3 $\times$ 10$^{-6}$
erg cm$^{-2}$ in the bandpass 50-300 keV
(estimated from the best-fit model of Paper I),
that exhibits a saturated line at $\approx$ 20 keV.
S2 is a 9 s spectrum with fluence
4.5 $\times$ 10$^{-6}$ erg cm$^{-2}$ exhibiting
harmonically spaced lines at $\approx$ 20 and 40 keV.
The midpoints of S1 and S2 are separated by 22.5 s.

The burst detector on {\it Pioneer Venus Orbiter (PVO)}
also observed GRB870303, allowing the use photon time-of-arrival
information to limit the possible source location to an annulus upon the sky.
Yoshida (private communication, correcting Yoshida et al.~1989)
reports this annulus to lie in the range
11.2$^{\circ}$ $\simless \theta_{\rm inc} \simless$ 57.6$^{\circ}$.
Since the shape and amplitude of a model counts spectrum that is derived
from a given photon spectrum depends sensitively on 
$\theta_{\rm inc}$, we treat this angle as a freely varying
model parameter in Paper I.  However,
testing the cyclotron scattering hypothesis for more than one
angle of incidence is not (currently) computationally feasible. 
Thus, we assume $\theta_{\rm inc}$ = 37.7$^{\circ}$ in this work;
this is the angle assumed by others who analyze
GBD data (e.g.~Murakami et al.~1988, Fenimore et al.~1988,
Wang et al.~1989a, Graziani et al.~1992, Graziani et al.~1993).

\subsection{Results: GRB870303 S1}

\label{sect:s1}

In Paper I, we describe fits to these data using exponentiated 
Gaussian absorption lines.
We fit the data in PC bins 8-15 and SC bins 2-31 with a
power-law model ($\frac{dN}{dE} \propto E^{-1.55}$), onto which is multiplied
one line with equivalent width $W_E \approx$ 10 keV.
For this model, $s_{\rm m}^2$ = 22.3 for 34 degrees of freedom (dof).
The frequentist significance of this line is 1.2 $\times$ 10$^{-5}$,
and the Bayesian odds favoring the model with a line is 120-1.  
(These numbers differ from those given above, and in Paper I,
in that $\theta_{\rm inc}$ is assumed to be 37.7$^{\circ}$.)

When we use the 1-0 geometry, the mode of the polar cap model places the
observer along ${\vec B}$, where
the first harmonic equivalent width is maximized, while
higher harmonic equivalent widths are minimized.
For this model, $s_{\rm m}^2$ = 33.0 (33 dof), with
$B_{\rm 12}$ = 1.78, $N_{\rm e,21}$ = 5, and 0.875 $\leq \mu \leq$ 1.
This fit compares unfavorably with the best phenomenological model
fit ($s_{\rm m}^2$ = 22.3).
We decrease ${\Delta}\mu$ and examine $\mu \approx$ 1,
to maximize the first harmonic strength and to minimize the effect of the
weak second harmonic.
If we use a bin-width ${\Delta}\mu = \frac{1}{256}$, $s_{\rm m}^2$ = 26.7,
with $N_{\rm e,21}$ = 10 (the upper limit)
(see Figure \ref{fig:s110}, and Tables 
\ref{tab:modcomp} and \ref{tab:s1bayes}).
Smaller bin widths do not lead to further reduction in $s_{\rm m}^2$.  
We note that tests using values of $N_{\rm e,21}$ above the upper limit
indicate that $s_{\rm m}^2$ increases,
due to increasingly apparent shoulders in
the red and blue wings of the line.
This fit indicates that we observe line formation from a privileged position 
above the polar cap if the line-formation region is illuminated isotropically.

The difference ${\Delta}s_{\rm m}^2$ between this best-fit and that of 
Paper I is due to the fact that at the polar cap, 
we cannot create a
first harmonic line with a sufficiently large equivalent width
($W_{\rm E} \approx$ 10 keV).
To determine the equivalent width of the best-fit line model,
we model the line profile using
three exponentiated Gaussians, two representing the shoulders and one
representing the line core:
\begin{equation}
{{dN}\over{dE}}(E)~=~C(E)~\times~\Pi_{\rm i=1}^{\rm i=3} \exp(-\beta_i \exp[-{{(E-E_{\rm c,i})^2}\over{2\sigma_i^2}} ] ) .
\end{equation}
$C(E)$ is the continuum photon flux at energy $E$, 
$\beta_{\rm 2} = \beta_{\rm core} >$ 0, and $\beta_{\rm 1}$ and
$\beta_{\rm 3} <$ 0.
The equivalent width of the line is
$W_{\rm E} \approx$ 5 keV, a value that depends only 
weakly upon $N_{\rm e}$ in the vicinity of the mode.  
(This value is roughly
2$\sigma$, or ${\Delta}s_{\rm m}^2 \approx$ 4, away from the best-fit
phenomenological value $W_{\rm E} \approx$ 10 keV.)
The equivalent width of the line core is $\approx$ 7 keV, while each of the
shoulders has equivalent width $\approx -1$ keV.

One can increase the equivalent width is to limit photon input into the 
line-forming region to a cone with axis parallel to ${\hat n}$.
This decreases the number of photons spawned into the first harmonic line
core by greatly reducing the angle-averaged cross-section of the second and
third harmonics.
Such a model would be consistent with the
hypothesis that photons are beamed during continuum formation
(Share et al.~1986;
Ho {\&} Epstein 1989; Ho, Epstein, {\&} Fenimore 1990; Dermer 1990).
A typical beam cone opening angle for 20-40 keV continuum photons is
$\theta_{\rm cone} \approx$ 45$^{\circ}$.
Such a model is also consistent with the 
hypothesis that the line-formation region is suspended far above the polar cap 
(Dermer \& Sturner 1991, Sturner \& Dermer 1994).\footnote{
Another, similar, hypothesis that cannot be directly tested with static
models is that the line-forming region
is driven off the
star in a wind by radiation pressure (Miller et al.~1991, 1992; 
Isenberg et al.~1998a).}
In this case, only photons which travel along the magnetic
polar axis will interact with the localized region of line formation
($\theta_{\rm cone} \rightarrow$ 0$^{\circ}$).
We compute spectra for these two test cases,
assuming $B_{\rm 12}$ = 1.76 and $N_{\rm e,21}$ = 0.6 (6\% of the best-fit
value for isotropic photon input).
We find that both hypotheses are consistent with the data;
in both cases, 
$s_{\rm m}^2$ = 23.0, and the line equivalent width is increased to
$\approx$ 10 keV.

We find that the general cyclotron scattering model, with $\Psi$ and $\phi$ as
additional free parameters, fits to the data better than the polar cap model,
with $s_{\rm m}^2$ = 23.6 (31 dof).
The best-fit for this model occurs for $\Psi = \frac{\pi}{2}$;
we may interpret this as meaning that
line formation occurs at the neutron star magnetic equator.
The observer is oriented along
${\vec B}$ (i.e., is looking along the surface of the slab);
the line-of-sight opacity in the first harmonic is thus maximized
(see Figures \ref{fig:10samp}a and \ref{fig:eqsamp}b),
and the equivalent width of the line can reach the desired value
$\approx$ 10 keV.

Comparing the polar cap and general models, we find that
the significance of the extra two parameters in the general model is
$\alpha_{\rm \chi^2MLR}$ = 0.21; we select the polar cap model.
It is however
of great theoretical interest that models where lines form far from
a polar cap can fit the data so well, a point to which we return in
{\S}\ref{sect:gal}.

For the 1-1 geometry, the polar cap and general models fit to the data with
$s_{\rm m}^2$ = 37.0 (33 dof) and 24.4 (31 dof), respectively.  
Polar cap model spectra do not fit well to the S1 data because of the 
presence of either line shoulders (which have maximum size at $\mu$ = 1) or a
strong second harmonic line (which has maximum strength at $\mu$ = 0).
Comparing the polar cap and general models, we find that
the significance of the extra two parameters of the general model is 
$\alpha_{\rm \chi^2MLR}$ = 1.6 $\times$ 10$^{-3}$; we
select the general model (see Figure \ref{fig:s111},
and Tables \ref{tab:modcomp} and \ref{tab:s1bayes}). 
The S1 data thus indicate strongly that line formation does not
occur at the magnetic polar cap, within the context of a semi-infinite
atmosphere.

\subsection{Results: GRB870303 S2}

\label{sect:s2}

In Paper I, we describe fits to these data using exponentiated 
Gaussian absorption lines.
We fit the data in PC bins 7-15 and SC bins 2-31 with a
power-law model that is cut off exponentially
($\frac{dN}{dE} \propto E^{-1.22} \times \exp[\frac{E({\rm keV})}{137.3}]$).
To model the lines, we multiply onto this continuum
two lines with equivalent width $W_E \approx$ 2.3 and 4.6 keV.
For this model, $s_{\rm m}^2$ = 33.6 (35 dof).
The frequentist significance of these lines is 4 $\times$ 10$^{-5}$,
and the Bayesian odds favoring the model with lines is 17-1.
(Again, these numbers differ from those given above, and in Paper I, 
in that they are derived assuming $\theta_{\rm inc}$ = 37.7$^{\circ}$).

For the 1-0 geometry, the polar cap and general models 
fit to the data with $s_{\rm m}^2$ = 33.6 (33 dof)
and 31.7 (31 dof), respectively. We find that the relatively weak lines
of S2 can form at any location on a magnetized neutron star, 
i.e.~for each $\Psi$, there exists some values of $\mu$ and $\phi$ such that
the data are fit well with the cyclotron scattering model.
The significance of the two extra parameters of the general model is
$\alpha_{\rm \chi^2MLR}$ = 0.39; we select the polar cap model (see 
Figure \ref{fig:s210}, and Tables \ref{tab:modcomp} and \ref{tab:s2bayes}).

For the 1-1 geometry, the polar cap and general models 
fit to the data with $s_{\rm m}^2$ = 33.9 (33 dof) and 32.9 (31 dof),
respectively. Again, as is the case for the 1-0 geometry, we find 
that the S2 lines can form anywhere on a magnetized neutron star.
The significance of the two extra parameters of the general model is
$\alpha_{\rm \chi^2MLR}$ = 0.61;
we select the polar cap model (see Figure \ref{fig:s211}, 
and Tables \ref{tab:modcomp} and \ref{tab:s2bayes}).
We note that $\mu \rightarrow$ 0 for this geometry, unlike for the
1-0 geometry; this is directly related
to the presence of shoulders in spectra as $\mu \rightarrow$ 1.

\subsection{Results: the Combined (S1+S2) Data}

\label{sect:j}

In Paper I, we describe fits to combined (S1+S2) data using exponentiated 
Gaussian absorption lines.  We select
a line model parameterized by the first harmonic energy ($E_{\rm c} \approx$
21.6 keV),
two first harmonic equivalent widths ($W_{\rm E,1,S1} \approx$ 
10.3 keV and $W_{\rm E,1,S2} \approx$ 2.5 keV), 
and one second harmonic equivalent width ($W_{\rm E,2,S1} = W_{\rm E,2,S2} 
\approx$ 3.1 keV).
For this model, $s_{\rm m}^2$ = 58.0 (68 dof).
The frequentist significance of the lines, evaluated jointly, is 
3.1 $\times$ 10$^{-8}$,
and the Bayesian odds favoring the model with lines is 8080-1.
(Again, these numbers differ from those given above, and in Paper I, 
in that they are derived assuming $\theta_{\rm inc}$ = 37.7$^{\circ}$).

As noted in {\S}\ref{sect:modcomp}, there are eight polar cap models for
joint fits to two datasets, which have a minimum of three, and maximum of
six, free parameters.  After selecting those models with the lowest values
of $s_{\rm m}^2$ for each possible number of free parameters, we find
$s_{\rm m}^2$ = 70.8 (69 dof), 66.5 (68 dof), 64.4 (67 dof), and 60.3 (66 dof),
respectively for the 1-0 geometry (Table \ref{tab:modfit}).
In no case is ${\Delta}s_{\rm m}^2$ sufficiently large so that we may 
reject the null hypothesis; the significance of the 3 additional parameters of
the most complex model is $\alpha_{\rm \chi^2MLR}$ = 0.015.  
The additional reduction in $s_{\rm m}^2$ that would be
necessary for $\alpha_{\rm \chi^2MLR}$ to 
reach our selection criterion of 0.01 is $\approx$ 0.9.
Because the variation of $s_{\rm m}^2$ between 
Monte Carlo manifestations of input models is $\simless$ 1,
there is only a small possibility that the 6-parameter model would 
be selected more often that the 3-parameter model
if we were create and fit an infinite number of Monte Carlo spectra.
Thus we select the simplest 3-parameter polar cap model
(see Tables \ref{tab:modcomp} 
and \ref{tab:jcred10}, and Figure \ref{fig:0jcred10}; 
the credible regions, which we do not present,
are similar to those in Figure \ref{fig:s210}).

Fits of the 32 general models 
(with minimum five, and maximum ten, free parameters)
to the combined (S1+S2) data for the 1-0
geometry yield
$s_{\rm m}^2$ = 69.5 (67 dof) for the simplest 5-parameter model, 
and 59.6 (66 dof) for the 6-parameter model for which
$\mu_{\rm S1} \neq \mu_{\rm S2}$ (Table \ref{tab:modfit}).
The significance of the
additional parameter is $\alpha_{\rm \chi^2MLR}$ = 1.7 $\times$ 10$^{-3}$;
we select the 6-parameter model.
No more complicated model is favored over this model
(see Tables \ref{tab:modcomp} and \ref{tab:jcred10},
and Figure \ref{fig:90jcred10}; in this figure, we show only the
credible regions for the parameter subset $[\mu_{\rm S1},\mu_{\rm S2}]$).

Comparing best-fit
polar cap and general models, we find that
${\Delta}s_{\rm m}^2$ = 11.2 for ${\Delta}P$
= 3.  The significance of the additional parameters 
is $\alpha_{\rm \chi^2MLR}$ = 0.01.  {\it We conclude that on the
basis of the current evidence, we cannot choose between these models.}

The application of the polar cap models to the combined (S1+S2) 
data for the 1-1 geometry yields best-fits
of $s_{\rm m}^2$ = 73.7 (69 dof) for the simplest model, through
$s_{\rm m}^2$ = 71.3 (66 dof) for the most complex model;
$\alpha_{\rm \chi^2MLR}$ = 0.49 (Table \ref{tab:modfit}).  
We select the simplest model 3-parameter
model (see Tables \ref{tab:modcomp} and \ref{tab:jcred11}, and 
Figure \ref{fig:0jcred11};
the credible regions, which we do not present,
are similar to those in Figure \ref{fig:s210}).

The application of the general class of models yields 
$s_{\rm m}^2$ = 71.6 (67 dof), 
65.0 (66 dof; $\mu_{\rm S1} \neq \mu_{\rm S2}$), 
and 59.9 (65 dof; $\mu_{\rm S1} \neq \mu_{\rm S2}$ and 
$\phi_{\rm S1} \neq \phi_{\rm S2}$), respectively; fitting models with eight
or more free parameters offers little further reduction in $s_{\rm m}^2$.
The significance
of the first additional parameter is $\alpha_{\rm \chi^2MLR}$ = 0.01; we
cannot choose between the 5- and 6-parameter models.  We describe the 
result of choosing each of these models in turn:
\begin{enumerate}
\item{If we choose the 5-parameter model, we would compare it to the 
7-parameter model and find $\alpha_{\rm \chi^2MLR}$ = 2.9 $\times$ 10$^{-3}$.
We would select the 7-parameter model.
We would compare this model to the best-fit 3-parameter polar cap 
model and determine that
${\Delta}s_{\rm m}^2$ = 13.8 for ${\Delta}P$ = 4; $\alpha_{\rm \chi^2MLR}$ =
8.0 $\times$ 10$^{-3}$.  
The amount by which ${\Delta}s_{\rm m}^2$ surpasses the
$\alpha_{\rm \chi^2MLR}$ = 0.01 criterion is $\approx$ 0.5. 
This is insufficient to conclude that the data select the more complex model,
and we would conclude that we cannot select between models.}
\item{If we choose the 6-parameter model, we would compare it to the 
7-parameter model 
and find $\alpha_{\rm \chi^2MLR}$ = 0.024.  We would select the
6-parameter model.
We would compare this
model to the best-fit 3-parameter polar cap
model and determine that
${\Delta}s_{\rm m}^2$ = 8.7 for ${\Delta}P$ = 3; $\alpha_{\rm \chi^2MLR}$ = 0.034.
We would select the 3-parameter polar cap model.}
\end{enumerate}
Thus, again, {\it we find that we cannot select between the 
simplest 3-parameter polar cap model and a 7-parameter general model
on the basis of the current evidence.}
(Best-fit values, credible intervals, etc., for the 7-parameter general model
may be found in Tables \ref{tab:modcomp} 
and \ref{tab:jcred11}, and Figure \ref{fig:90jcred11};
note that in this figure,
we show only the credible regions for the parameter subset
$[\mu_{\rm S1},\mu_{\rm S2},\phi_{\rm S1},\phi_{\rm S2}]$.)

\subsection{Limits on Neutron Star Rotation Period}

\label{sect:rot}

The combined (S1+S2) data indicate that if
$\Psi$ = $\frac{\pi}{2}$, the best-fit values of $\mu$ and/or $\phi$ 
change during the 22.5 s between S1 and S2 for both the 1-0 and 1-1
geometries.
In other words,
the orientation of the observer relative to the line-forming region
changes with time.  The simplest way to interpret this 
result is to invoke neutron star rotation
(see, e.g., Lamb, Wang, \& Wasserman 1992).

For the 1-0 geometry, the best-fit model
has $\mu$ changing as a function of time,
but not $\phi$ (Table \ref{tab:jcred10}).
Because of the symmetries which exist for the particular
case of $\Psi = \frac{\pi}{2}$, $\phi \in [-\frac{\pi}{8},\frac{\pi}{8}]$ or 
$\in [\frac{7\pi}{8},\frac{9\pi}{8}]$.  However, if $\phi \neq 0$ or
$\pi$, then $\phi$ must change as $\mu$ changes, violating the
original model assumption that $\phi$ does not change.
This condition leads us to posit a model of the rotating neutron star
in which:
(1) the rotation axis is perpendicular to the observer's
line-of-sight;
(2) the star may rotate in either the clockwise or
counter-clockwise directions;
(3) the magnetic axis lies within the neutron star-observer
plane; and
(4) line formation is localized and occurs only where the magnetic equator
intersects the neutron star-observer plane.
This model is shown pictorially in Figure \ref{fig:nsrot}.
We assume that the neutron star rotates
less than once between S1 and S2, since otherwise we could 
expect to see evidence for line(s) during the time period separating the two
spectra.

The rotation period is given by
\begin{equation}
t_{\rm rot}~=~\frac{2{\pi}}{\vert \theta_{\rm S1} - \theta_{\rm S2} \vert} \times 22.5 {\rm s} \,,
\end{equation}
where $\theta_{\rm S1} = \cos^{-1}(\mu_{\rm S1})$ and
$\theta_{\rm S2} = \cos^{-1}(\mu_{\rm S2})$.
Generally, there are two
possible rotation periods that we may derive, depending upon the direction
of the star's rotation.
However, for $\Psi = \frac{\pi}{2}$, there are four possible
times (which we hereafter denote as $t_{\rm rot,n}$),
because of the symmetry between $\phi = 0$ and $\phi = \pi$ (see
Figure \ref{fig:nsrot}).

To define credible intervals for the rotation periods,
we repeatedly sample values of the cosines
$\mu_{\rm S1}$ and $\mu_{\rm S2}$ from the 
posterior distribution
$p(\mu_{\rm S1},\mu_{\rm S2} \vert {\Psi}=\frac{\pi}{2},{\phi}=0~{\rm or}~{\pi},D)$
and determine for each sampled cosine pair
the possible rotation periods $t_{\rm rot,n}$.
(This distribution is 
slightly different from that shown in Figure \ref{fig:90jcred10}, which 
includes marginalization over $\phi$.)
We determine probability distributions from 
10$^6$ values of $t_{\rm rot,n}$, and use these distributions
to estimate the 1, 2, and 3$\sigma$ credible intervals given
in Table \ref{tab:rot}.

For the 1-1 geometry, both
$\mu$ and $\phi$ change as a function of time,
which greatly complicates the derivation of rotation periods.
In particular, the orientations of the magnetic
and rotation axes, ${\hat H}$ and ${\hat R}$, can be arbitrary.
In the Appendix, we describe a Bayesian method which we use to
derive rotation periods once we sample values of $\mu$ and $\phi$ for
S1 and S2.
We determine the probability distributions from
10$^4$ values of $t_{\rm rot,n}$, and use these distributions
to estimate the 1, 2, and 3$\sigma$ credible intervals given
in Table \ref{tab:rot}.

\section{Discussion}

\label{sect:disc}

In this paper, we demonstrate that
Monte Carlo models of cyclotron scattering in the strong
magnetic field ($B \sim$ 10$^{12}$ G) of a galactic neutron star
can successfully account for the positions and strengths
of the lines exhibited at $\approx$ 20 keV in GRB870303 S1 and 
$\approx$ 20 and 40 keV in GRB870303 S2.
Our results are robust to changes in slab geometry and magnetic field 
orientation.
Given that physically rigorous models of line formation within
the cosmological burst environment do not yet exist
(Stanek et al.~1993 and Ulmer \& Goodman 1995, e.g., invoke femtolensing),
our results, when
paired with the successful fits of cyclotron scattering models to
the data of GRB880205 by Wang et al.~(1989a) and Freeman et al.~(1992),
provide strong support for the hypothesis that
that some (though not all) GRBs are galactic in origin.

\subsection{Galactic Source Population}

\label{sect:gal}

\subsubsection{Static Line Formation at the Polar Cap}

We model static line formation at the magnetic polar cap 
of a neutron star with a simple dipole field 
by setting ${\vec B}$ parallel to the slab normal ${\hat n}$, i.e.~by 
setting $\Psi$ = 0.
We find that while we can easily fit this model to the data of GRB870303 S2, 
acceptable fits to the data of GRB870303 S1 can be made only with difficulty 
(1-0 geometry), or cannot be made at all (1-1 geometry).
For the 1-0 geometry, we conclude that either the observer
is oriented directly along ${\vec B}$ ($\mu \rightarrow$ 1), 
or that the line-formation region is
levitating far above the polar cap, where it presents a small solid
angle to continuum photons, with the result that the equivalent widths
of the second and third harmonics are reduced relative to the width of the 
first harmonic.
The latter conclusion is consistent with the suggestion by
Dermer \& Sturner (1991), and Sturner \& Dermer (1994),
that a scattering atmosphere with a geometry similar to the 1-0 may be 
present in both accretion-powered pulsars and GRBs, with layers that are
optically thick to line scattering but thin to continuum
scattering forming within a few stellar radii of the neutron star center.
For the 1-1 geometry, we find that prominent emission-like
shoulders on either side of the first harmonic prevent a
good fit to the S1 data if the observer is oriented along ${\vec B}$
($\mu =$ 1); by decreasing $\mu$, we can reduce the magnitude of the shoulders,
but second and third harmonics form, preventing an acceptable fit to 
these data.\footnote{
However, we note that shoulders are not always inconsistent with
observed GRB line profiles:
Freeman et al.~(1992) show that 
the 1-1 polar cap model, with shoulders, fits to the data of GRB880205
better than a 1-0 polar cap model without shoulders; $s_{\rm m}^2$ 
falls from 43.9 to 36.6.}

The results of fits to the S1 data argue against static line formation
at the magnetic polar cap.  If line formation does
indeed occur above the polar cap in an outflowing plasma, 
the neutron star probably
resides within the galactic halo at a distance of $\simgreat$ 50 kpc.
Two lines of reasoning point to this conclusion.
First, as pointed out by Lamb et al.~(1990), a static
polar cap line formation region
will be disrupted on timescales $\sim$ 10$^{-6}$ s unless the
the burst flux is below the magnetic Eddington limit.
At the non-magnetic Eddington limit, the radiative pressure 
exerted by both line and continuum photons 
upon electrons in the line-forming layer
balances the gravitational force upon the protons
which are electrostatically coupled to the electrons;
the magnetic Eddington limit is determined by replacing the
Thomson cross-section of the electron with the resonant cyclotron scattering 
cross-section of the first harmonic.
This lowers the luminosity limit from $\sim$ 10$^{38}$
erg s$^{-1}$ to $\sim$ 10$^{36}$ erg s$^{-1}$.
The upper limit on the distance to the burst source 
with an electron-proton line-forming layer is thus $\sim$ 100 pc.

Second, Loredo \& Wasserman (1998b) determine that the
data of the 3B catalog (Meegan et al.~1996)
is consistent with 
the hypothesis that there is a component of the source population
residing within the galaxy, with that component either being comprised
of dim local halo sources at distances $\simless$ 1 kpc, or
luminous halo sources at distances $\simgreat$ 50 kpc.
They make this conclusion by using Bayesian methods 
(developed in Loredo \& Wasserman 1995) to
compare a model in which
sources residing in a Bahcall-Soneira halo with core size 2 kpc 
(Bahcall \& Soneira 1980) are mixed with cosmological GRB sources,
with the best-fit cosmological GRB source model from
Loredo \& Wasserman (1998a). 
Each class is assumed to contain standard candle bursters, with the
further assumptions for the cosmological bursters that the
comoving burst rate is homogeneous and independent of redshift.
(Loredo \& Wasserman 1998a demonstrate that
this simple cosmological model fits the data as well as
more complicated models with either an inhomogeneous burst rate or
a power-law burst luminosity function.)
The best fit of the halo model is
better than that of the purely cosmological model, 
for both dim local and luminous halo bursters.
The best-fit fractions of
dim local halo sources are 0.59 (64 ms 3B catalog data) and 0.36 
(1024 ms data), and the bursters are limited to distances
$\simless$ 1 kpc;
the respective fractions for luminous halo sources are 0.11 and 0.073,
with inferred distances $\simgreat$ 50 kpc.
However, the 3B data lack the power to tightly constrain the model
parameters, and
thus the credible intervals for each fraction include
zero at a significance level $\simless$ 2$\sigma$, or 95.5\%.
(Despite the lack of constraining power,
the data do not support the bursters residing at distances
intermediate between 1 and 50 kpc.)
Because of this lack of constraining power,
the Bayesian odds favoring the galactic component models are
not large: 2.5 and 6.7 (64 and 1024 ms) for dim local halo sources,
and 0.45 and 0.25 for luminous halo sources.  An odds of
$>$ 10-20 would be considered strong
evidence in favor of an alternative model (see the review by
Kass \& Raftery 1995 and references therein).

We note that if lines 
exhibited by accretion-powered pulsar (APP) spectra are also formed
via cyclotron scattering, the lack of shoulders in
the 12 known APPs with lines (Mihara 1995)
places a lower limit on column densities
that is much larger than the column densities determined in our analyses,
within the context of the 1-1 geometry.
This result is shown by Isenberg et al.~(1998b), who
consider cylindrical line-formation regions with
photon injection along the axis, which represent the canonical model of
the emission region of accretion-powered pulsars.
They show that if $N_{e,21} \simgreat$ 1000, the shoulders disappear.
This implies that even if the 1-1 geometry is an acceptable approximation
for both APPs and GRBs, the mechanisms underlying line formation must
be significantly different.

\subsubsection{Static Line Formation at the Magnetic Equator}

The most theoretically intriguing result of our analyses is the possibility
that line-formation may occur away from the magnetic polar cap (i.e.
for $\Psi \neq$ 0).
The S1 data provide strong evidence that, for the 1-1 geometry, 
line formation does not occur at the polar cap, with the best-fit location
being the magnetic equator ($\Psi$ = $\frac{\pi}{2}$).  The combined (S1+S2) 
data provide marginal evidence, for both
geometries, favoring equatorial line formation.  Line formation at a
magnetic equator may be static even if the burst luminosity greatly exceeds
the critical Eddington luminosity in a strong magnetic field, since the
closed magnetic field lines there may effectively trap the plasma (see, e.g.,
Zheleznyakov \& Serber 1994, 1995).
Thus the assumption of line formation in static layers may be
acceptable even if the progenitor neutron star
resides in the Galactic halo at distances $\simgreat$ 50 kpc.

\subsubsection{Line Formation in an Outflow}

If line formation occurs at the magnetic polar cap of a luminous halo
burster, the line formation region will initially
flow outwards along ${\vec B}$.
In an outflow, the variation of the magnetic field and plasma velocity with
altitude will tend to broaden the lines.  Thus the relevant question
is whether we can still observe narrow lines in the spectra
of distant halo GRBs.  The answer appears to be yes.
Miller et al.~(1991,1992) calculated
the properties of the second and third harmonics, approximating Raman
scattering with cyclotron absorption (see {\S}\ref{sect:cycabs}).
They showed that narrow lines can
be formed at these harmonics, and they
successfully fit outflow absorption models
to the second harmonic of GRB880205.  Chernenko \& Mitrofanov (1995) calculate
the properties of the first harmonic line in an outflow, but they assume 
the line to be formed via
absorption; they find that narrow first harmonic lines are possible.
Isenberg et al.~(1998a) show that the
cyclotron scattering model can create narrow lines at all harmonics.
They treat this problem using a variant of
the Monte Carlo code applied in this work.  They assume a hot spot at the
magnetic pole of a neutron star, and apply the radiation force calculation
of Mitrofanov \& Tsygan (1982) to determine the plasma outflow velocity.
They conclude that cyclotron scattering 
lines can form within a relativistic outflow with properties similar to
those of GRB870303 and GRB880205, provided the hot spot is a small fraction
of the stellar surface ($r_{\rm hot} \simless$ 0.1$R_{\rm NS}$).

A natural consequence of the fact that the line-forming layer may move along
field lines is that the height of the line-forming layer may
change with time, as, for example, the burst intensity or hot spot radius
changes.  A change in height would be accompanied by a change in inferred
magnetic field strength (i.e.~by an inferred shift in line-centroid energy).
Yoshida et al.~(1992) report that the magnetic field 
strength shows a declining trend within the 9 s interval in which the 
line candidates of GRB880205 have maximum significance, which indicates
that the line-forming region is flowing outwards from the neutron star.
We find that
the combined (S1+S2) data are nearly consistent with the hypothesis
that the line-forming region moves {\it inward} during the $\approx$
20 s between line epochs.
Within the context of the most complex 6-parameter polar cap model,
which the data almost favor over the simplest 3-parameter model,
$B_{\rm 12}$ changes from 1.76 to 1.96 from S1 to S2, indicating that S2
line formation might occur closer to the polar cap than S1 line formation.

\subsection{Cyclotron Scattering versus Cyclotron Absorption}

\label{sect:cycabs}

Many authors (e.g.~Fenimore et al.~1988, Graziani et al.~1992) fit data with
the cyclotron absorption model, motivated by its computational 
simplicity and by the fact
that approximating cyclotron Raman scattering by cyclotron
absorption is approximately valid for the second and higher harmonics.  
The cyclotron absorption model spectrum is
\begin{equation}
CA(E)~=~C(E) \exp\left[-\sum_{n=1}^m A_n G_n(E)\right] \,,
\end{equation}
where $C$ is the continuum spectrum, $m$ is the number of harmonics, and
\begin{equation}
G_n(E)~=~\frac{1}{\sqrt{\pi}{\Delta}E_n} \exp\left(-\frac{(E-E_n)^2}{{\Delta}E_n^2}\right)
\end{equation}
is the absorption line profile for the $n^{\rm th}$ harmonic.
The line widths ${\Delta}E_n$ are given by
$E_n\sqrt{\frac{2kT_{e,\parallel}\mu'^2}{m_ec^2}}$,
where $\mu'$ is the cosine of the 
angle between the line of sight and ${\vec B}$.
The line amplitudes $A_n$ are given by
$N_{e,n}^{\rm los}\alpha_n$, where 
$N_{e,n}^{\rm los}$ is the column density along the line of sight, and
\begin{equation}
\alpha_n \approx \frac{5.5\times10^{-20}}{(n-1)!} \left(\frac{n^2}{2}\frac{B}{B_c}\right)^{n-1} (1+\mu'^2) (1-\mu'^2)^{n-1}~{\rm keV~cm}^{-2}
\end{equation}
is the absorption coefficient of the $n^{\rm th}$ harmonic.
Because modeling the first harmonic line with cyclotron absorption is not
valid, we must allow different temperatures and column densities when
fitting to the first, and higher, harmonics.
If, for example, we fit two harmonic lines to the data,
the relationships between the cyclotron absorption model parameters and
the physical parameters of the line-forming region are $E_1 \propto B$,
$A_1 \propto N_{\rm e,1}^{\rm los}$(1+$\mu'^2$) and
$A_2 \propto B N_{\rm e,2}^{\rm los}$(1$-\mu'^4$), and
${\Delta}E_1 = E_1 \sqrt{\frac{kT_{e,\parallel,\rm 1}\mu'^2}{m_ec^2}}$ and
${\Delta}E_2 = 2 E_1 \sqrt{\frac{kT_{e,\parallel,\rm 2}\mu'^2}{m_ec^2}}$.
$N_{\rm e,1}^{\rm los}$ and $\frac{kT_{e,\parallel,\rm 1}\mu'^2}{m_ec^2}$,
however, have no direct physical meaning.

We may use the cyclotron absorption model
to describe the line-formation region if at least both the
second and third harmonics are strong, i.e.~the data request
that each of these
harmonics be fit with a line shape parameterized by centroid energy,
equivalent width, and full-width (see Paper I for a description of this
particular parameterization).  Only then is the number of required fit
parameters, five (we assume $E_3 = \frac{3}{2}E_2$), larger than the
number of physical parameters needed to describe the region
($B$, $T_{e,\parallel,2} = T_{e,\parallel,3}$, 
$N_{e,2}^{\rm los} = N_{e,3}^{\rm los}$, $\mu'$).
Otherwise, such as when only the 
first and second harmonics are observed, radiative transfer calculations are 
required.  No simple description can be used to explain the first
harmonic line, the appearance of which depends critically on the outcome
of the multiple resonant scatters required in order for individual photons
to escape, as well as on the introduction of spawned photons at energies
near that of the first harmonic.

The cyclotron scattering model can describe the line-forming region
using fewer free
parameters than the cyclotron absorption model (because of the
relationship between $T_{e,\parallel}$ and $B$ in the scattering
model), and can use the first harmonic data to help forge the 
description.  However, we have
found that moderate resolution GRB data may
require an even smaller number of free parameters than the minimum
three of the cyclotron scattering model.
In Table \ref{tab:disc},
we list the number of model parameters required in fits to the
GRB870303 data with the
exponentiated Gaussian absorption line model (Paper I), and the cyclotron
absorption and scattering models.
Each harmonic in these data can be adequately modeled with 
a saturated line shape parameterized by centroid energy and
equivalent width; the full-width is set to be proportional to the
equivalent width.
The number of free parameters is then $N_{\rm harm}$ + 1, where
$N_{\rm harm}$ is the number of observed harmonics ($N_{\rm harm}$
line width parameters and one harmonic line energy).
This is an upper limit.  For instance,
the S2 data is adequately fit using two, rather than three, parameters
(by setting $W_{\rm E,2}$=2$W_{\rm E,1}$),
because the second harmonic data lack constraining power.

Lamb (1992) argues that physically-based models which have parameters not
required by the data will (1) adequately fit the data, but (2) have reduced
diagnostic power because of a lack of constraint on the individual
model parameters.
This is certainly true for the cyclotron absorption model.
For both GRB870303 (Graziani et al.~1992) and GRB880205 
(Fenimore et al.~1988), 
the cyclotron absorption model determines
the magnetic field well (since it is a function of only $E_{\rm 1}$), but
$\mu'$ is undetermined and $kT_{e,\parallel}$
and $N_{\rm e}^{\rm los}$ are therefore poorly constrained.
However, we determine in this work that photon spawning
and the presence of line shoulders give the 
cyclotron scattering model greater diagnostic power than we 
would have predicted.
(Note that such the absorption model fits adequately because large shoulders
are not observed in GRB870303 and GRB880205;
if large shoulders were present, no 
model using absorption-like profiles would adequately fit the data.)
Spawned photons fill the first harmonic in
the polar cap model, decreasing its ability to adequately fit the S1 data in
the 1-0 geometry; the addition of shoulders in the 1-1 geometry makes no
adequate fit possible.  While polar cap models adequately fit the S2 data,
the line shoulders which appear in the 1-1 geometry greatly reduce the
range of $\mu$ consistent with the data (cf.~the 1-0 result, for which there
is no constraint on $\mu$ at the 3$\sigma$ limit).

Fits to the combined (S1+S2) data in Paper I clearly
indicate that the data prefer a four-parameter 
model in which $W_{\rm E,1}$
changes between S1 and S2.  A cyclotron scattering model
having different $\mu$ and/or $N_{\rm e}$ values for S1 and S2,
and at least four parameters overall, would thus be expected
to provide the best fit to the data.
Fits of the general model to the joint data fulfill that expectation,
with six and seven parameter models chosen.
Surprisingly, however, the simplest three-parameter
polar cap models adequately fit the data for both
the 1-0 and 1-1 geometries; spawned photons and line shoulders prevent the
data from selecting more complex models over the simplest model, even 
if up to six parameters are allowed to float freely (Table \ref{tab:modfit}).  

\subsection{Neutron Star Rotation}

\label{sect:nsrot}

The use of cyclotron scattering models allows us to place rigorous limits
upon the rotation period of the neutron star source of GRB870303.
Previously, Graziani et al.~(1992) interpreted the results of fits
to GRB870303 S1 and S2 to {\it qualitatively} place limits on the
rotation period.  They use the observation of one harmonic during
S1 and two harmonics during S2 to suggest that line formation 
is observed along ${\vec B}$ for S1 and perpendicular to ${\vec B}$ for S2.  
Invoking both localized
line formation and line formation along an equatorial arc, they derive
a rotation period 45 s $\simless t_{\rm rot} \simless$ 180 s.
Lamb et al.~(1992) 
invokes similar arguments to explain the changes in
line-centroid energy, strength, and width of the line candidate exhibited by
the {\it HEAO-1} A4 data of GRB780325 (Hueter 1987), and to predict
40 s $\simless t_{\rm rot} \simless$ 80 s.

Examining the real-time data of GRB870303, Yoshida et al.~(1989) find an
additional peak in the PC time history beyond the two seen in burst mode
(Figure \ref{fig:time}).  The peaks have periodicity $\approx$ 30~s.
They conclude that the data in the third peak can be adequately fit
an $\approx$ 10 keV
thermal bremmstrahlung spectrum, an $\approx$ 1.6 keV blackbody spectrum,
or an $\approx$ 0.3 keV thermal cyclotron spectrum are consistent
with these data.  No spectral features are apparent, but the spectrum is
perhaps too soft for lines to be unambiguously detected.
Such a periodicity is perhaps barely consistent with $t_{\rm rot,3}$ in the
1-0 geometry, but is consistent with both $t_{\rm rot,3}$ and $t_{\rm rot,4}$
in the 1-1 geometry.  If we hypothesize that the neutron star completes more
than half a rotation between S1 and S2, it follows that a localized region
of line formation disappears behind the stellar disk between the two epochs.
Graziani et al.~(1993) consider a model in which lines appear at time
$t_{\rm 1}$, disappear at time $t_{\rm 2}$, and are constant in between.  They
find that the uncertainties $\sigma_{\rm t_1}$ and 
$\sigma_{\rm t_2}$ in the times at which the lines appear and disappear
are large, and they cannot exclude the possibility that lines are present
throughout GRB870303.

Rotation periods $\simgreat$ 22.5~s are longer than those which
would be expected for a neutron star born with rotation period 1~s and which
has undergone magnetic braking for a Hubble time ($P_{\rm max} \approx$ = 5~s).
Mass-transfer from companions slows some 
accretion-powered X-ray pulsars, which have rotation periods up to 
$\approx$ 800~s (see, e.g., Nagase 1989).
Similarly, accretion of matter onto high-velocity
neutron stars on the way to or within a Galactic corona may act to
both slow a neutron star and provide fuel for GRBs.  Repeated GRBs
(Graziani et al.~1998) may in turn cause 
losses in angular momentum, slowing the neutron star, if mass is ejected
from the source.

\acknowledgements

The authors would like to thank the referee, David Band,
for his careful reading of the text and his many helpful comments.
We also wish to thank Mike Isenberg, Carlo Graziani, and Tom Loredo
for helpful discussions.  This work was
supported in part by NASA Graduate Traineeship NGT-50778 and NASA Grant
NAGW-830.

\appendix

\section{General Derivation of the Neutron Star Rotation Period}

\label{sect:nsrotapp}

The orientation of the observer relative to the slab where line formation
takes place is described by a polar cosine, $\mu$, and azimuth angle,
$\phi$ (see Figure \ref{fig:coord}).  If both $\mu$ and $\phi$ change
with time, we cannot use the simple model described in {\S}\ref{sect:rot}, 
in which $\phi$ is constant in time, to place limits
on the rotation period(s) of the underlying neutron star.
Here, we outline a general Bayesian method for deriving credible intervals
for the rotation period(s).

We first write out a marginalized posterior probability distribution for
$\mu$ and $\phi$ for the best-fit model, adding as free parameters
the magnetic field axis orientation at the epoch of the first spectrum,
${\hat H_1}$, and the rotation axis, ${\hat R}$:
\begin{eqnarray}
p(\mu_1,\mu_2,\phi_1,\phi_2,{\hat H_1},{\hat R}{\vert}D,I,\Psi=\frac{\pi}{2})&=&p(\mu_1,\mu_2,\phi_1,\phi_2,{\hat H_1},{\hat R}{\vert}I) \times \nonumber \\
&&\int\int dB dN_{\rm e} p(B{\vert}I) p(N_{\rm e}{\vert}I) \times \\
\label{eqn:a1}
&&\frac{{\like}(\mu_1,\mu_2,\phi_1,\phi_2,{\hat H_1},{\hat R},B,N_{\rm e},\Psi=\frac{\pi}{2})}{p(D{\vert}I)} \nonumber .
\end{eqnarray}
(Here, we assume that $\Psi = \frac{\pi}{2}$; we do not marginalize over
this parameter.)
We determine the rotation periods $t_{\rm rot,n}$ 
by sampling from this posterior distribution.
(There are four possible rotation periods that can be derived, for reasons
given in {\S}\ref{sect:rot} and illustrated in Figure \ref{fig:nsrot}.)
However, we must deal with the Bayesian prior, the first factor
on the right-hand side of eq.~(\ref{eqn:a1}).  We expand it:
\begin{eqnarray}
p(\mu_1,\mu_2,\phi_1,\phi_2,{\hat H_1},{\hat R} \vert I)&=&p({\hat R} \vert I) p({\hat H_1} \vert {\hat R},I) \cdot\cdot\cdot p(\phi_2 \vert \phi_1,\mu_2,\mu_1,{\hat H_1},{\hat R},I) \nonumber \\
&=&p({\hat R} \vert I) p({\hat H_1} \vert I) p(\mu_1 \vert I) p(\mu_2 \vert \mu_1,{\hat H_1},{\hat R},I) \times \\
&&p(\phi_1 \vert \mu_1,{\hat H_1},I) p(\phi_2 \vert \mu_2,{\hat H_1},{\hat R},I) \nonumber.
\end{eqnarray}
We randomly sample $\mu_1$, $\mu_2$, ${\hat H_1}$, and ${\hat R}$ from uniform
distributions.  Because of polar symmetry,
the values $\mu_1$ and $\mu_2$ 
define circles centered on the visible disk of the neutron star
where line formation may take place.
Given ${\hat H_1}$, we can define the magnetic equator,
which
either intersects the line-formation circle twice, or not at all.
Since we assume $\Psi = \frac{\pi}{2}$, the vector pointing out of the
star at the intersection points is simply ${\hat n_1}$ 
(Figure \ref{fig:coord}).
At the intersection points, ${\hat O}\cdot{\hat n_1}$ = $\mu_1$,
where ${\hat O}$ is the vector pointing to the observer.  If we define
${\hat O}$ as (0,-1,0), $\mu_1$ = $-n_{1,y}$.  The only 
unknown is then $n_{1,x}$, which we can solve for numerically.
Given the location of line-formation,
we project ${\hat O}$ onto the local surface tangent
(i.e.~onto a vector which is perpendicular to ${\hat n_1}$).
The dot product of this projected vector
with the vector $-{\hat H_1}$ 
(the direction of the field at the magnetic equator) 
is the cosine of $\phi_1$.

We may determine
${\hat n}_2$ by using the information that ${\hat R}\cdot{\hat n}$ is 
constant and ${\hat O}\cdot{\hat n}_2$ = $\mu_2$.
Either two solutions to these equations exist, or none at all.
Given a solution,
we next determine the possible local field direction $-{\hat H_2}$.
The vectors ${\hat R}$, ${\hat n}_1$, and ${\hat n}_2$
allow us to 
use the law of cosines to determine $\beta$, the angle through which
the neutron star rotates between the two epochs.
We transform the Cartesian coordinate system such that ${\hat R}'$ = (0,0,1),
rotate the system through angle $\beta$, and undo the transformation.
We then solve for $\phi_2$, as above.

The likelihood ${\like}$ does not depend upon
${\hat H_1}$ or ${\hat R}$, so the integrated expression in eq.~\ref{eqn:a1}
is proportional
to the four-dimensional posterior probability distribution
$p(\mu_1,\mu_2,\phi_1,\phi_2 \vert
D,I,\Psi = \frac{\pi}{2})$. 
For the particular case of interest in this paper, this distribution is
known: two dimensional ``slices" of it are shown in Figure \ref{fig:90jcred11}.
We use this distribution as a rejection 
function: we assess the relative probability of our solution with
respect to the best-fit parameters of the model, 
sample a random number $r$, and accept the
solution if $r < \frac{p_{\rm solution}}{p_{\rm best~fit}}$.  If we accept
the solution, we then solve for the rotation period $t_{\rm rot,n}$.

\newpage

%Figure 1
\begin{figure}
\caption{
An illustration showing
the coordinate system used in this work.
The infinite plane-parallel slab is threaded by a magnetic field ${\vec B}$
oriented at an angle $\Psi$ relative to the slab normal ${\hat n}$.
The angles $\theta$ (or angle cosine $\mu = \cos{\theta}$) and
$\phi$ specify the direction of escaping photons relative
to ${\vec B}$ and ${\hat n}$.}
\label{fig:coord}
\end{figure}

%Figure 2
\begin{figure}
\caption{
An illustration showing
the two plane-parallel slab geometries which we examine in this work,
denoted ``1-0'' and ``1-1.''
The numbers represent the relative electron
column densities above and below the continuum photon source plane.
Injected photons travel through the slab, and emerge from one of
the plane-parallel faces of the slab.  The observer is situated above the
source plane.}
\label{fig:geom}
\end{figure}

%Figure 3
\begin{figure}
\caption{
Sample cyclotron scattering line spectra generated using the
GRB870303 S2 continuum spectrum, with $B_{12}$ = 1.7 and $N_{\rm e,21}$ = 0.6,
for the 1-0 geometry and $\Psi$ = 0.
In spectrum (a), the observer is oriented directly above the slab, while
in spectrum (b), the observer looks along the slab surface.
The line-of-sight column density and
the scattering cross-section $\sigma_N \propto
(1+\mu_{\rm B})(1-\mu_{\rm B}^2)^{N-1}$, where $\mu_{\rm B}$ is the cosine
of the angle between observer and field, dictate the
strength of the $N^{\rm th}$ harmonic line.}
\label{fig:10samp}
\end{figure}

%Figure 4
\begin{figure}
\caption{
Sample cyclotron scattering line spectra generated under the same
conditions as Figure \ref{fig:10samp}, but for the 1-1 geometry.
Shoulders appear at the
first harmonic in spectrum (a) for reasons described in
{\S}\ref{sect:samp}.
}
\label{fig:11samp}
\end{figure}

%Figure 5
\begin{figure}
\caption{
Sample cyclotron scattering line spectra generated for similar
conditions as Figure \ref{fig:10samp}, but for $\Psi$ = $\frac{\pi}{2}$.
As indicated by the diagram at left,
in spectrum (a), the observer is oriented directly above the slab;
in spectrum (b), the observer looks along the slab surface and along
the magnetic field; and in spectrum (c),
the observer looks along the slab surface, perpendicular to the 
magnetic field.
}
\label{fig:eqsamp}
\end{figure}

%Figure 6
\begin{figure}
\caption{
A flow chart showing the method which we use to select the best
fitting polar cap ($\Psi$ = 0) and general ($\Psi \neq$ 0)
cyclotron scattering models, when jointly fitting two (or more)
datasets.
We then directly compare the selected polar cap and general models with the
$\chi^2$ Maximum Likelihood Ratio model comparison test.
See {\S}\ref{sect:modcomp} for details.
}
\label{fig:flow}
\end{figure}

%Figure 7
\begin{figure}
\caption{
{\it Ginga} Proportional Counter (PC; top) and Scintillation
Counter (SC; bottom) time histories of GRB870303.
The PC data is presented in 1 s bins, while the SC data 
is presented in 0.5 s bins.
The burst triggered the recording of {\it Ginga} GBD burst-mode
data at $\approx$ 16 s; the preceding 16 s of burst-mode data,
in memory at the time of trigger, were recorded and not overwritten.
Overall, 64 s of burst-mode data were recorded.
Epochs S1 (4 s) and S2 (9 s) are shown;
the midpoints of S1 and S2 are separated by
22.5 s.}
\label{fig:time}
\end{figure}

%Figure 8
\begin{figure}
\caption{
{\it Ginga} GBD count-rate spectra for intervals S1 and S2 of GRB870303,
normalized by energy-loss bin width.
S1 exhibits a single line at $\approx$ 20 keV, while the spectrum
S2 exhibits harmonically spaced lines at $\approx$ 20 keV and 40 keV.}
\label{fig:data}
\end{figure}

%Figure 9
\begin{figure}
\caption{
Left: The data and predicted counts (upper panel),
photon spectrum (middle panel), and residuals of the fit in units of
$\sigma$ (lower panel), for the best cyclotron scattering
line model fit to the data of GRB870303 S1 for the 1-0 geometry.
Right: Two-dimensional 1$\sigma$, 2$\sigma$, and 3$\sigma$
Bayesian credible regions for this fit, as a function of the stated
parameters.}
\label{fig:s110}
\end{figure}

%Figure 10
\begin{figure}
\caption{
Left: The data and predicted counts (upper panel),
photon spectrum (middle panel), and residuals of the fit in units of
$\sigma$ (lower panel), for the best cyclotron scattering
line model fit to the data of GRB870303 S1 for the 1-1 geometry.
Center and Right:
Two-dimensional 1$\sigma$, 2$\sigma$, and 3$\sigma$
Bayesian credible regions for the fit.}
\label{fig:s111}
\end{figure}

%Figure 11
\begin{figure}
\caption{
Left: The data and predicted counts (upper panel),
photon spectrum (middle panel), and residuals of the fit in units of
$\sigma$ (lower panel), for the best cyclotron scattering
line model fit to the data of GRB870303 S2 for the 1-0 geometry.
Right: Two-dimensional 1$\sigma$, 2$\sigma$, and 3$\sigma$
Bayesian credible regions for this fit, as a function of the stated
parameters.}
\label{fig:s210}
\end{figure}

%Figure 12
\begin{figure}
\caption{
Left: The data and predicted counts (upper panel),
photon spectrum (middle panel), and residuals of the fit in units of
$\sigma$ (lower panel), for the best cyclotron scattering
line model fit to the data of GRB870303 S2 for the 1-1 geometry.
Right: Two-dimensional 1$\sigma$, 2$\sigma$, and 3$\sigma$
Bayesian credible regions for this fit, as a function of the stated
parameters.}
\label{fig:s211}
\end{figure}

%Figure 13
\begin{figure}
\caption{
The data and predicted counts (upper panel),
photon spectrum (middle panel), and residuals of the fit in units of
$\sigma$ (lower panel), for the best cyclotron scattering
line model fit to the data of GRB870303 S1+S2 for the 1-0 geometry
and $\Psi$ = 0.  Left: The fit to S1.  Right: The fit to
S2.  
}
\label{fig:0jcred10}
\end{figure}

%Figure 14
\begin{figure}
\caption{
The data and predicted counts (upper panel),
photon spectrum (middle panel), and residuals of the fit in units of
$\sigma$ (lower panel), for the best cyclotron scattering
line model fit to the data of GRB870303 S1+S2 for the 1-0 geometry
and $\Psi$ = $\frac{\pi}{2}$.
Left: The fit to S1.  Center: The fit to S2. 
Right:
Two-dimensional 1$\sigma$, 2$\sigma$, and 3$\sigma$ Bayesian credible regions
for the parameter subset $(\mu_{\rm S1},\mu_{\rm S2})$.}
\label{fig:90jcred10}
\end{figure}

%Figure 15
\begin{figure}
\caption{
The data and predicted counts (upper panel),
photon spectrum (middle panel), and residuals of the fit in units of
$\sigma$ (lower panel), for the best cyclotron scattering
line model fit to the data of GRB870303 S1+S2 for the 1-1 geometry
and $\Psi$ = 0.  Left: The fit to S1.  Right: The fit to S2.
}
\label{fig:0jcred11}
\end{figure}

%Figure 16
\begin{figure}
\caption{
The data and predicted counts (upper panel),
photon spectrum (middle panel), and residuals of the fit in units of
$\sigma$ (lower panel), for the best cyclotron scattering
line model fit to the data of GRB870303 S1+S2 for the 1-1 geometry
and $\Psi$ = $\frac{\pi}{2}$.
Left: The fit to S1.  Left Center: The fit to S2.
Right Center and Right:
Two-dimensional 1$\sigma$, 2$\sigma$, and 3$\sigma$ Bayesian credible regions
for the parameter subset
$(\mu_{\rm S1},\mu_{\rm S2},\phi_{\rm S1},\phi_{\rm S2})$.}
\label{fig:90jcred11}
\end{figure}

%Figure 17
\begin{figure}
\caption{
A model of neutron star rotation for the 1-0 geometry, for
$\Psi = \frac{\pi}{2}$.  The rotation
axis ${\hat \Omega}$ is perpendicular to the observer's line-of-sight, and
the star rotates in either the clockwise or counter-clockwise directions.
The magnetic field axis lies within the neutron star-observer plane.
Line formation is localized and occurs where the magnetic equator intercepts
the neutron star-observer plane.  Because of the symmetry between 
$\phi$ = 0 and $\pi$, there are four angles through which the neutron
star may rotate during the 22.5 s between S1 and S2, and thus four
rotation periods may be derived for each choice of $\mu_{\rm S1}$ and
$\mu_{\rm S2}$.  (The same four times would be derived if instead of
rotation to S2, $\phi =$ 0, we assume S2, $\phi = \pi$.)
}
\label{fig:nsrot}
\end{figure}

\newpage

\clearpage

\begin{deluxetable}{cc}
\tablenum{1}
\tablecaption{Cyclotron Scattering Model Parameter Values}
\tablewidth{0pt}
\tablehead{
\colhead{Parameter} & \colhead{Values} }
\startdata
$B_{\rm 12}$ (G) & (1.55,1.83,2.10,2.40)\tablenotemark{a}\\
$\Psi$ & (0,$\frac{\pi}{6}$,$\frac{\pi}{3}$,$\frac{\pi}{2}$)\\
log $N_{\rm e,21}$ (cm$^{-2}$) & $(-2.20,-1.50,-1.10,-0.5,-0.2,0.1,0.4,0.7,1.0)$\\
$\mu$ & Bin Size ${\Delta}\mu$ = 0.125 ($\mu$ $\in$ [0,1])\\
$\phi$ (rad) & Bin Size ${\Delta}\phi$ = $\frac{\pi}{8}$ \\
 & ($\phi~\in$ [0,$\pi$] if $\Psi \neq$ 0 or $\frac{\pi}{2}$,\\
 & or $\phi~\in$ [0,$\frac{\pi}{2}$] if $\Psi$ = $\frac{\pi}{2}$)\\
\enddata
\tablenotetext{a}{We can shift $B_{\rm 12}$ by $\pm$ 10\% during fits with
no loss in model accuracy.}
\label{tab:param}
\end{deluxetable}

\clearpage

\begin{deluxetable}{ccccccc}
\tablenum{2}
\tablecaption{Frequentist Model Comparison}
\tablewidth{0pt}
\tablehead{
 & & \colhead{1-0} & & & \colhead{1-1} & \\
\colhead{Spectrum} & \colhead{$s_{\rm m}^2$ Polar Cap} & \colhead{$s_{\rm m}^2$ General} & \colhead{$\alpha_{\rm \chi^2MLR}$} & \colhead{$s_{\rm m}^2$ Polar Cap} & \colhead{$s_{\rm m}^2$ General} & \colhead{$\alpha_{\rm \chi^2MLR}$}}
\startdata
S1 & 26.7 (33)$^{\ast}$ & 23.6 (31) & 0.21 & 37.7 (33) & 24.4 (31)$^{\ast}$ & 1.3$\times$10$^{-3}$ \\
S2 & 33.6 (33)$^{\ast}$ & 31.7 (31) & 0.39 & 33.9 (33)$^{\ast}$ & 32.9 (31) & 0.61 \\
S1+S2$^{\ast}$ & 70.8 (69)$^{\ast}$ & 59.6 (66)$^{\ast}$ & 0.01 & 73.7 (69)$^{\ast}$ & 59.9 (65)$^{\ast}$ & 8.0$\times$10$^{-3}$ \\
\tablecomments{The number of degrees of freedom is given in parentheses.}
\enddata
\tablenotetext{\ast}{Selected model.  For both the 1-0 and 1-1 geometries, the data do not conclusively select either the polar cap model or general model.}
\label{tab:modcomp}
\end{deluxetable}

\clearpage

\begin{deluxetable}{ccccccc}
\tablenum{3}
\tablecaption{S1: Bayesian Credible Intervals}
\tablewidth{0pt}
\tablehead{
 & & \colhead{1-0} & & & \colhead{1-1} & \\
Selected Model: & & \colhead{Polar Cap} & & & \colhead{General ($\Psi$ = $\frac{\pi}{2}$)\tablenotemark{a}} & \\
\colhead{Parameter} & \colhead{1$\sigma$} & \colhead{2$\sigma$} & \colhead{3$\sigma$} & \colhead{1$\sigma$} & \colhead{2$\sigma$} & \colhead{3$\sigma$} }
\startdata
$B_{\rm 12}$ (G) & 1.76$_{-0.14}^{+0.16}$ & 1.76$_{-0.27}^{+0.25}$ & 1.76$_{-0.39}^{+0.51}$ & 1.83$_{-0.11}^{+0.12}$ & 1.83$_{-0.25}^{+0.28}$ & 1.83$_{-0.36}^{+0.45}$ \\
${\log}N_{\rm e,21}$ (cm$^{-2}$) & 1.00$_{-1.47}^{+0.00}$\tablenotemark{b,c} & 1.00$_{-2.35}^{+0.00}$ & 1.00$_{-3.14}^{+0.00}$ & -1.52$_{-0.29}^{+0.75}$ & -1.52$_{-0.63}^{+1.58}$ & -1.52$_{-0.73}^{+2.08}$ \\
$\mu$ & 0.990$_{-0.990}^{+0.007}$\tablenotemark{d,e} & 0.990$_{-0.990}^{+0.008}$\tablenotemark{f} & 0.990$_{-0.990}^{+0.008}$\tablenotemark{g} & 0.06$_{-0.06}^{+0.05}$\tablenotemark{h} & 0.06$_{-0.06}^{+0.17}$ & 0.06$_{-0.06}^{+0.25}$\\
$\phi$ (rad) & & & & 0.20$_{-0.20}^{+0.20}$\tablenotemark{i} & 0.20$_{-0.20}^{+0.66}$ & 0.20$_{-0.20}^{+1.07}$ \\
\enddata
\tablenotetext{a}{The credible region does not include marginalization over $\Psi$}
\tablenotetext{b}{${\log}N_{\rm e,21}$ = 1 is a parameter boundary}
\tablenotetext{c}{The 1$\sigma$ credible region does not include ${\log}N_{\rm e,21}$ = (0.32,0.47)}
\tablenotetext{d}{$\mu_{\rm best-fit} \in$ [0.988,0.992]}
\tablenotetext{e}{The 1$\sigma$ credible region does not include $\mu$ = (0.09,0.61)}
\tablenotetext{f}{The 2$\sigma$ credible region does not include $\mu$ = (0.18,0.31)}
\tablenotetext{g}{The 3$\sigma$ credible region does not include $\mu$ = (0.187,0.194)}
\tablenotetext{h}{$\mu_{\rm best-fit} \in$ [0.000,0.125]}
\tablenotetext{i}{$\phi_{\rm best-fit} \in$ [0,0.39 ($\frac{\pi}{8}$)]}
\label{tab:s1bayes}
\end{deluxetable}

\clearpage

\begin{deluxetable}{ccccccc}
\tablenum{4}
\tablecaption{S2: Bayesian Credible Intervals}
\tablewidth{0pt}
\tablehead{
 & & \colhead{1-0} & & & \colhead{1-1} & \\
Selected Model: & & \colhead{Polar Cap} & & & \colhead{General ($\Psi$ = 0)} & \\
\colhead{Parameter} & \colhead{1$\sigma$} & \colhead{2$\sigma$} & \colhead{3$\sigma$} & \colhead{1$\sigma$} & \colhead{2$\sigma$} & \colhead{3$\sigma$} }
\startdata
$B_{\rm 12}$ (G) & 1.96$_{-0.09}^{+0.07}$ & 1.96$_{-0.19}^{+0.13}$ & 1.96$_{-0.33}^{+0.21}$ & 1.96$_{-0.06}^{+0.07}$ & 1.96$_{-0.14}^{+0.15}$ & 1.96$_{-0.24}^{+0.23}$ \\
${\log}N_{\rm e,21}$ (cm$^{-2}$) & -0.22$_{-0.56}^{+0.22}$ & -0.22$_{-0.99}^{+0.55}$ & -0.22$_{-1.48}^{+0.91}$ & -0.22$_{-0.52}^{+0.22}$ & -0.22$_{-0.94}^{+0.62}$ & -0.22$_{-1.64}^{+1.01}$ \\
$\mu$ & 0.31$_{-0.22}^{+0.09}$\tablenotemark{a} & 0.31$_{-0.31}^{+0.49}$\tablenotemark{b} & 0.31$_{-0.31}^{+0.67}$ & 0.06$_{-0.06}^{+0.12}$\tablenotemark{c} & 0.06$_{-0.06}^{+0.32}$ & 0.06$_{-0.06}^{+0.45}$ \\
\enddata
\tablenotetext{a}{$\mu_{\rm best-fit} \in$ [0.250,0.375].}
\tablenotetext{b}{The 2$\sigma$ credible region does not include $\mu$ = (0.56,0.58).}
\tablenotetext{c}{$\mu_{\rm best-fit} \in$ [0.000,0.125].}
\label{tab:s2bayes}
\end{deluxetable}

\clearpage

\begin{deluxetable}{cccccccc}
\tablenum{5}
\tablecaption{S1+S2: Frequentist Model Fits}
\tablewidth{0pt}
\tablehead{
 & & \colhead{1-0} & & & \colhead{1-1} & \\
\colhead{$P$\tablenotemark{a}} & \colhead{$s_{\rm m}^2$ Polar Cap} && \colhead{$s_{\rm m}^2$ General} & \colhead{$s_{\rm m}^2$ Polar Cap} && \colhead{$s_{\rm m}^2$ General} & \colhead{$s_{\rm m}^2$ Paper I\tablenotemark{b}} }
\startdata
3 & 70.8$^{\ast}$ &&   -        & 73.7$^{\ast}$ &&   -        &      \\
4 & 66.5       &&   -        & 72.2       &&   -        &~~~~~~58.0 \\
5 & 64.4       && 69.5       & 71.9       && 71.6       &      \\
6 & 60.3       && 59.6$^{\ast}$ & 71.3       && 65.0       &      \\
7 &   -        &&   -        &   -        && 59.9$^{\ast}$ &      \\
\enddata
\tablenotetext{\ast}{Selected model.}
\tablenotetext{a}{The number of free parameters.}
\tablenotetext{b}{Using exponentiated Gaussian absorption line profiles (Paper I).}
\label{tab:modfit}
\end{deluxetable}

\clearpage

\begin{deluxetable}{ccccccc}
\tablenum{6}
\tablecaption{S1+S2: Bayesian Credible Intervals for 1-0 Geometry}
\tablewidth{0pt}
\tablehead{
 & & \colhead{Polar Cap} & & & \colhead{General ($\Psi$ = $\frac{\pi}{2})$\tablenotemark{a}} & \\
\colhead{Parameter} & \colhead{1$\sigma$} & \colhead{2$\sigma$} & \colhead{3$\sigma$} & \colhead{1$\sigma$} & \colhead{2$\sigma$} & \colhead{3$\sigma$} }
\startdata
$B_{\rm 12}$ (G) & 1.95$_{-0.13}^{+0.05}$ & 1.95$_{-0.23}^{+0.11}$ & 1.95$_{-0.32}^{+0.15}$ & 1.87$_{-0.05}^{+0.08}$ & 1.87$_{-0.11}^{+0.14}$ & 1.87$_{-0.19}^{+0.20}$ \\
${\log}N_{\rm e,21}$ (cm$^{-2}$) & -0.22$_{-0.45}^{+0.25}$ & -0.22$_{-0.82}^{+0.55}$ & -0.22$_{-1.22}^{+0.97}$ & -0.22$_{-0.19}^{+0.44}$ & -0.22$_{-0.82}^{+0.58}$ & -0.22$_{-1.41}^{+0.67}$ \\
$\mu$ & 0.31$_{-0.12}^{+0.50}$\tablenotemark{b,c} & 0.31$_{-0.31}^{+0.68}$\tablenotemark{d} & 0.31$_{-0.31}^{+0.69}\tablenotemark{e}$ & & & \\
$\mu_{\rm S1}$ & & & & 0.19$_{-0.07}^{+0.14}$\tablenotemark{f} & 0.19$_{-0.14}^{+0.27}$ & 0.19$_{-0.15}^{+0.35}$ \\
$\mu_{\rm S2}$ & & & & 0.69$_{-0.10}^{+0.07}$\tablenotemark{g} & 0.69$_{-0.25}^{+0.12}$ & 0.69$_{-0.35}^{+0.19}$ \\
$\phi$ (rad) & & & & $0.20_{-0.20}^{+0.18}$\tablenotemark{h} & $0.20_{-0.20}^{+0.69}$ & $0.20_{-0.20}^{+0.97}$ \\
\enddata
\tablenotetext{a}{The credible region does not include marginalization over $\Psi$}
\tablenotetext{b}{$\mu_{\rm best-fit} \in$ [0.250,0.375]}
\tablenotetext{c}{The 1$\sigma$ credible region does not include $\mu$ = (0.49,0.60)}
\tablenotetext{d}{The 2$\sigma$ credible region does not include $\mu$ = (0.94,0.98)}
\tablenotetext{e}{The 3$\sigma$ credible region does not include $\mu$ = (0.97,0.98)}
\tablenotetext{f}{$\mu_{\rm S1,best-fit} \in$ [0.125,0.250]}
\tablenotetext{g}{$\mu_{\rm S2,best-fit} \in$ [0.000,0.125]}
\tablenotetext{h}{$\phi_{\rm best-fit} \in$ [0.00,0.39 ($\frac{\pi}{8}$)]}
\label{tab:jcred10}
\end{deluxetable}

\clearpage

\begin{deluxetable}{ccccccc}
\tablenum{7}
\tablecaption{S1+S2: Bayesian Credible Intervals for 1-1 Geometry}
\tablewidth{0pt}
\tablehead{
 & & \colhead{Polar Cap} & & & \colhead{General ($\Psi$ = $\frac{\pi}{2}$)\tablenotemark{a}} & \\
\colhead{Parameter} & \colhead{1$\sigma$} & \colhead{2$\sigma$} & \colhead{3$\sigma$} & \colhead{1$\sigma$} & \colhead{2$\sigma$} & \colhead{3$\sigma$} }
\startdata
$B_{\rm 12}$ (G) & 1.96$_{-0.07}^{+0.06}$ & 1.96$_{-0.14}^{+0.12}$ & 1.96$_{-0.23}^{+0.19}$ & 1.94$_{-0.09}^{+0.05}$ & 1.94$_{-0.17}^{+0.12}$ & 1.94$_{-0.25}^{+0.20}$ \\
${\log}N_{\rm e,21}$ (cm$^{-2}$) & -0.52$_{-0.38}^{+0.50}$ & -0.52$_{-0.53}^{+0.87}$ & -0.52$_{-0.88}^{+1.17}$ & -0.22$_{-0.70}^{+0.17}$ & -0.22$_{-1.44}^{+0.28}$ & -0.22$_{-1.88}^{+0.32}$ \\
$\mu$ & 0.06$_{-0.06}^{+0.04}$\tablenotemark{b} & 0.06$_{-0.06}^{+0.22}$ & 0.06$_{-0.06}^{+0.39}$ & & & \\
$\mu_{\rm S1}$ & & & & 0.19$_{-0.15}^{+0.02}$\tablenotemark{c} & 0.19$_{-0.19}^{+0.08}$ & 0.19$_{-0.19}^{+0.16}$ \\
$\mu_{\rm S2}$ & & & & 0.31$_{-0.13}^{+0.07}$\tablenotemark{d} & 0.31$_{-0.26}^{+0.10}$ & 0.31$_{-0.31}^{+0.12}$ \\
$\phi_{\rm S1}$ (rad) & & & & $0.20_{-0.20}^{+0.16}$\tablenotemark{e} & $0.20_{-0.20}^{+0.65}$ & $0.20_{-0.20}^{+1.08}$ \\
$\phi_{\rm S2}$ (rad) & & & & $0.59_{-0.11}^{+0.53}$\tablenotemark{f} & $0.59_{-0.30}^{+0.78}$ & $0.59_{-0.35}^{+0.86}$ \\
\enddata
\tablenotetext{a}{The credible region does not include marginalization over $\Psi$}
\tablenotetext{b}{$\mu_{\rm best-fit} \in$ [0.000,0.125]}
\tablenotetext{c}{$\mu_{\rm S1,best-fit} \in$ [0.125,0.250]}
\tablenotetext{d}{$\mu_{\rm S2,best-fit} \in$ [0.250,0.375]}
\tablenotetext{e}{$\phi_{\rm S1,best-fit} \in$ [0.00,0.39 ($\frac{\pi}{8}$)]}
\tablenotetext{f}{$\phi_{\rm S2,best-fit} \in$ [0.39 ($\frac{\pi}{8}$),0.78 ($\frac{\pi}{4}$)]}
\label{tab:jcred11}
\end{deluxetable}

\clearpage

\begin{deluxetable}{ccccccc}
\tablenum{8}
\tablecaption{S1+S2: Limits on Neutron Star Rotation Periods}
\tablewidth{0pt}
\tablehead{
 & & \colhead{1-0 ($\Psi$ = $\frac{\pi}{2}$)} & & & \colhead{1-1 ($\Psi$ = $\frac{\pi}{2}$)} & \\
\colhead{} & \colhead{1$\sigma$} & \colhead{2$\sigma$} & \colhead{3$\sigma$} & \colhead{1$\sigma$} & \colhead{2$\sigma$} & \colhead{3$\sigma$} }
\startdata
$t_{\rm rot,1}$ (s) & 282$_{-59}^{+119}$ & 282$_{-97}^{+480}$ & 282$_{-133}^{+\infty}$ & 717$_{-350}^{+1160}$ & 717$_{-555}^{+10200}$ & 717$_{-654}^{+\infty}$ \\
$t_{\rm rot,2}$ (s) & 65.9$_{-4.4}^{+5.2}$ & 65.9$_{-11.1}^{+12.3}$ & 65.9$_{-14.3}^{+23.1}$ & 55.0$_{-6.5}^{+29.3}$ & 55.0$_{-9.4}^{+103}$ & 55.0$_{-10.0}^{+268}$ \\
$t_{\rm rot,3}$ (s) & 34.1$_{-4.0}^{+5.8}$ & 34.1$_{-2.6}^{+4.1}$ & 34.1$_{-4.0}^{+5.8}$ & 38.0$_{-7.4}^{+2.9}$ & 38.0$_{-12}^{+6.4}$ & 38.0$_{-14.5}^{+45.7}$ \\
$t_{\rm rot,4}$ (s) & 24.4$_{-0.6}^{+0.7}$ & 24.4$_{-1.3}^{+1.3}$ & 24.4$_{-1.9}^{+2.2}$ & 23.2$_{-0.5}^{+0.7}$ & 23.2$_{-0.7}^{+2.9}$ & 23.2$_{-0.7}^{+11.7}$ \\
\enddata
\label{tab:rot}
\end{deluxetable}

\clearpage

\begin{deluxetable}{cccccc}
\tablenum{9}
\tablecaption{Number of Model Parameters in Fits}
\tablewidth{0pt}
\tablehead{
 & \colhead{Exponentiated} \\
 & \colhead{Gaussian} & \colhead{Cyclotron} & \colhead{Cyclotron} \\
\colhead{Spectrum} & \colhead{Absorption\tablenotemark{a}} & \colhead{Absorption} & \colhead{Scattering} }
\startdata
S1 & 2 & 3(1)\tablenotemark{b} & 3 (1-0)\\
   &   &                       & 5 (1-1)\\
S2 & 2 & 5(3)\tablenotemark{b} & 3 (1-0)\\
   &   &                       & 3 (1-1)\\
S1+S2 & 4 & 7(3)\tablenotemark{b} & 3 or 6\tablenotemark{c} (1-0)\\
      &   &                       & 3 or 7\tablenotemark{c} (1-1)\\
\enddata
\tablenotetext{a}{See Paper I for description.}
\tablenotetext{b}{The number in parentheses represents the number of 
physically-relevant parameters.}
\tablenotetext{c}{The data do not
conclusively select either the polar cap model or general model.}
\label{tab:disc}
\end{deluxetable}

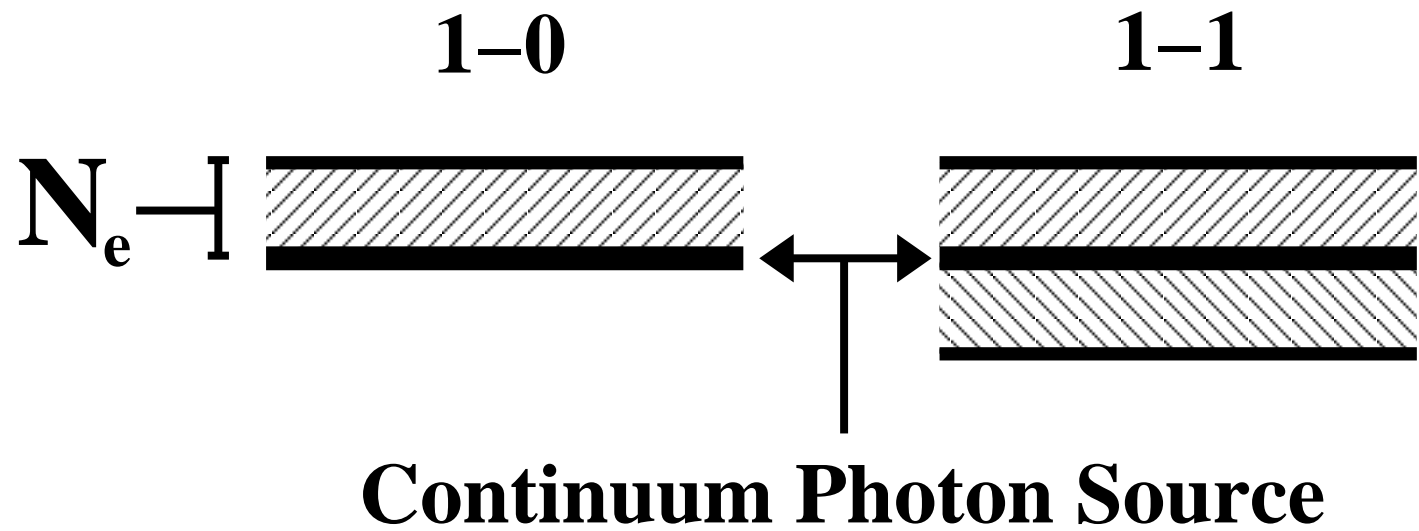
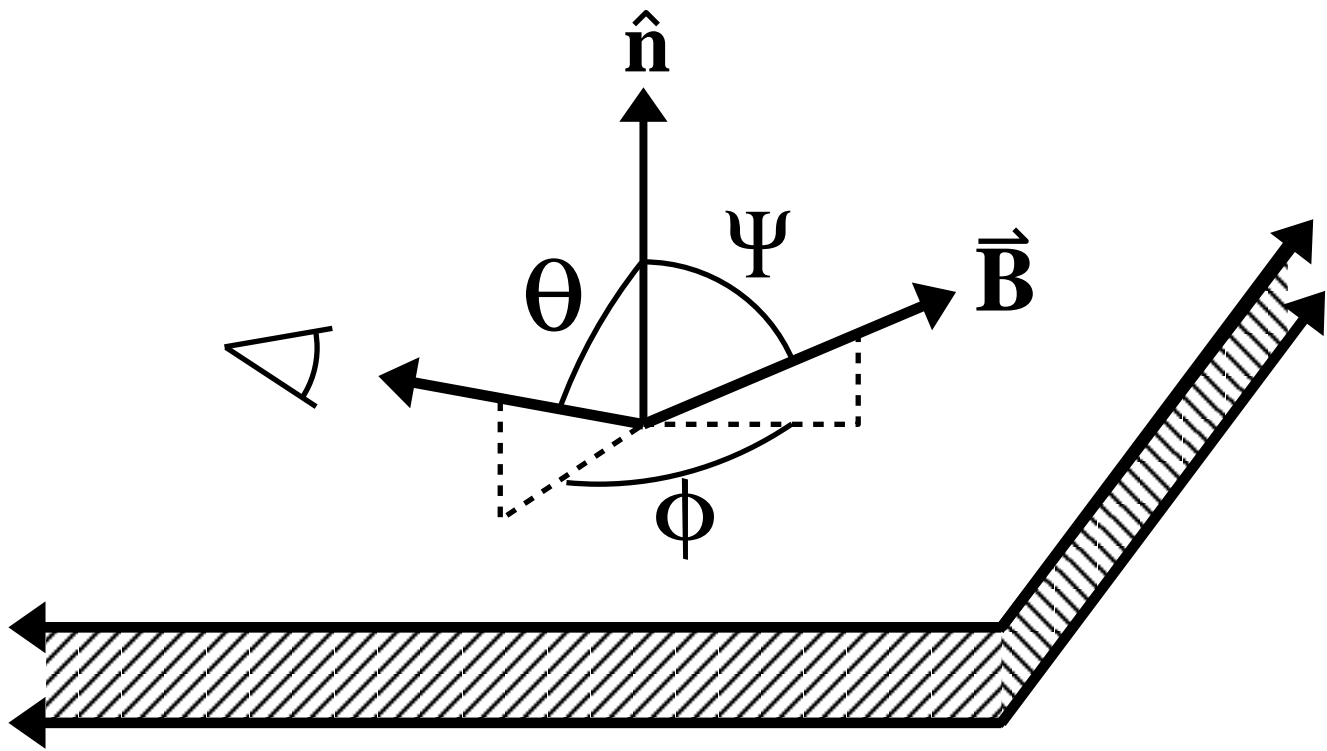
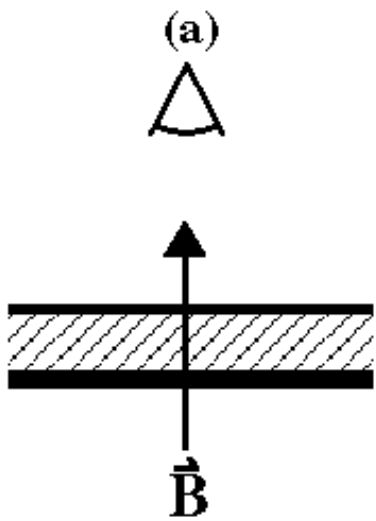
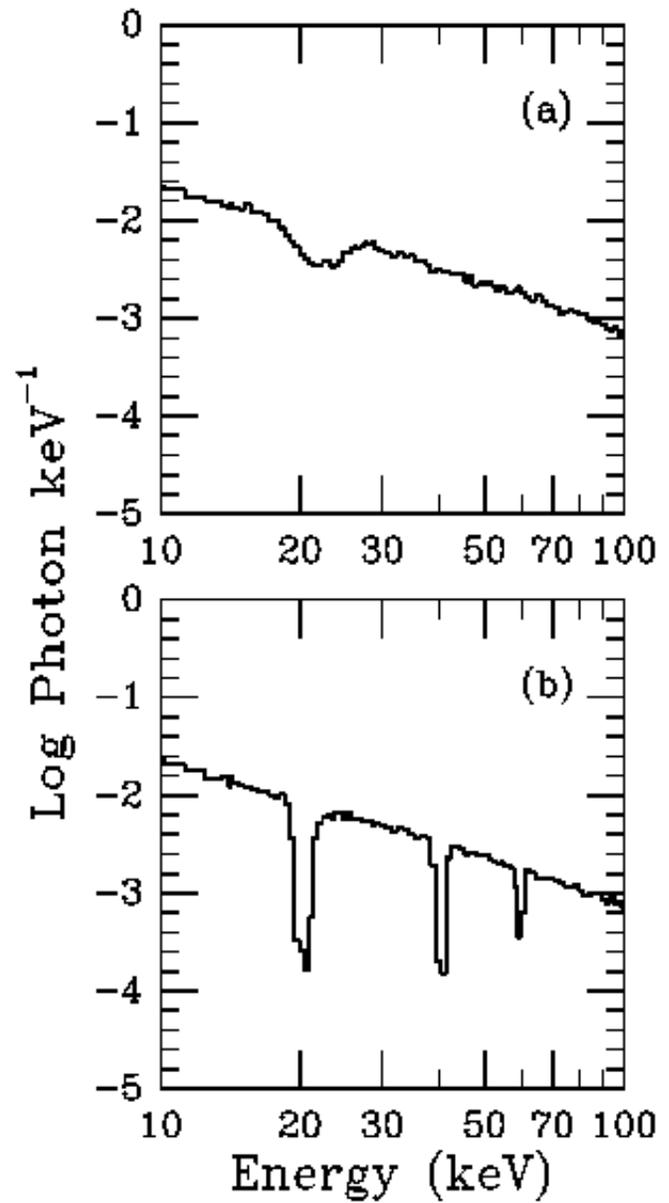
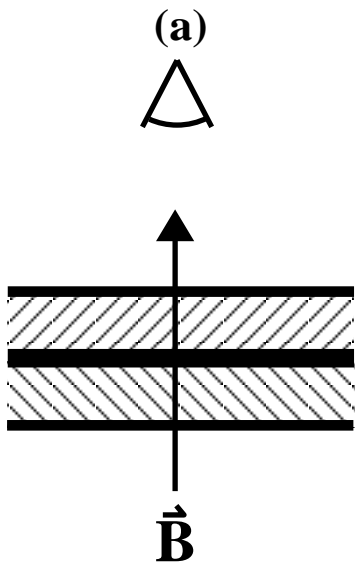
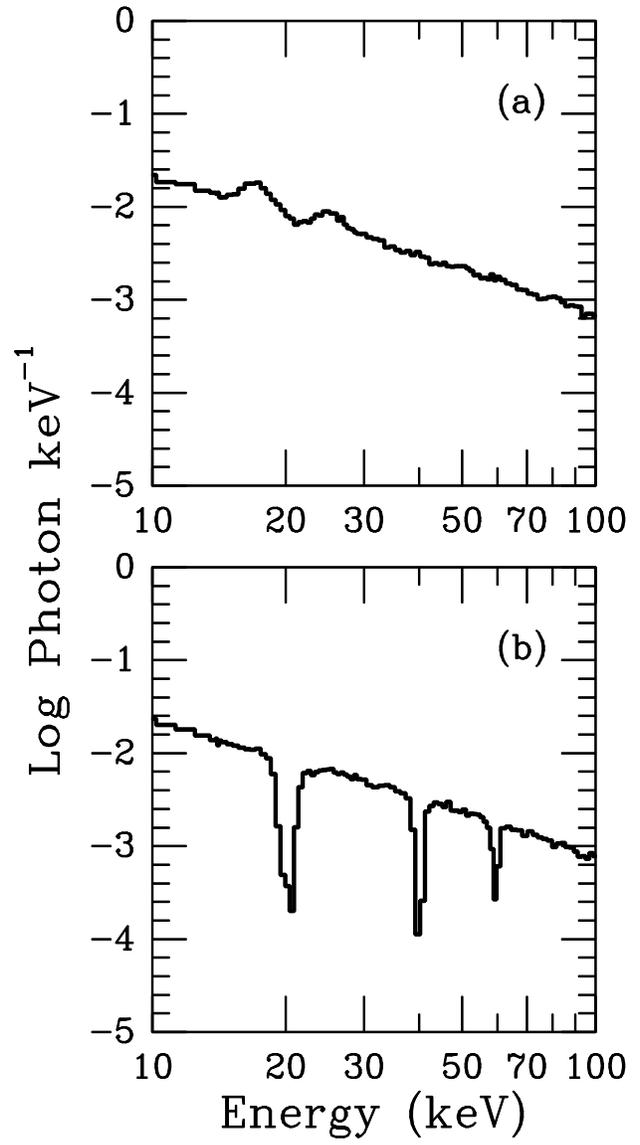
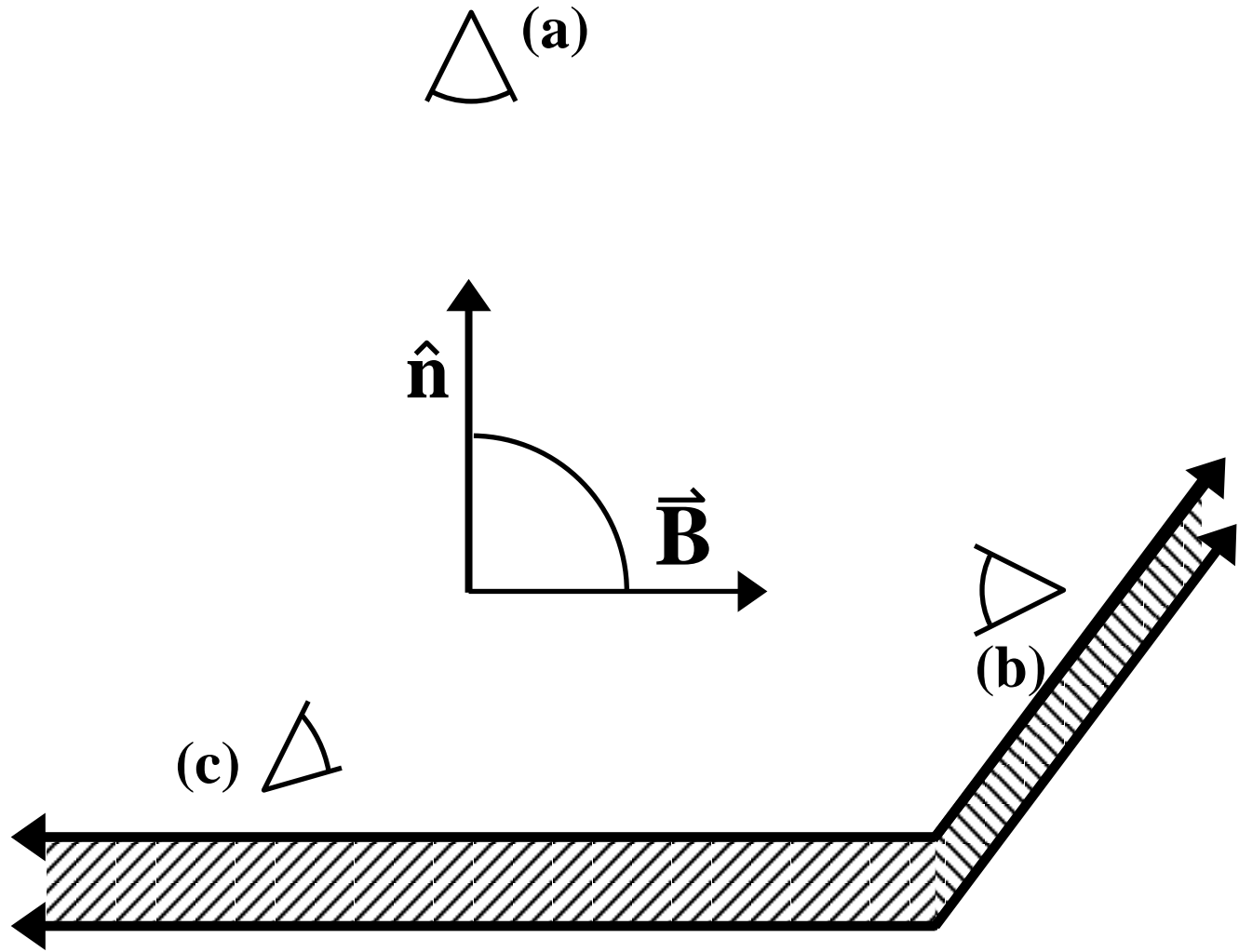
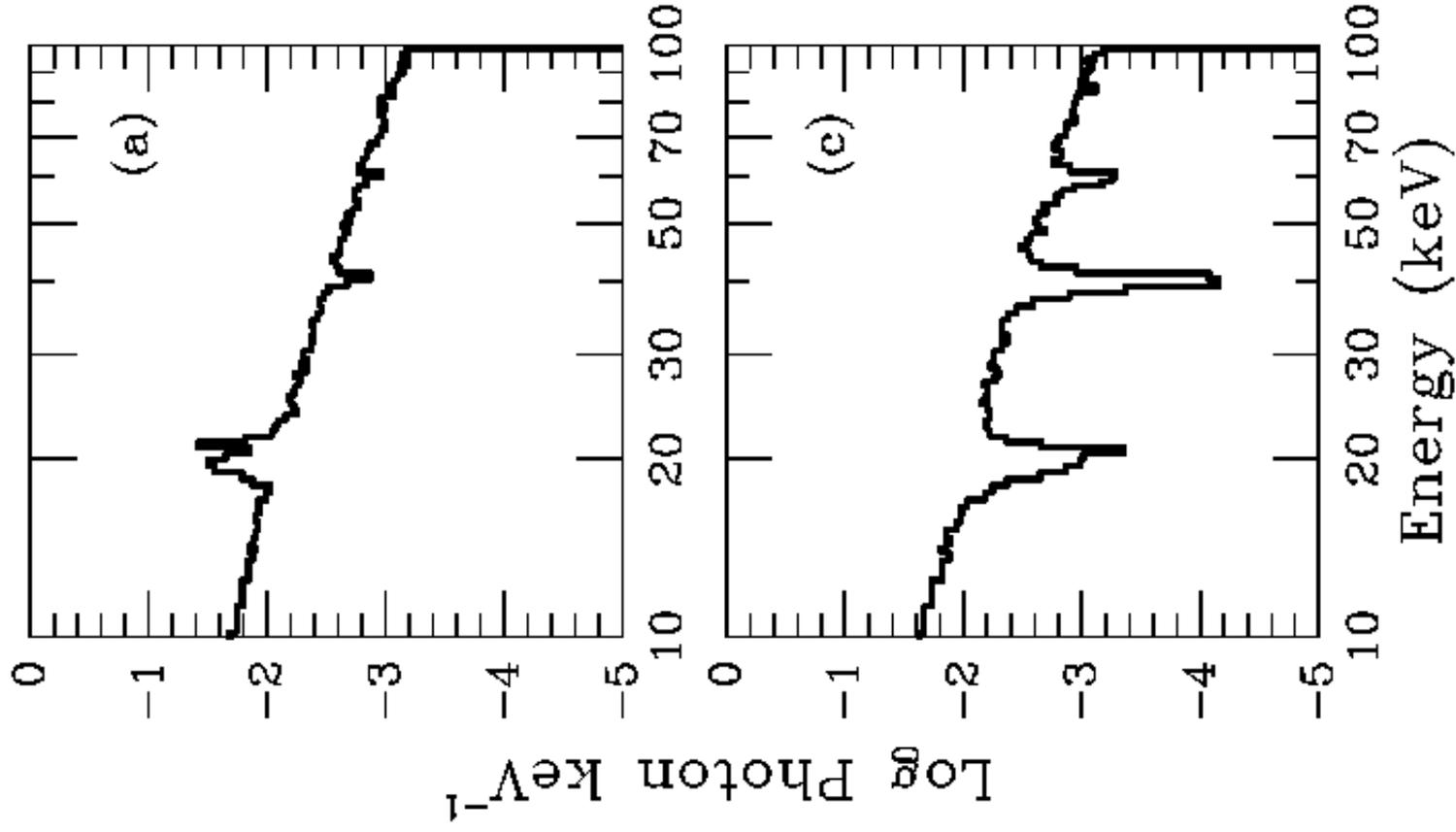
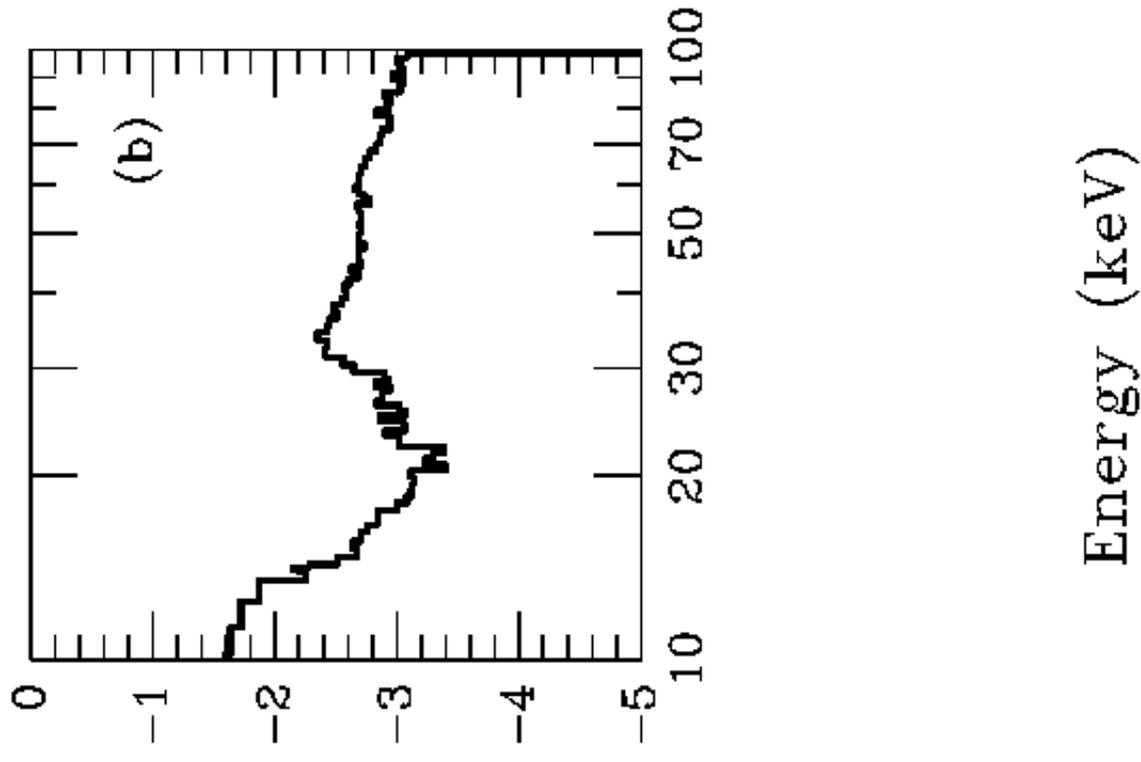
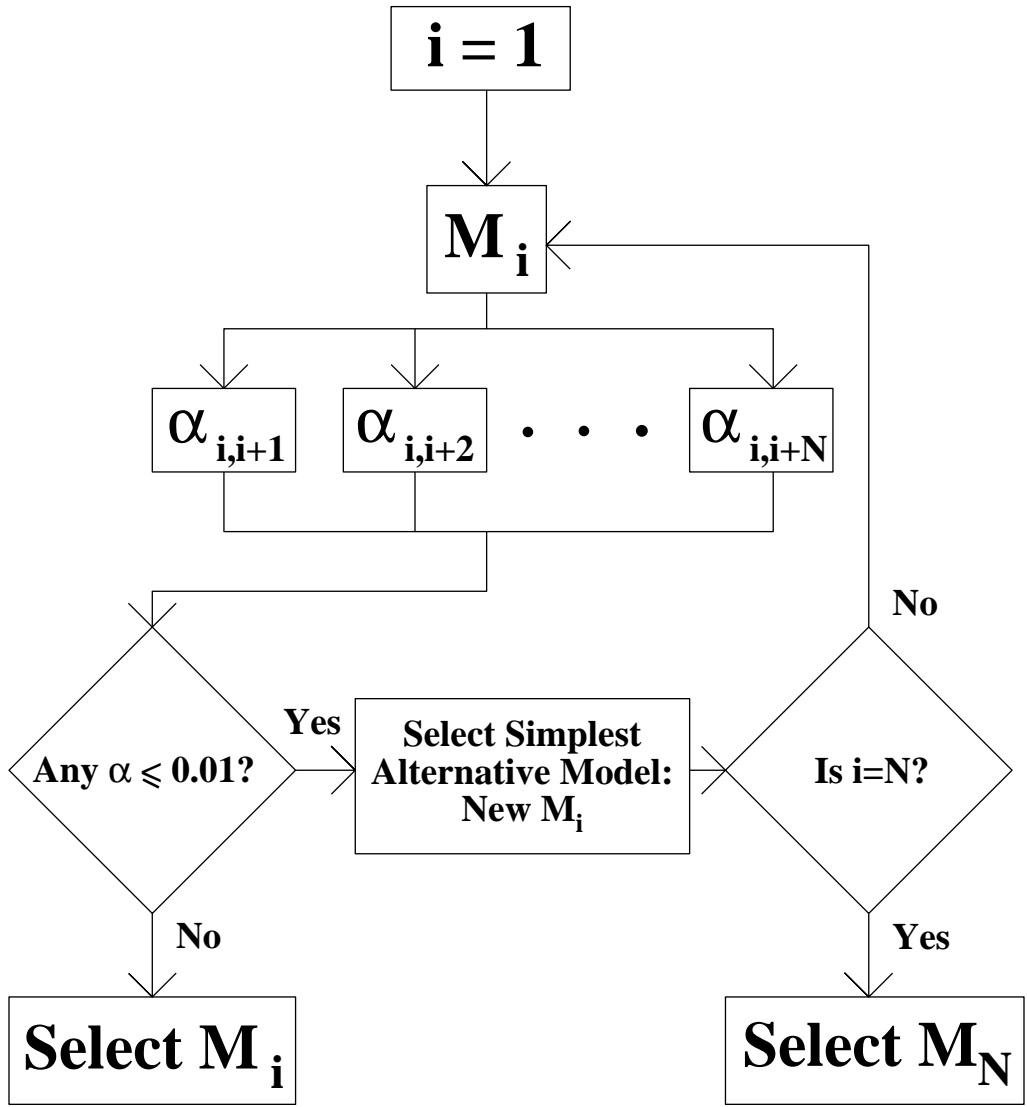
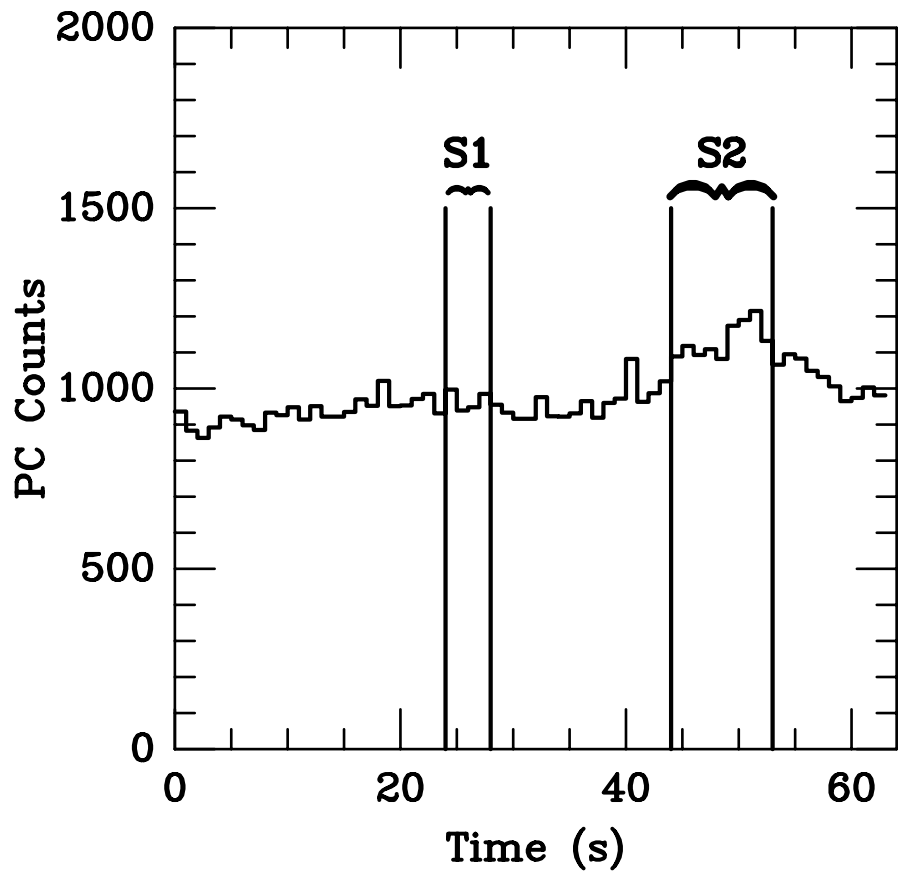
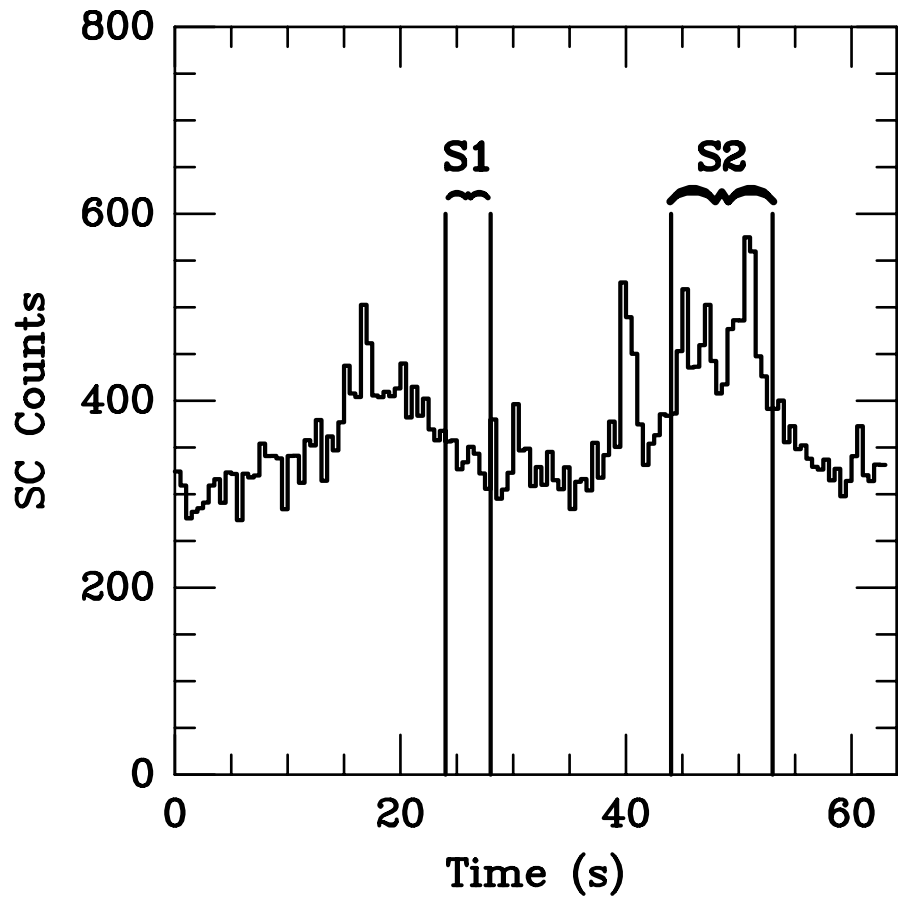
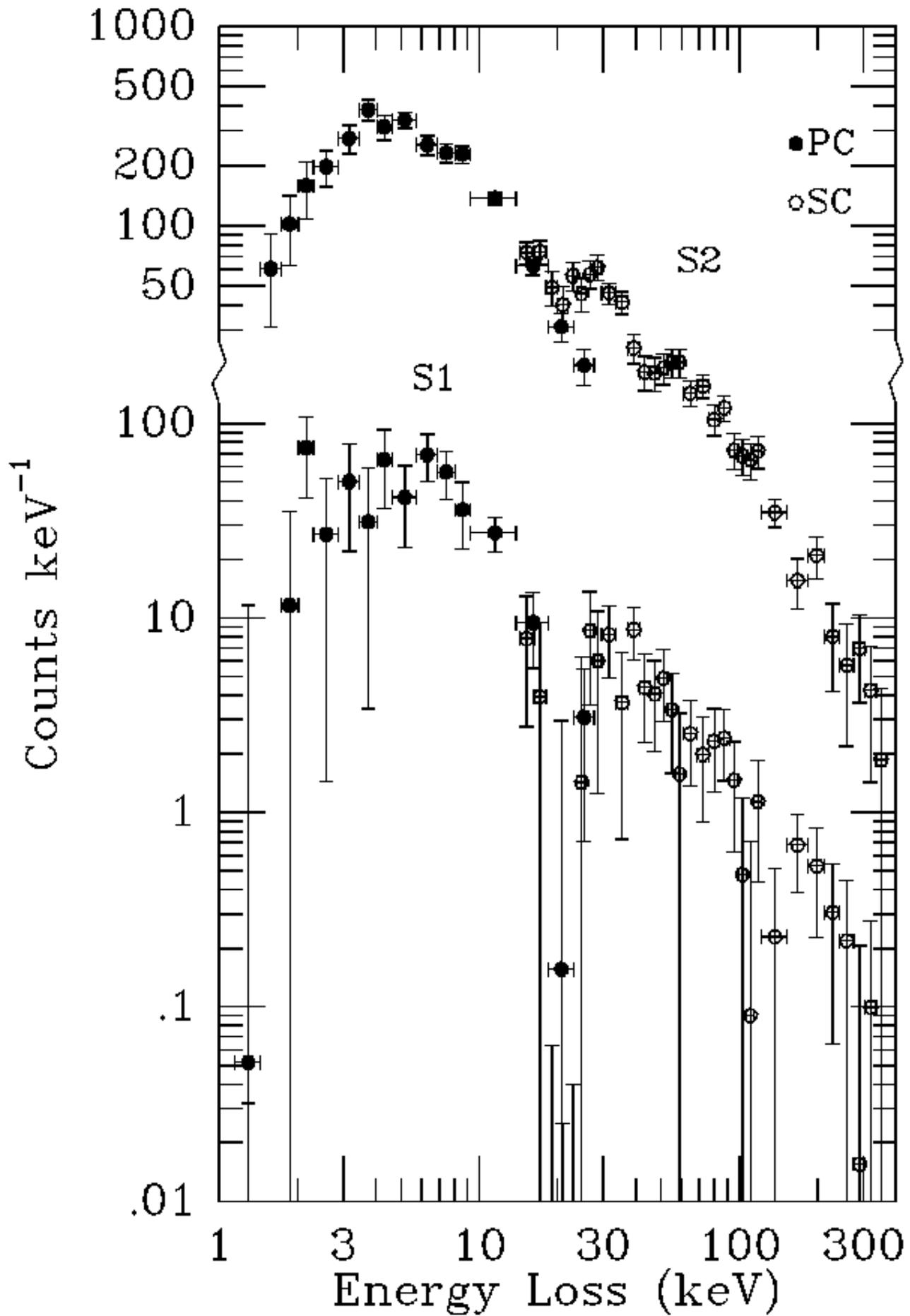
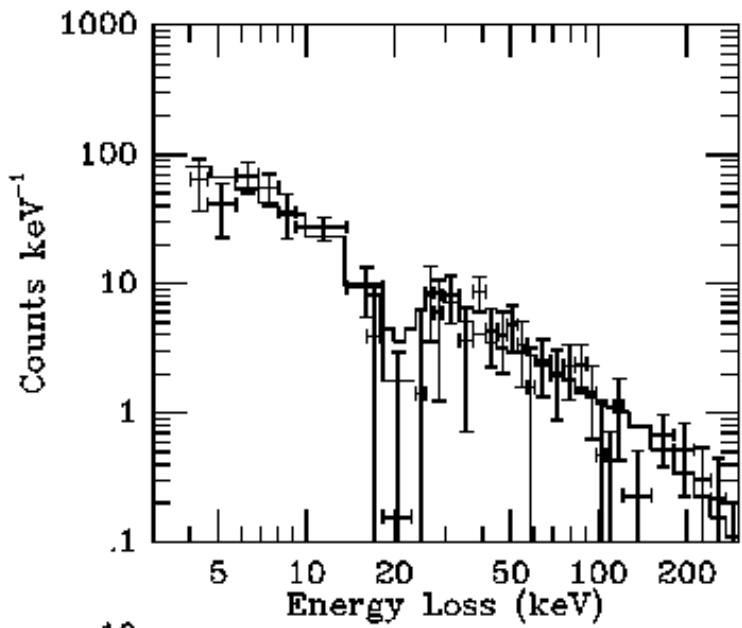
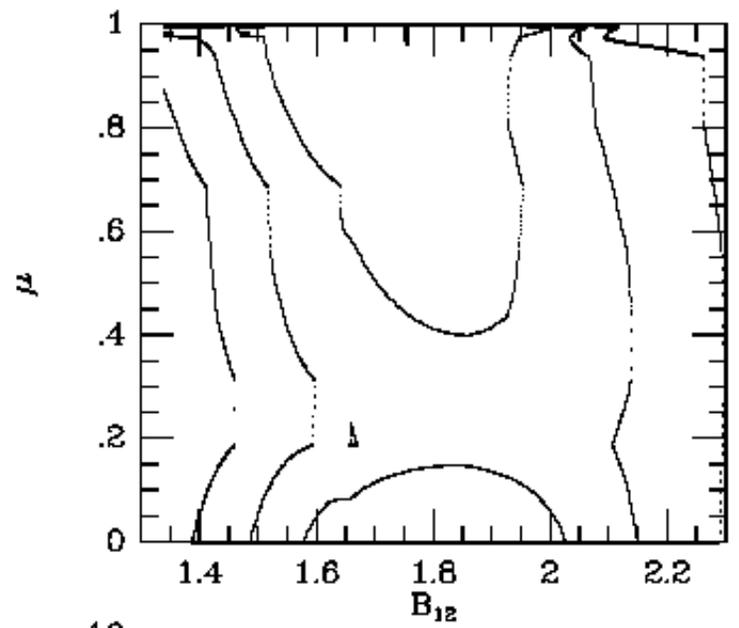
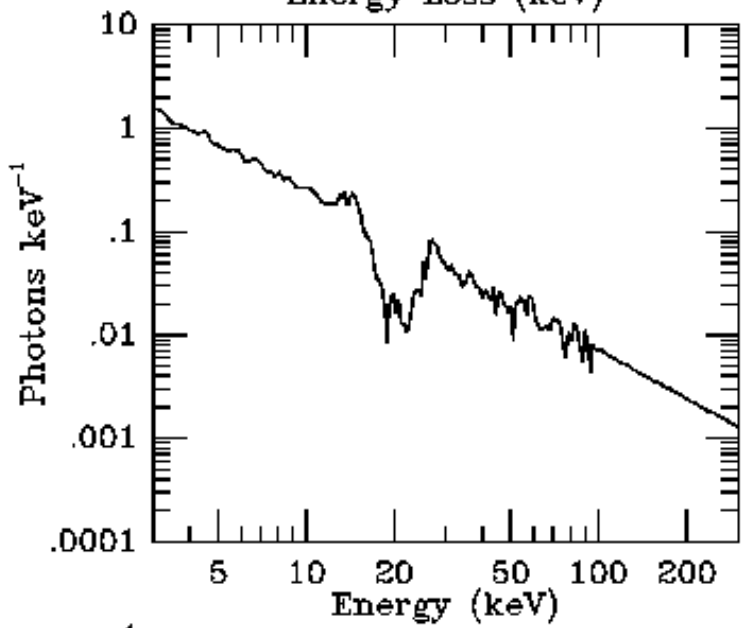
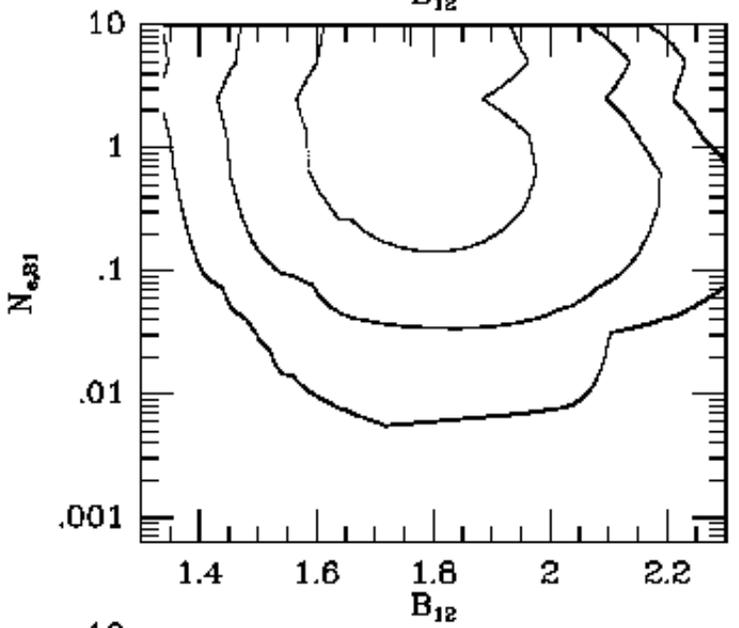
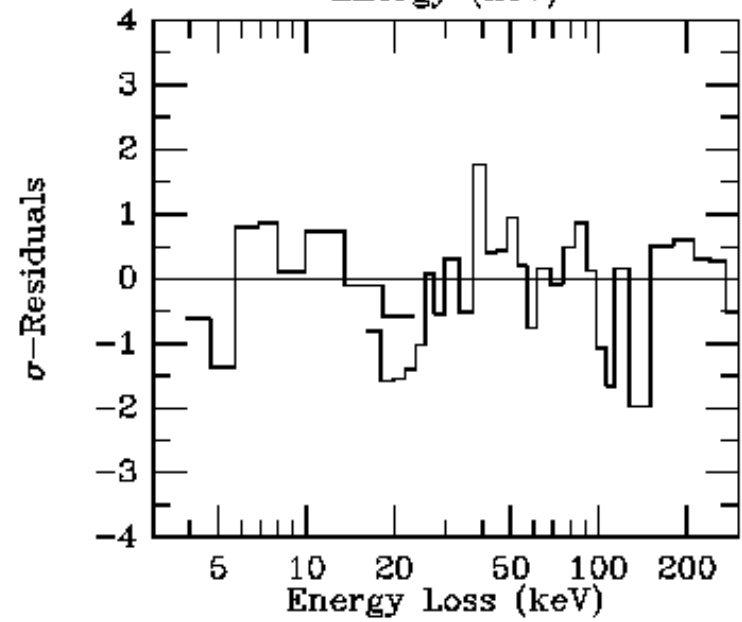
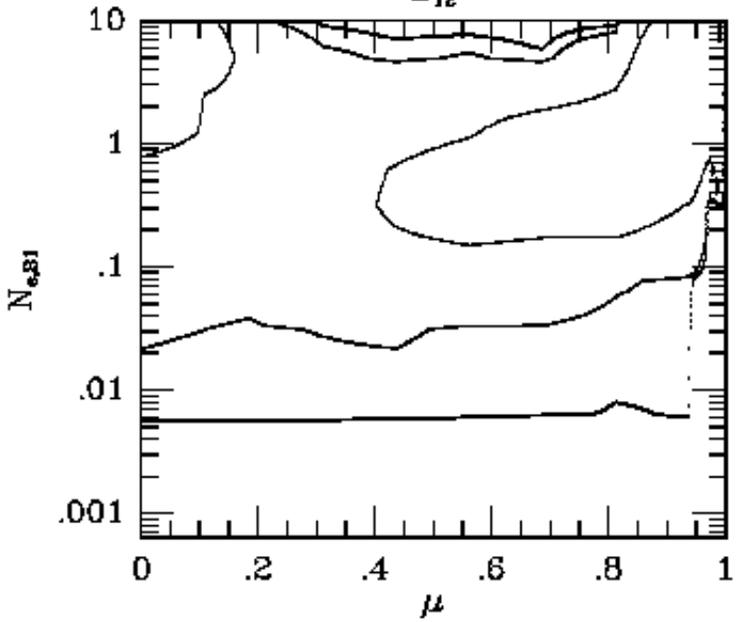
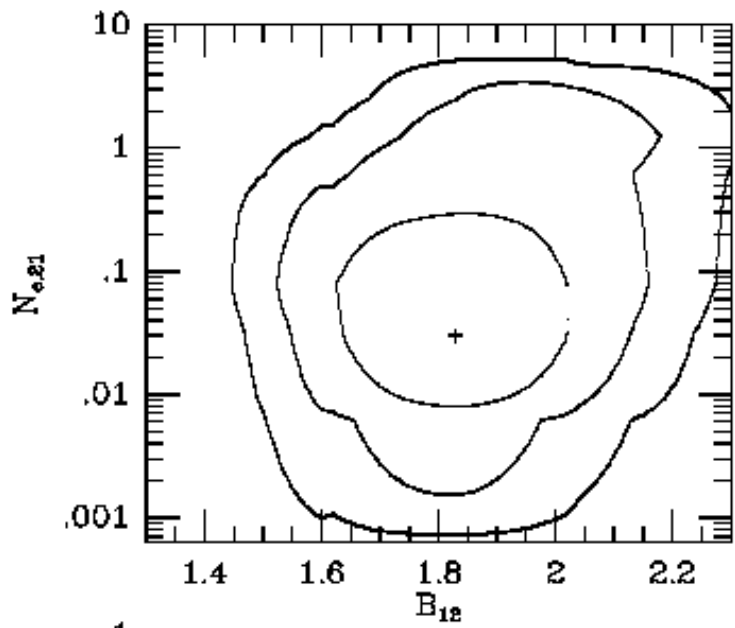
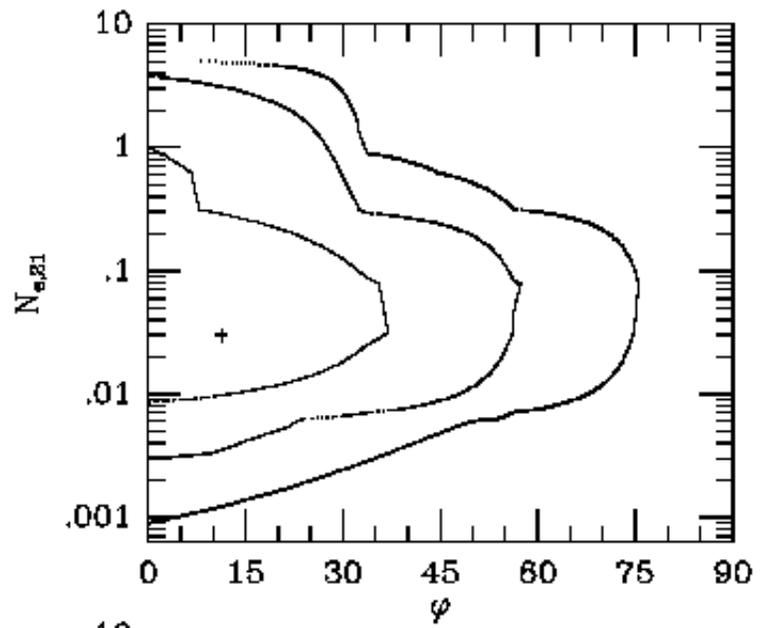
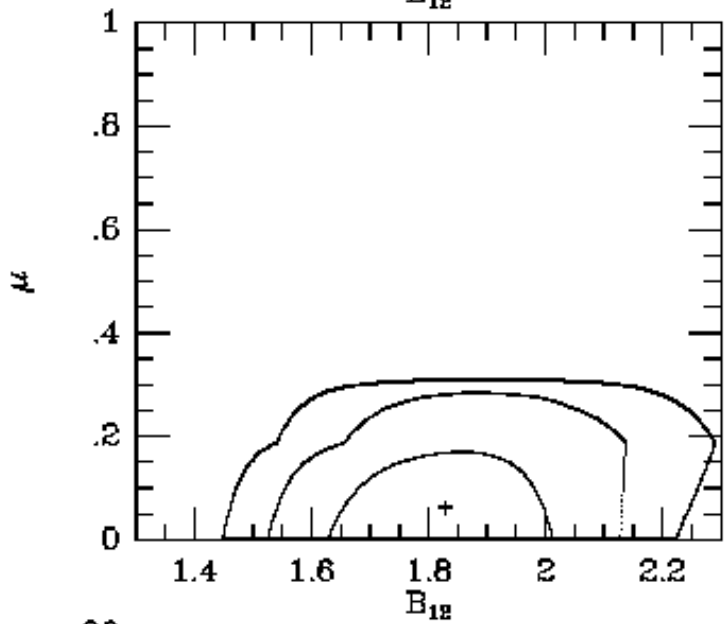
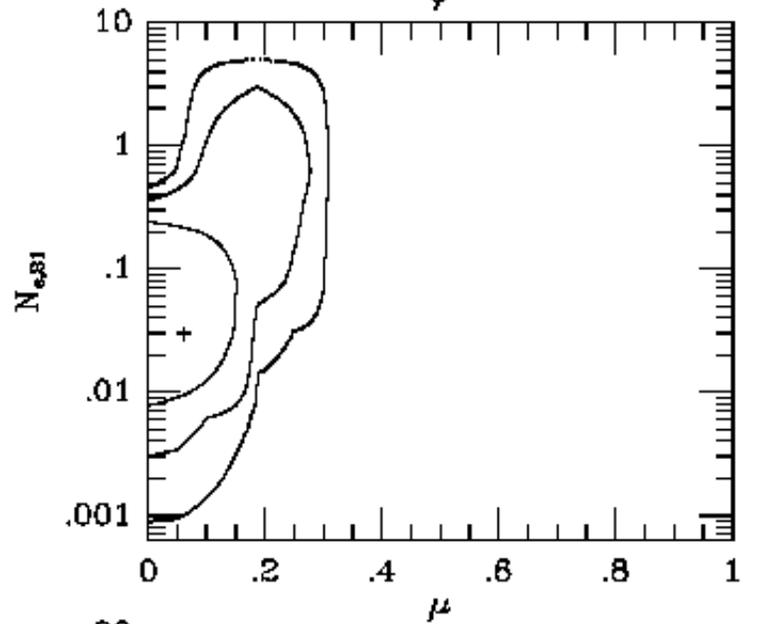
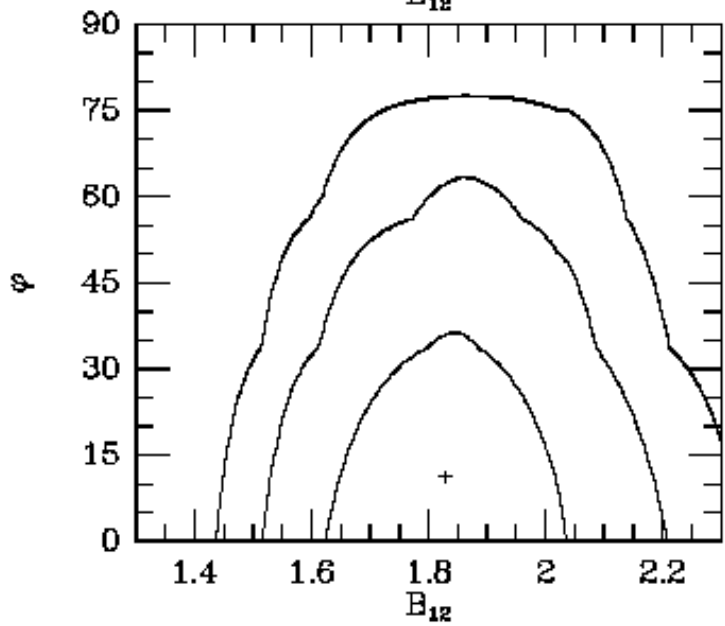
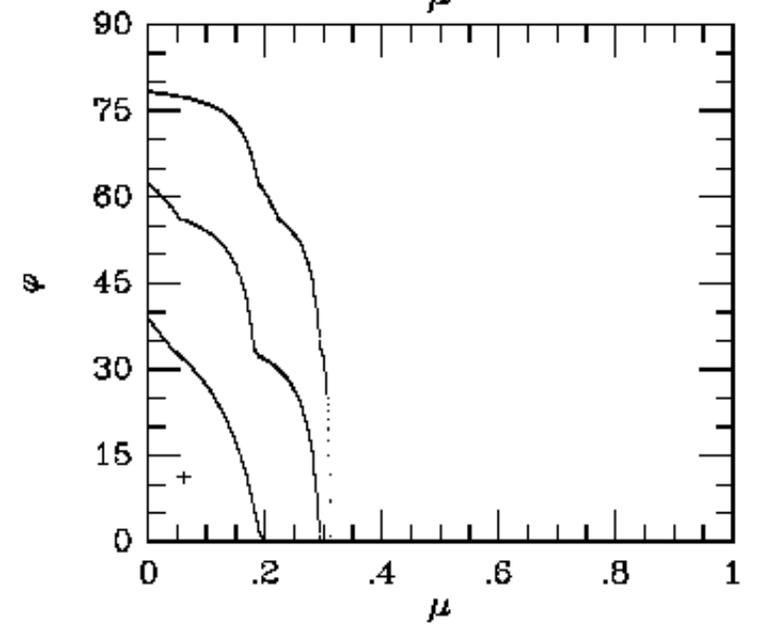
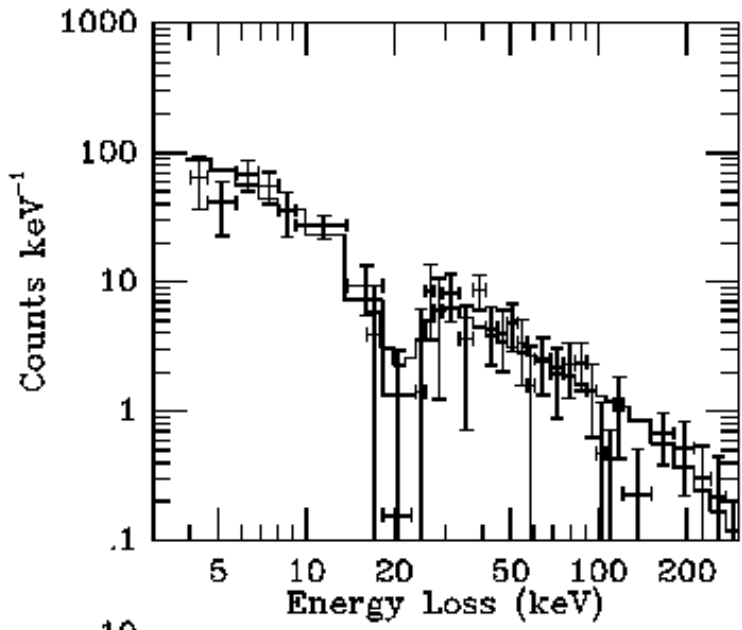
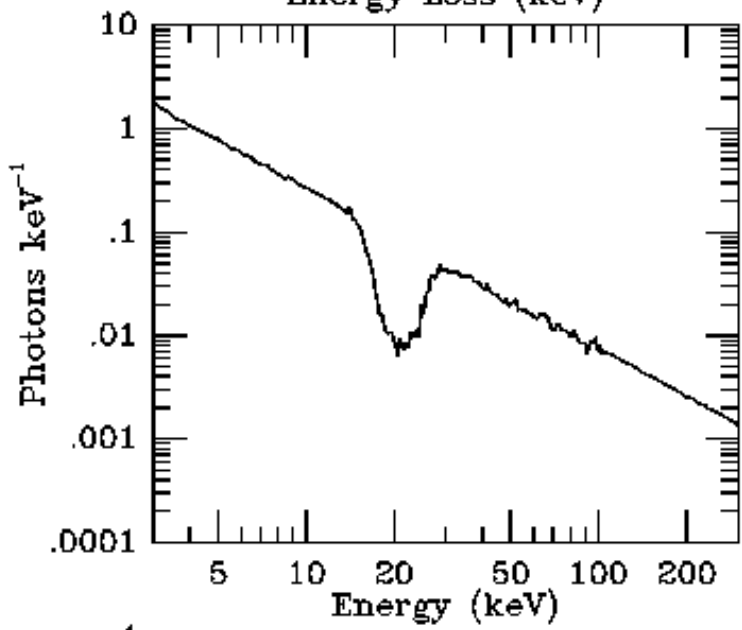
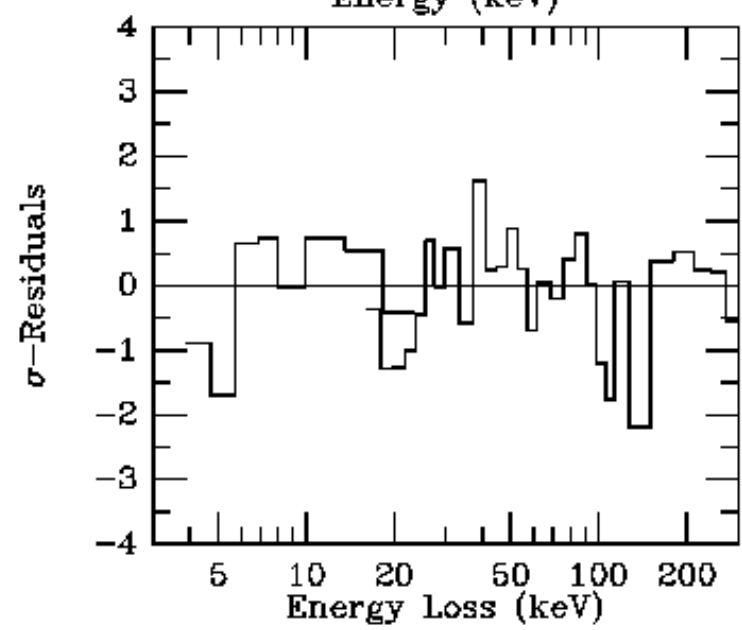
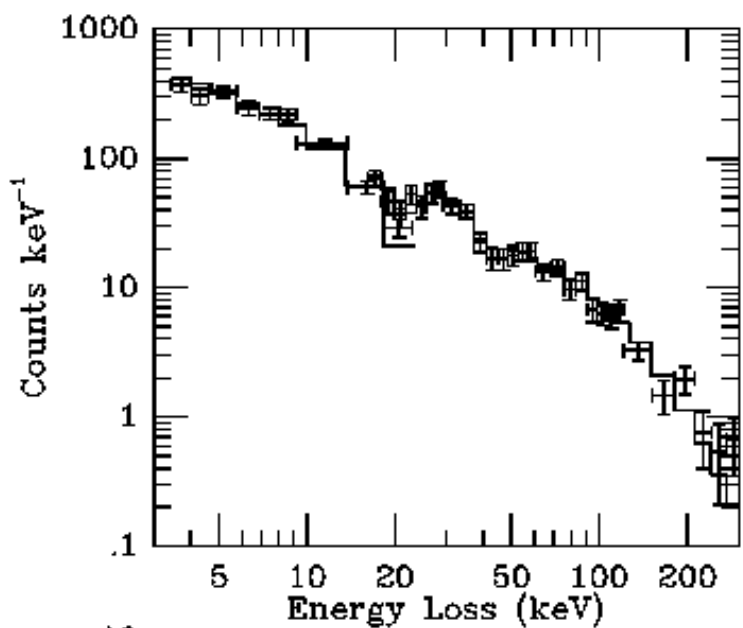
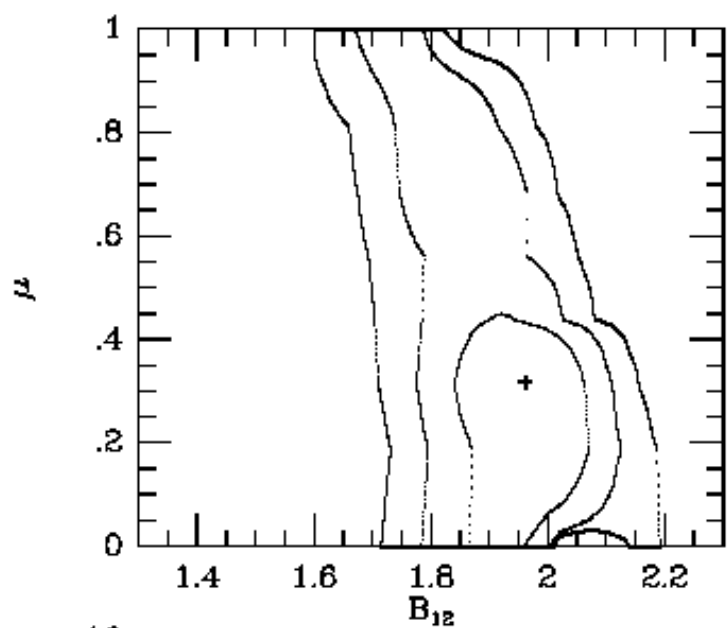
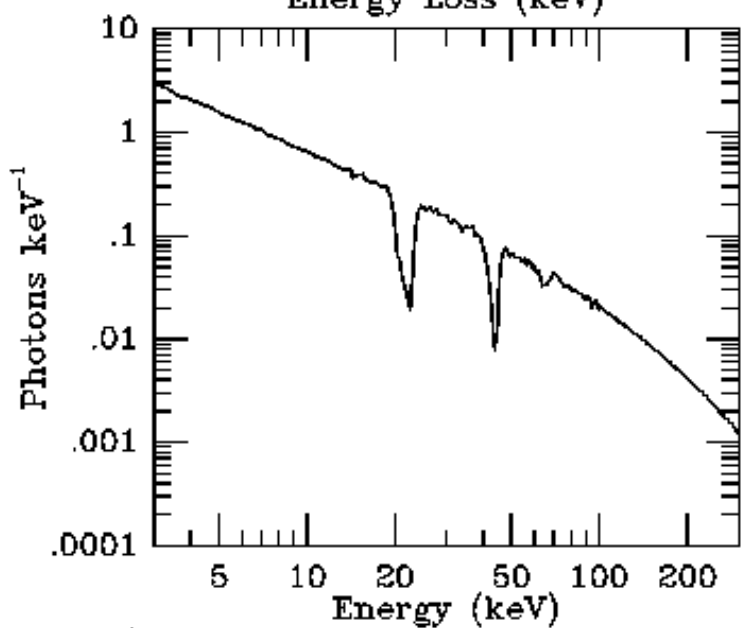
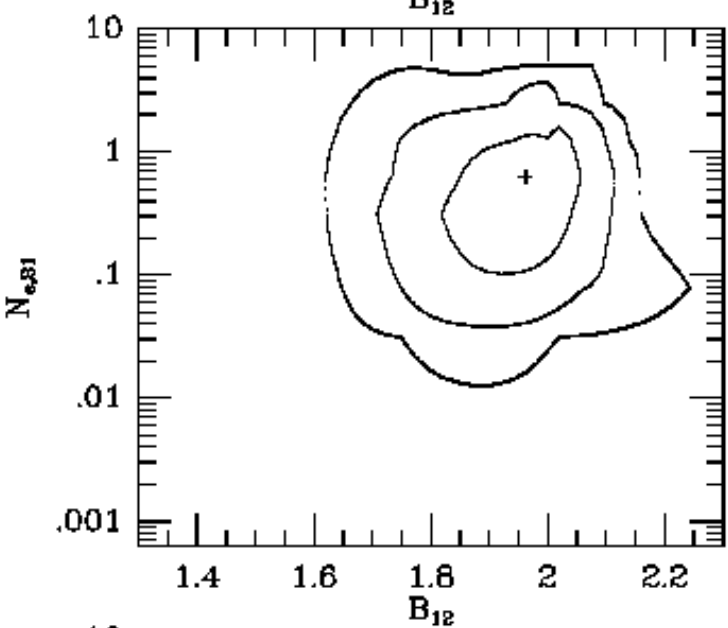
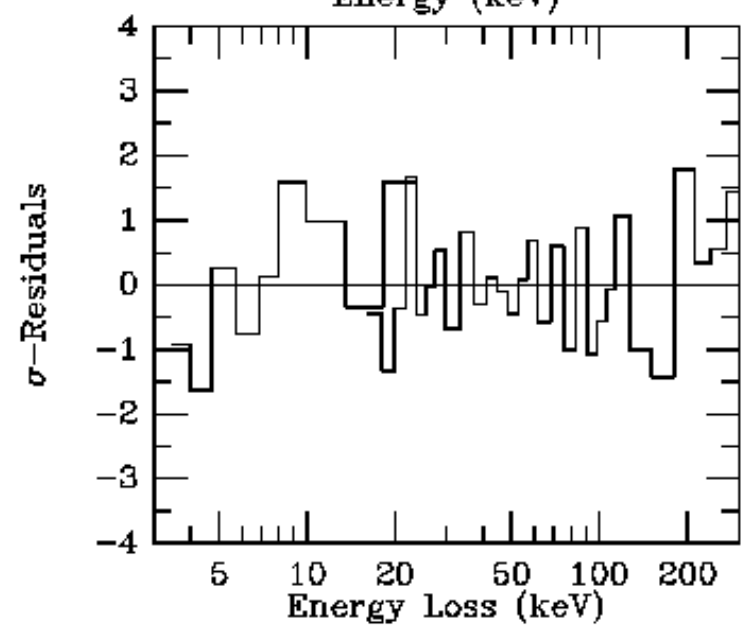
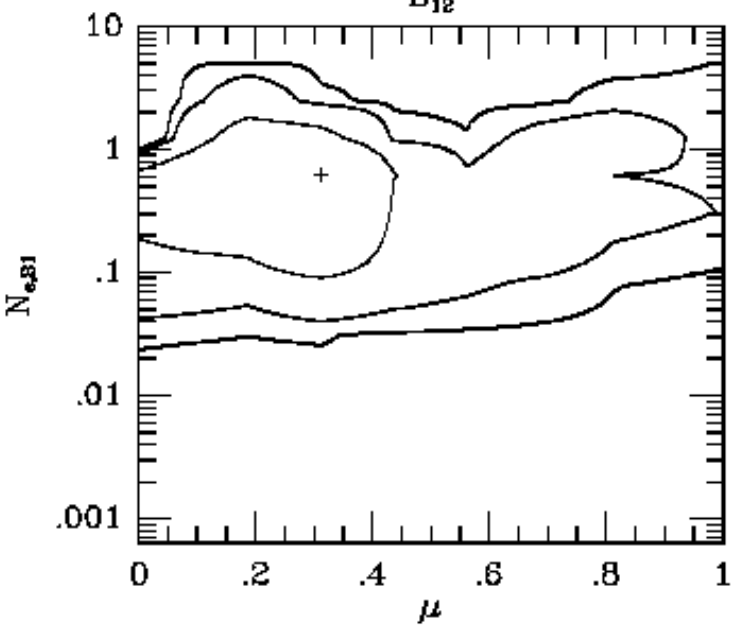
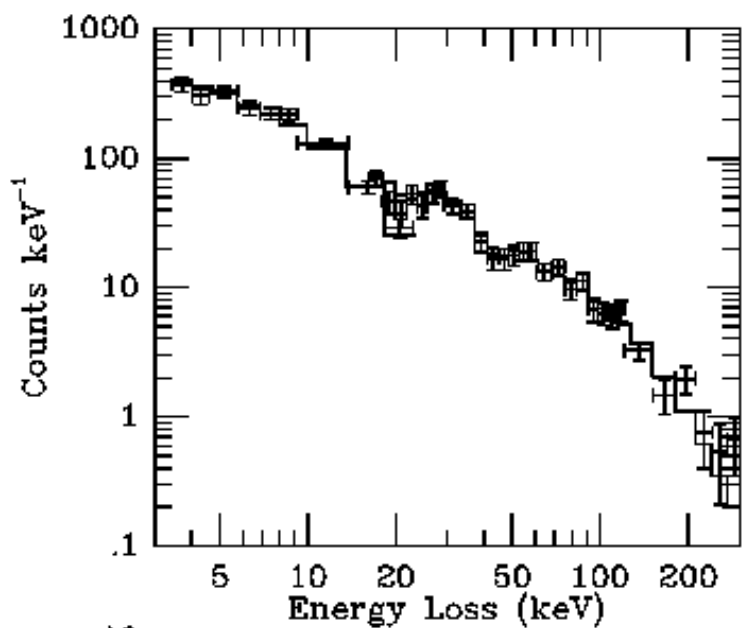
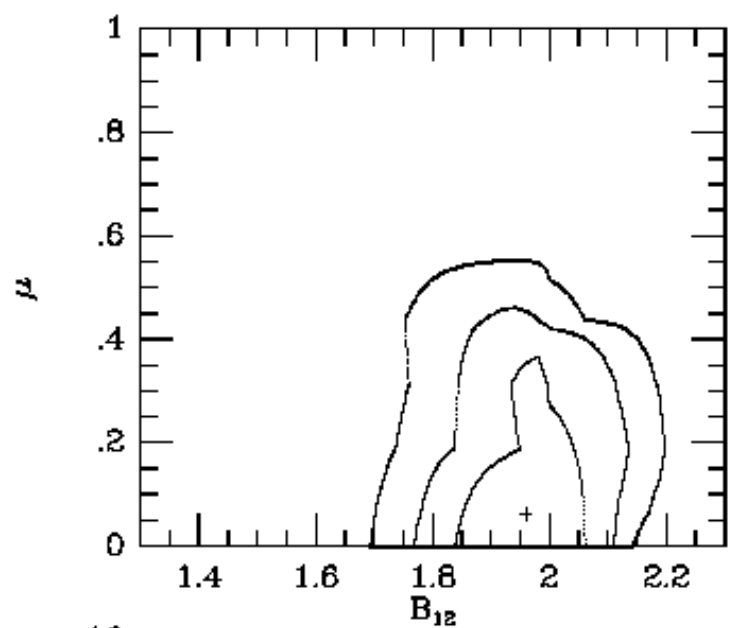
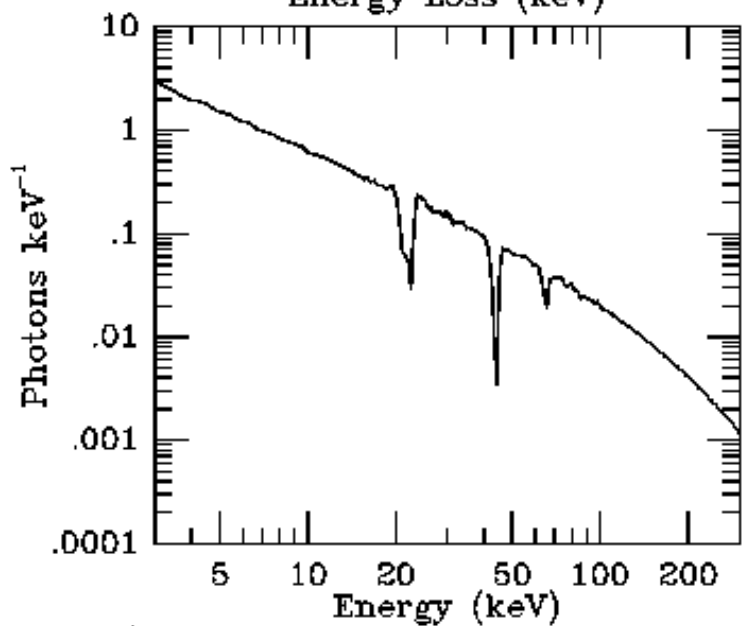
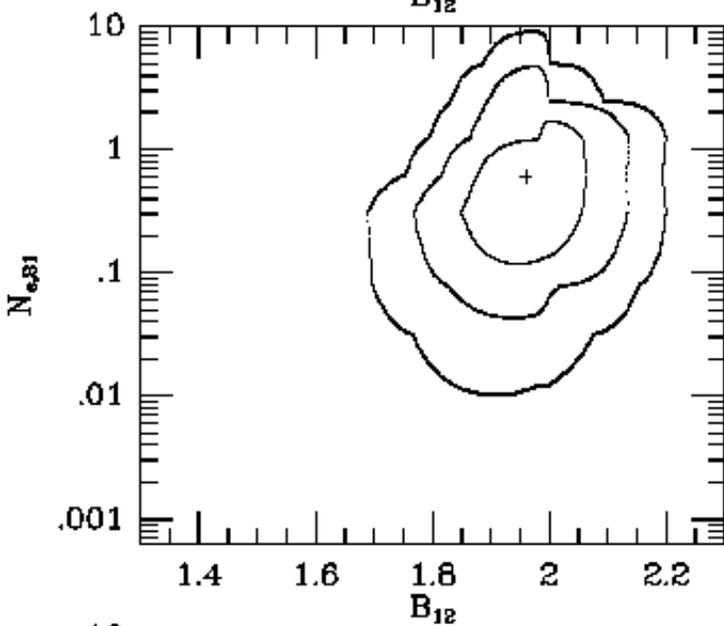
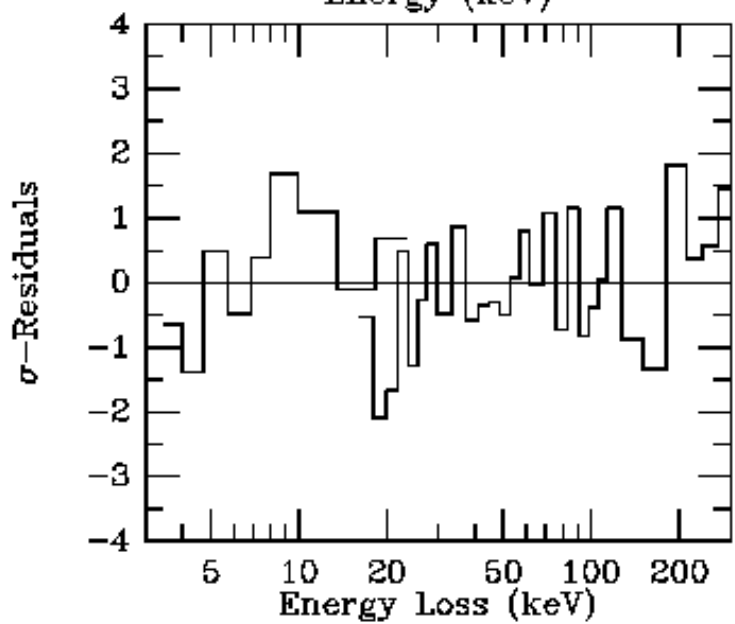
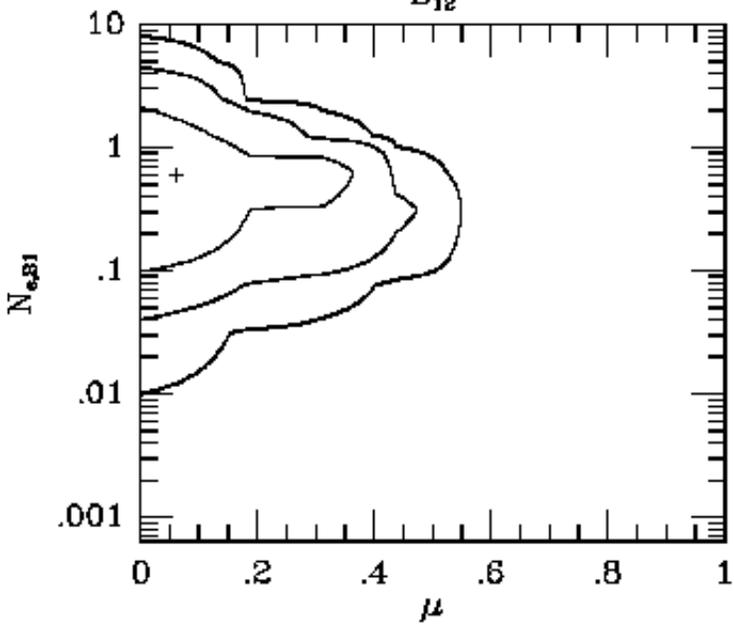
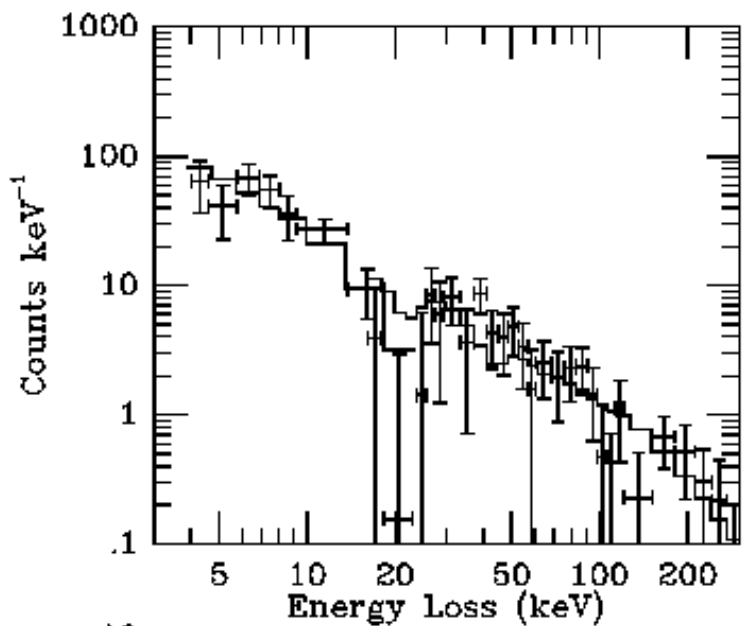
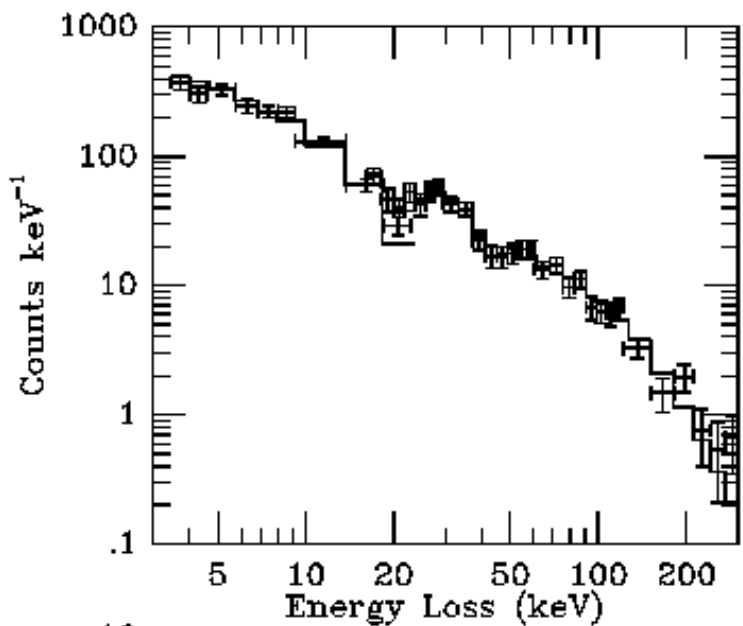
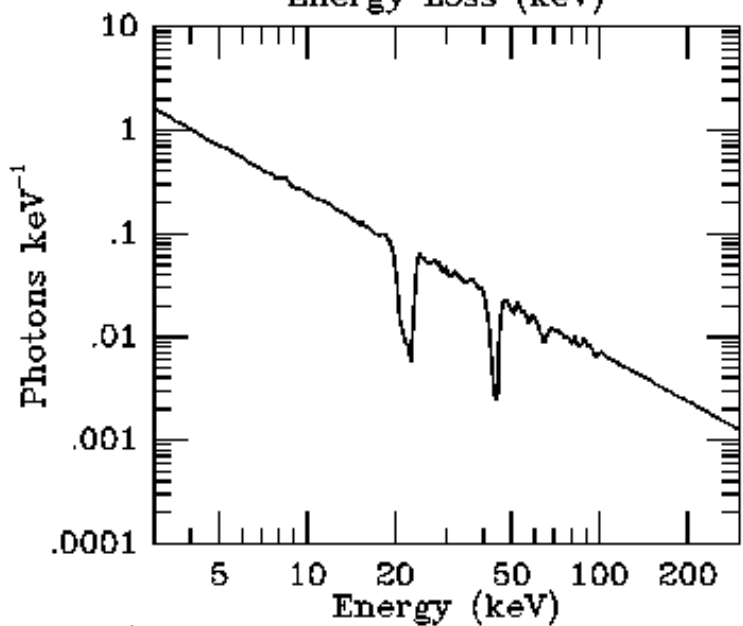
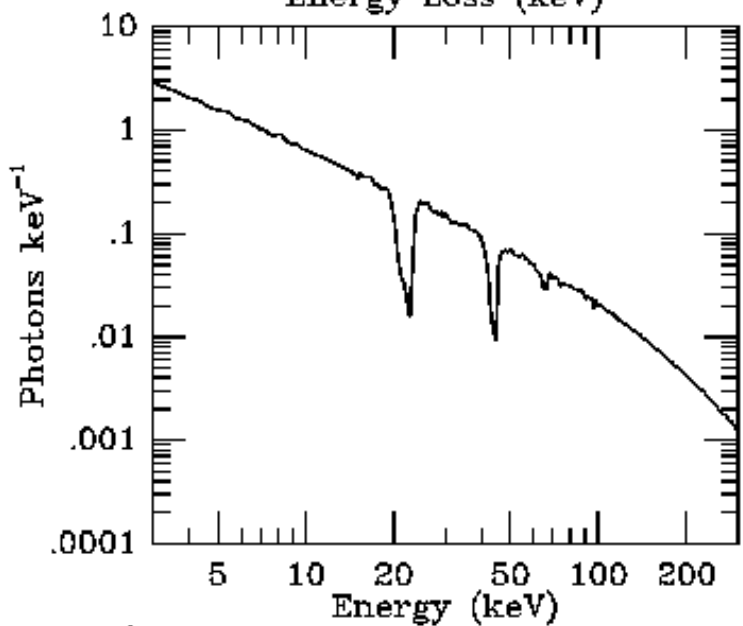
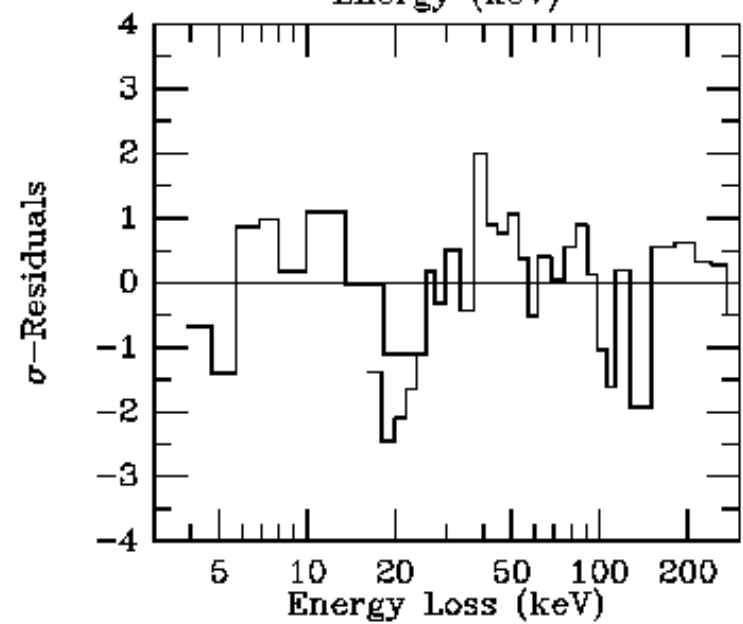
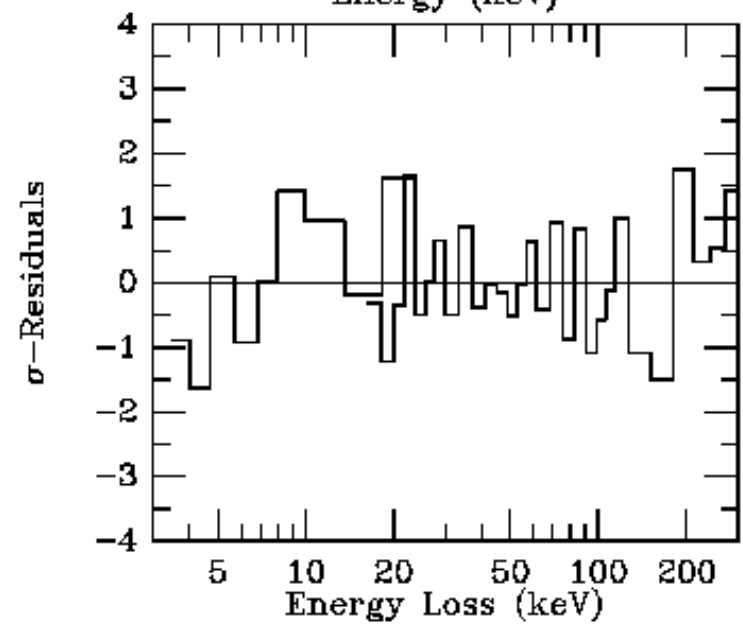
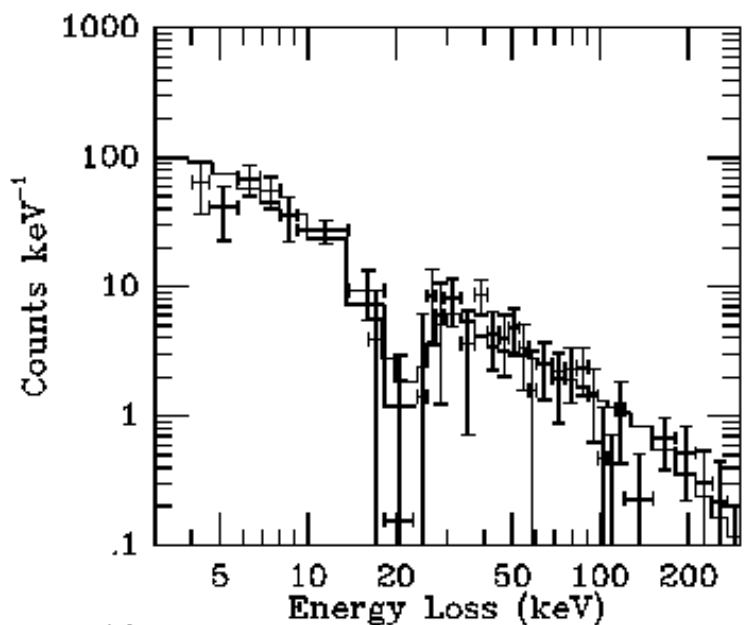
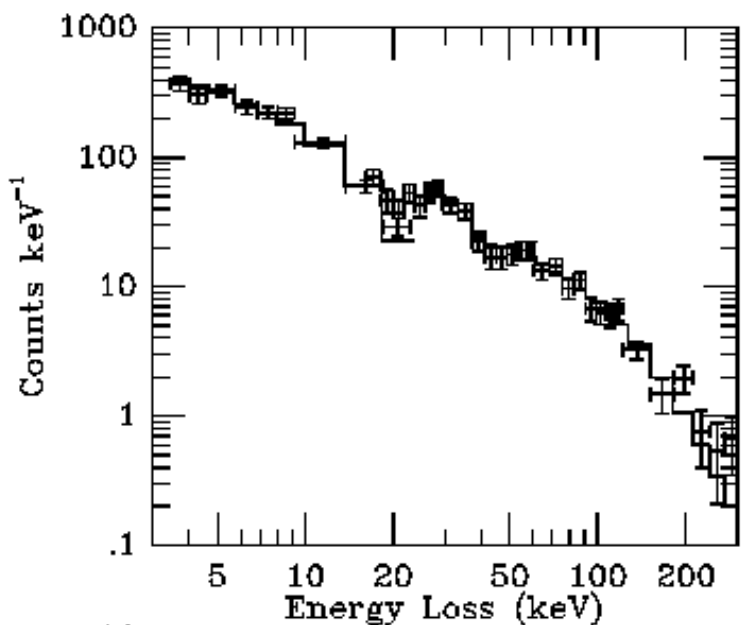
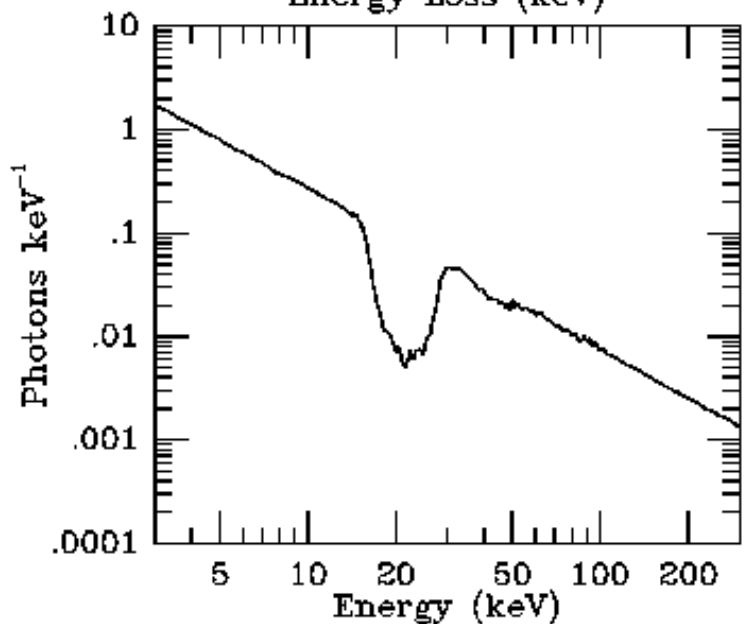
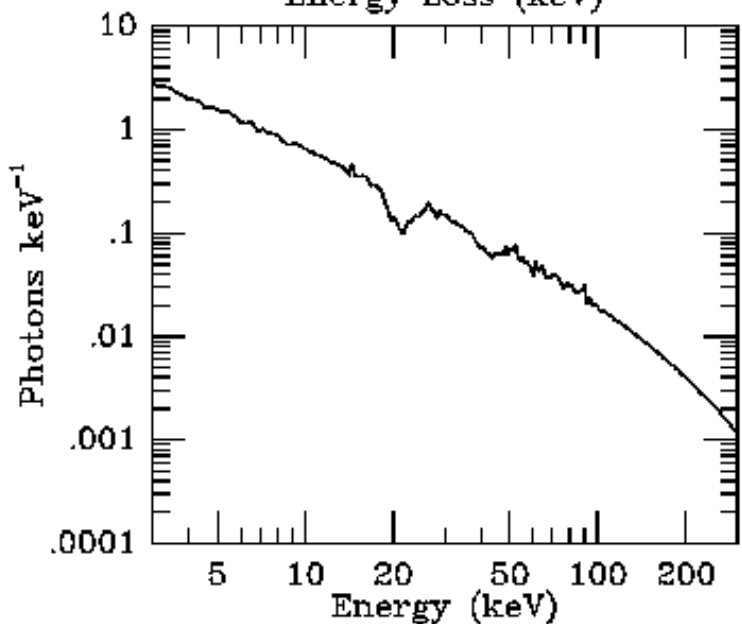
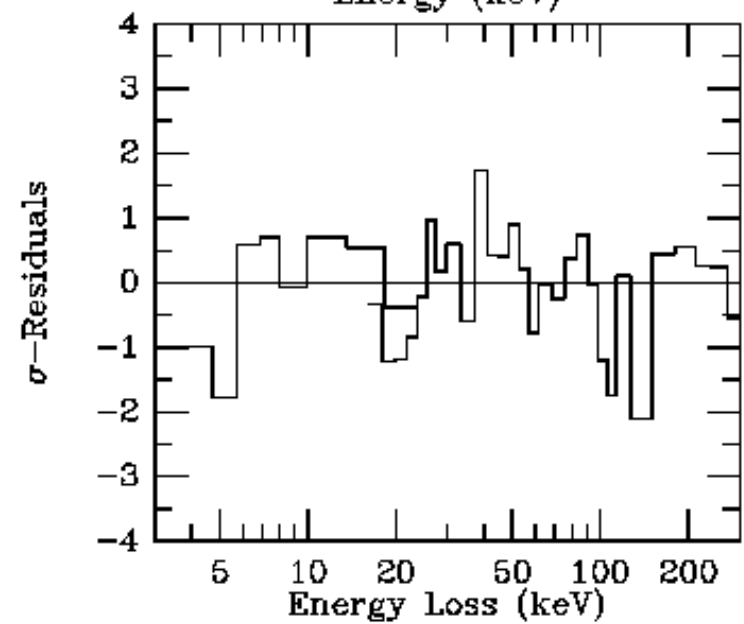
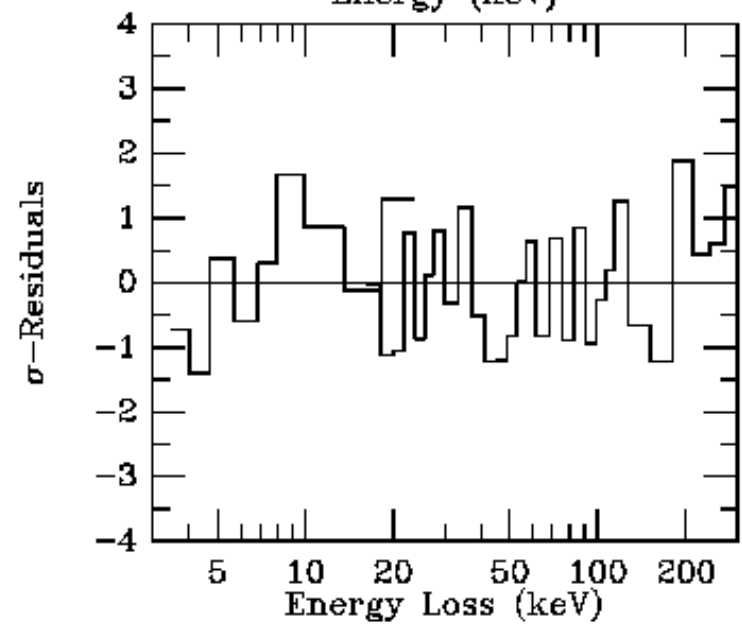
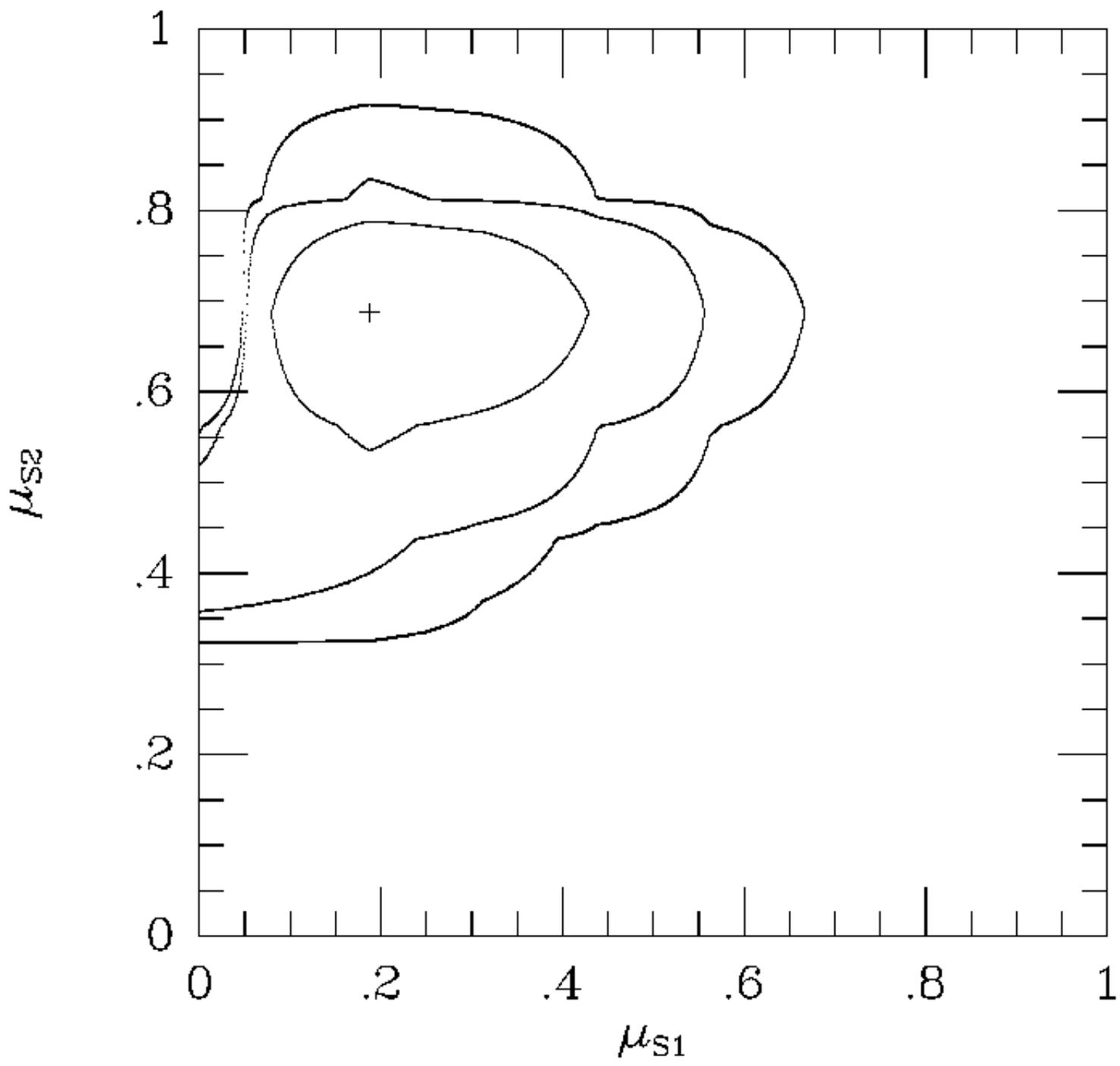
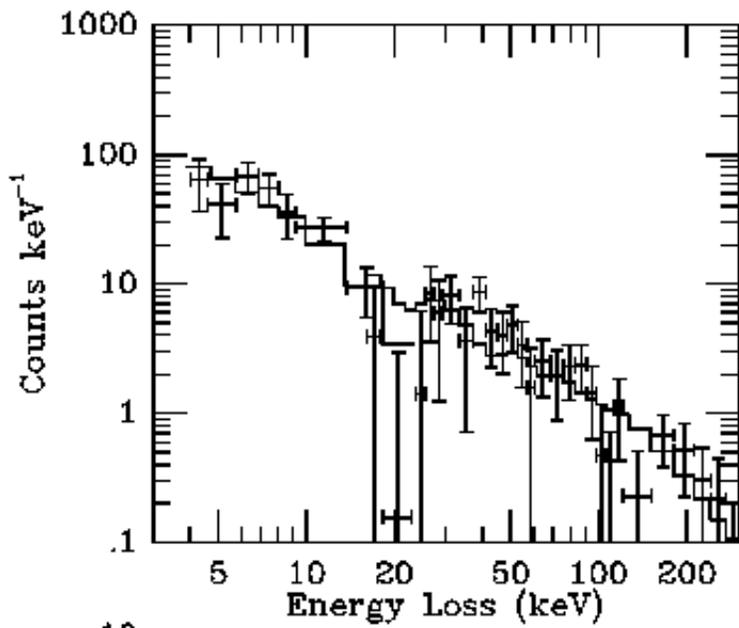
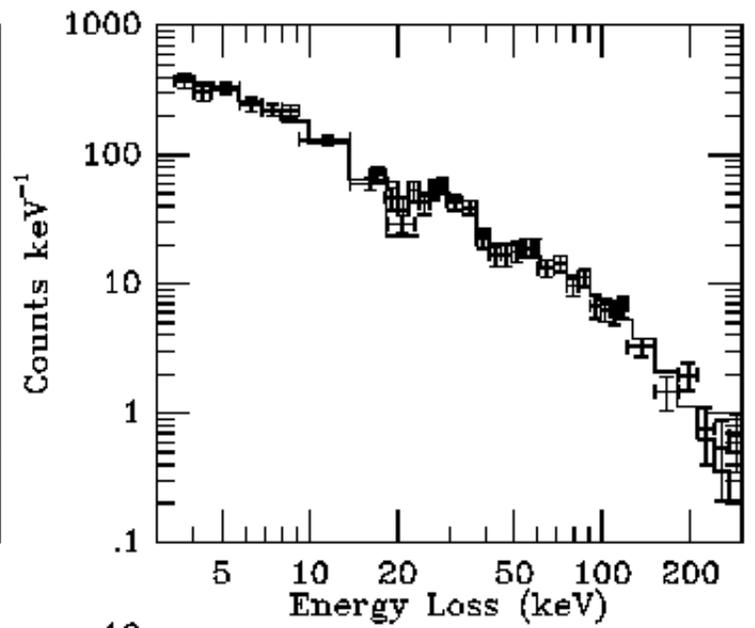
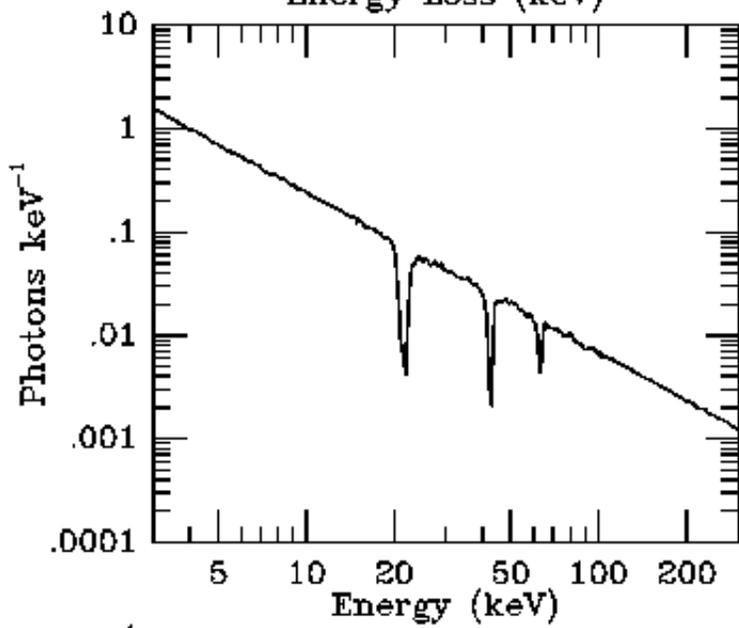
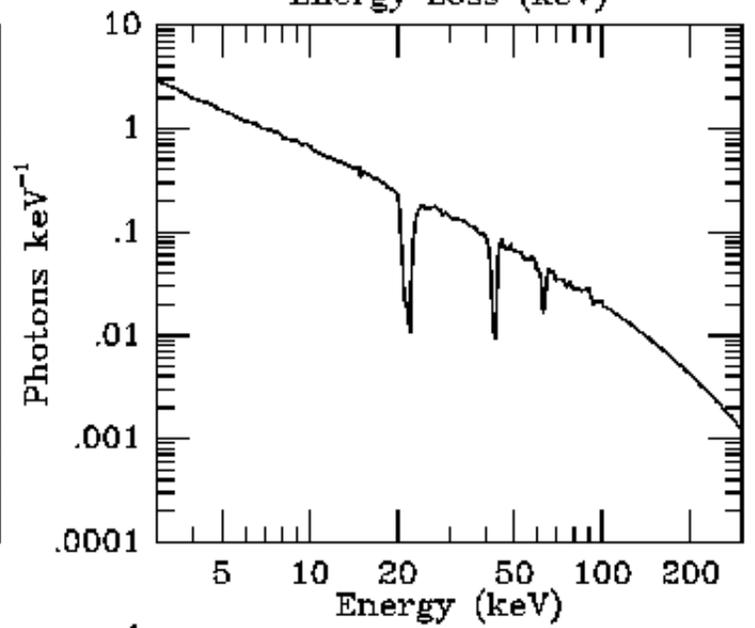
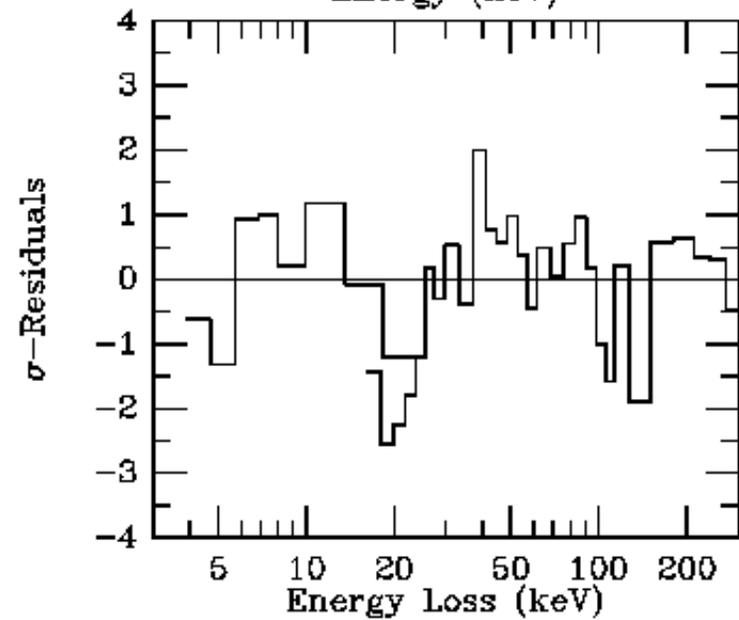
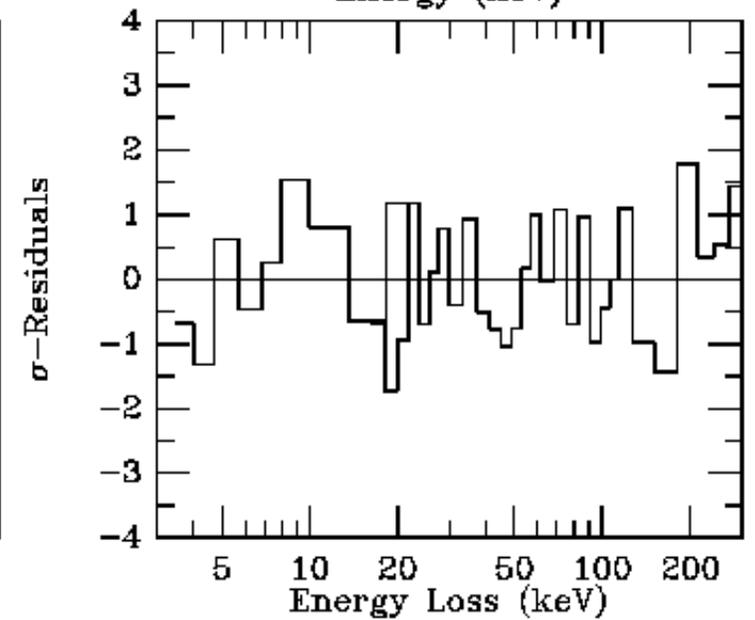
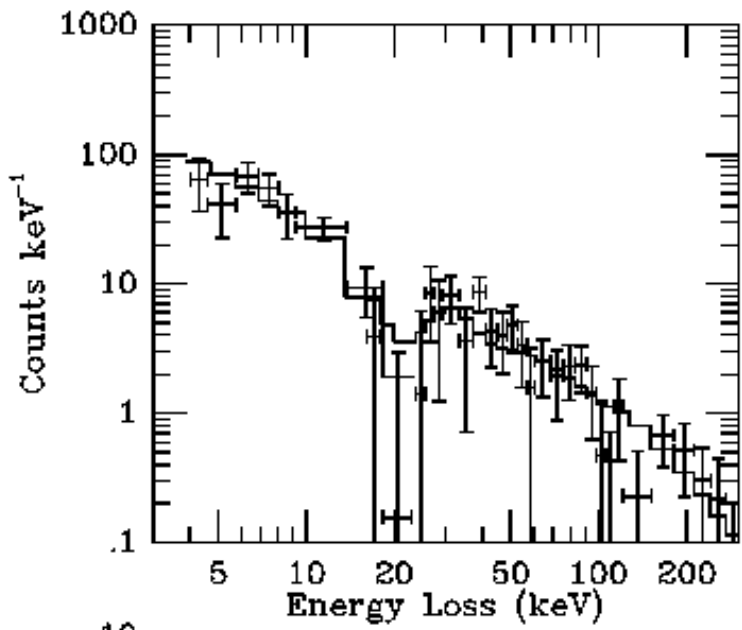
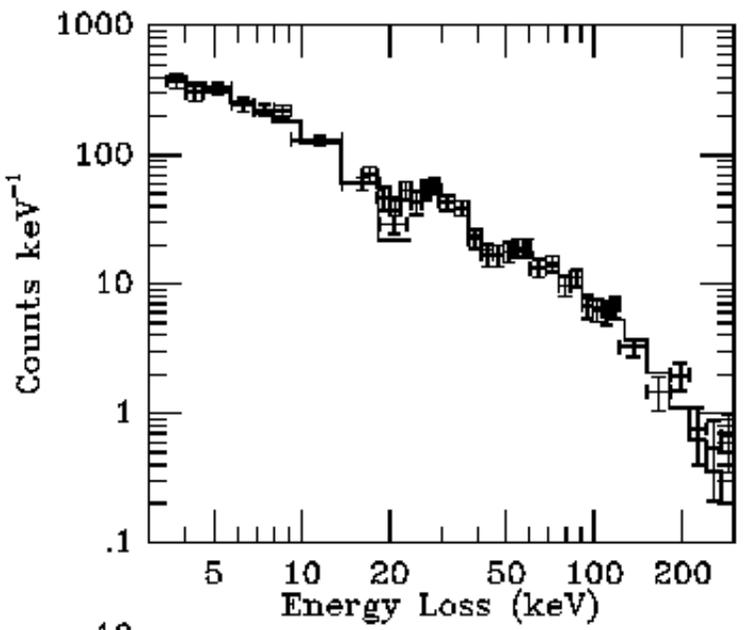
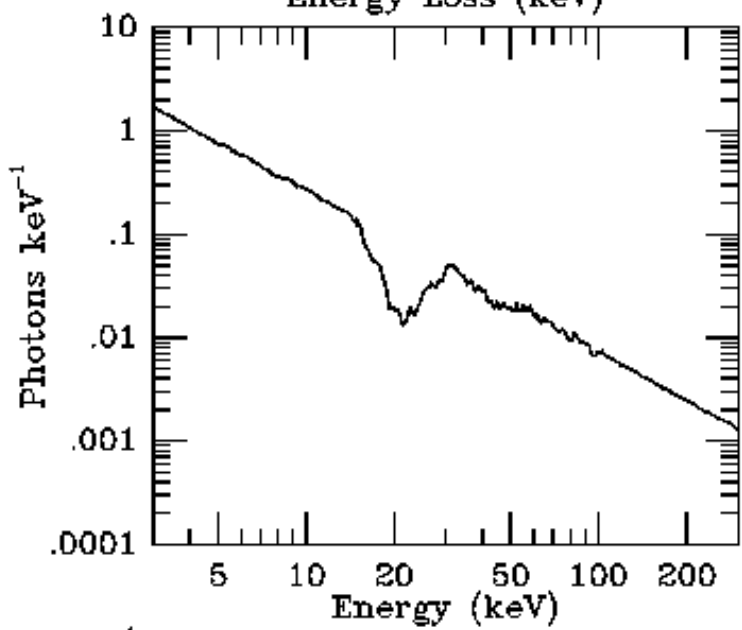
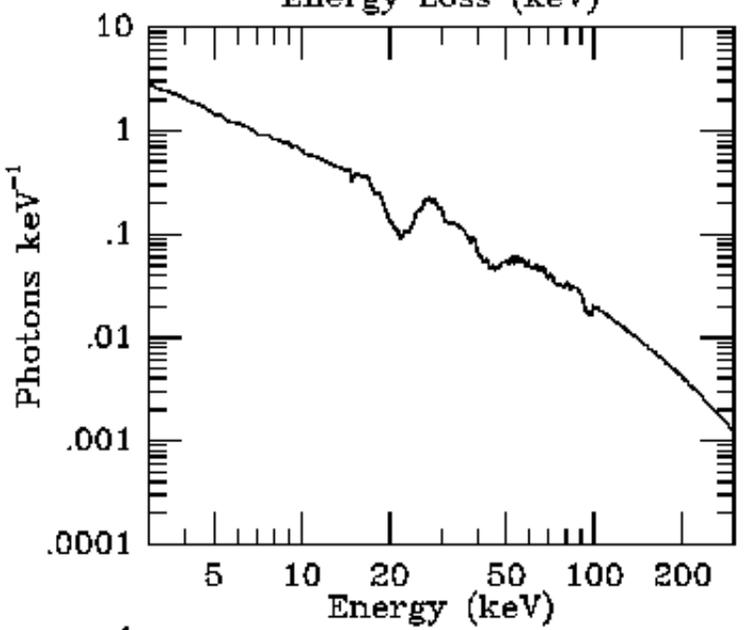
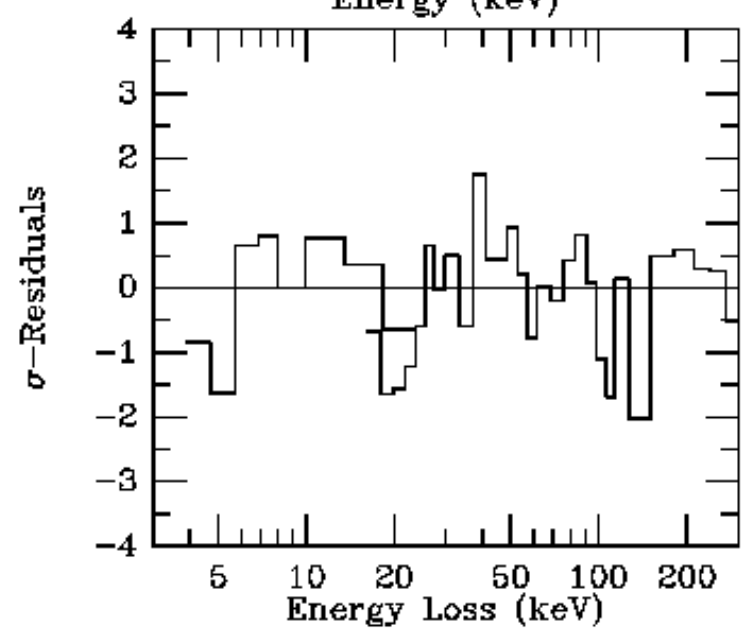
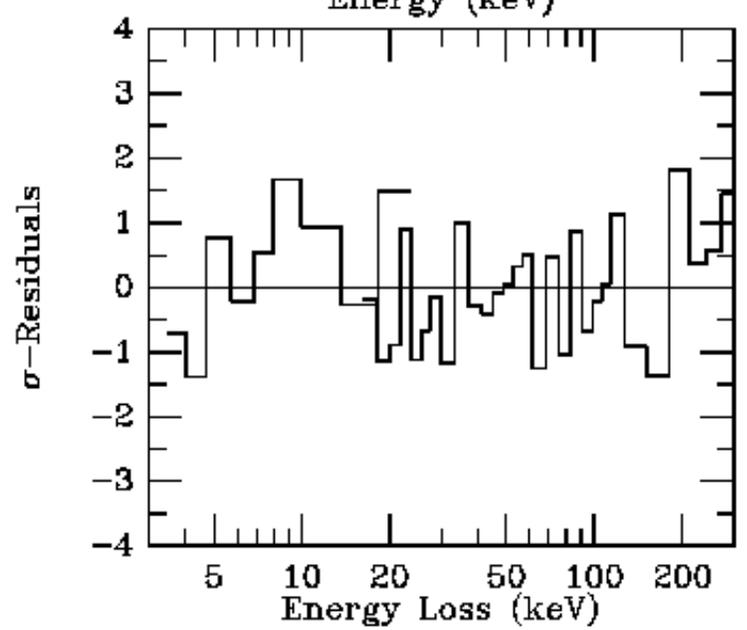
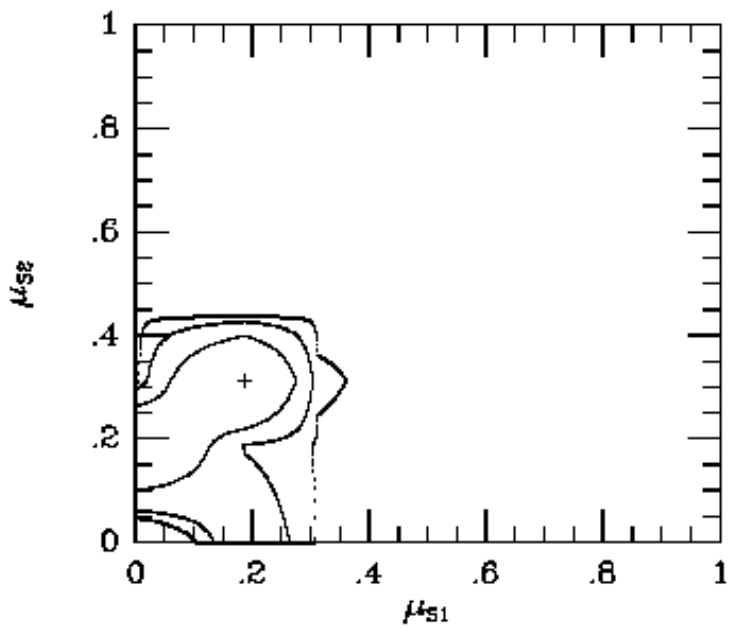
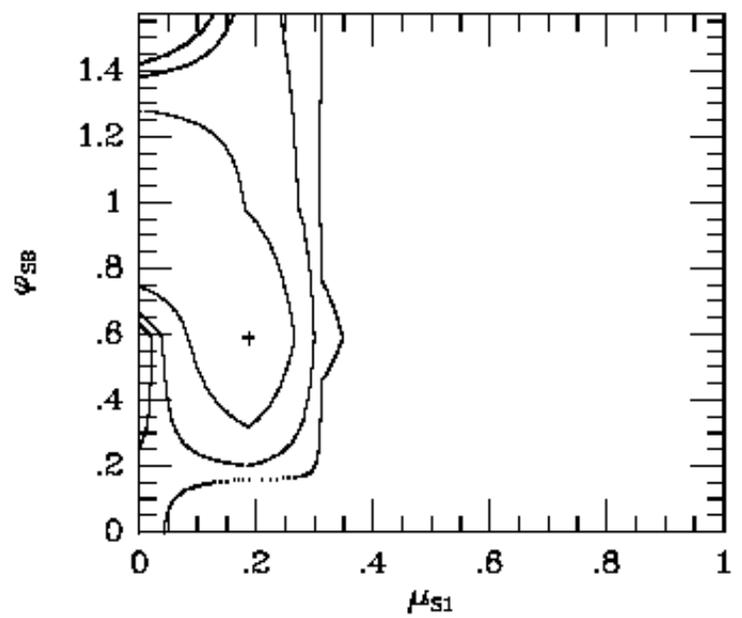
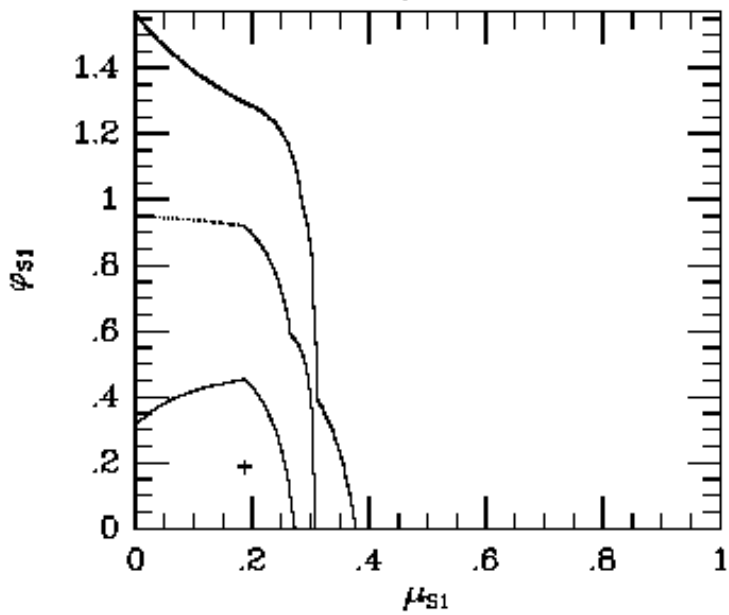
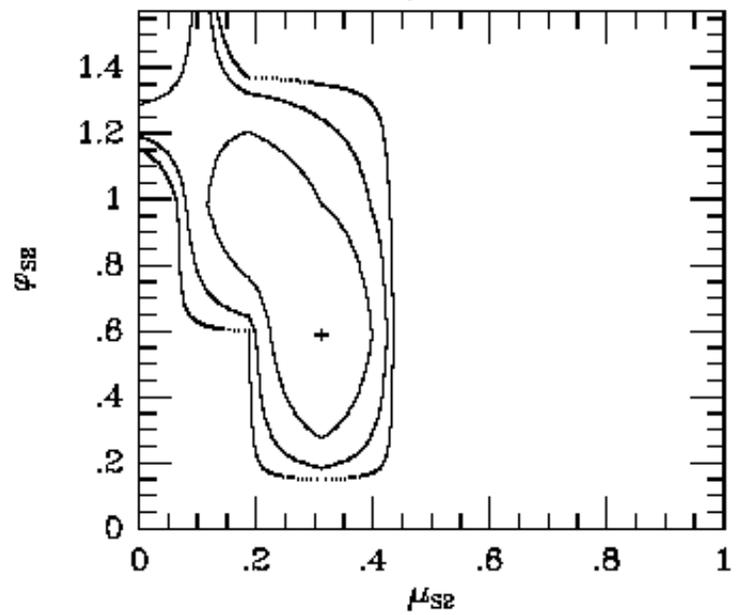
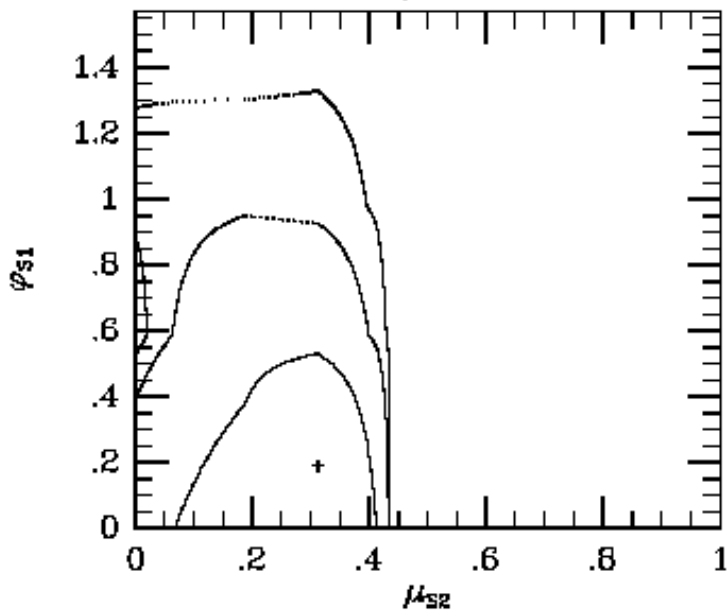
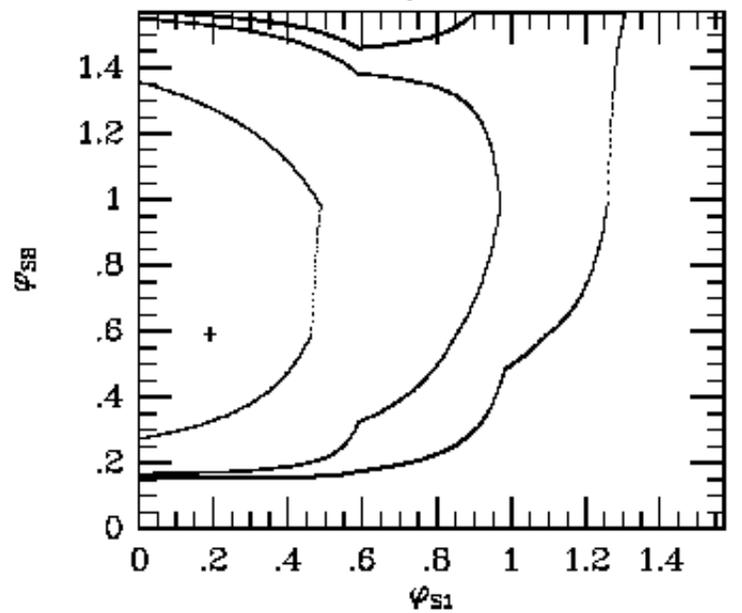
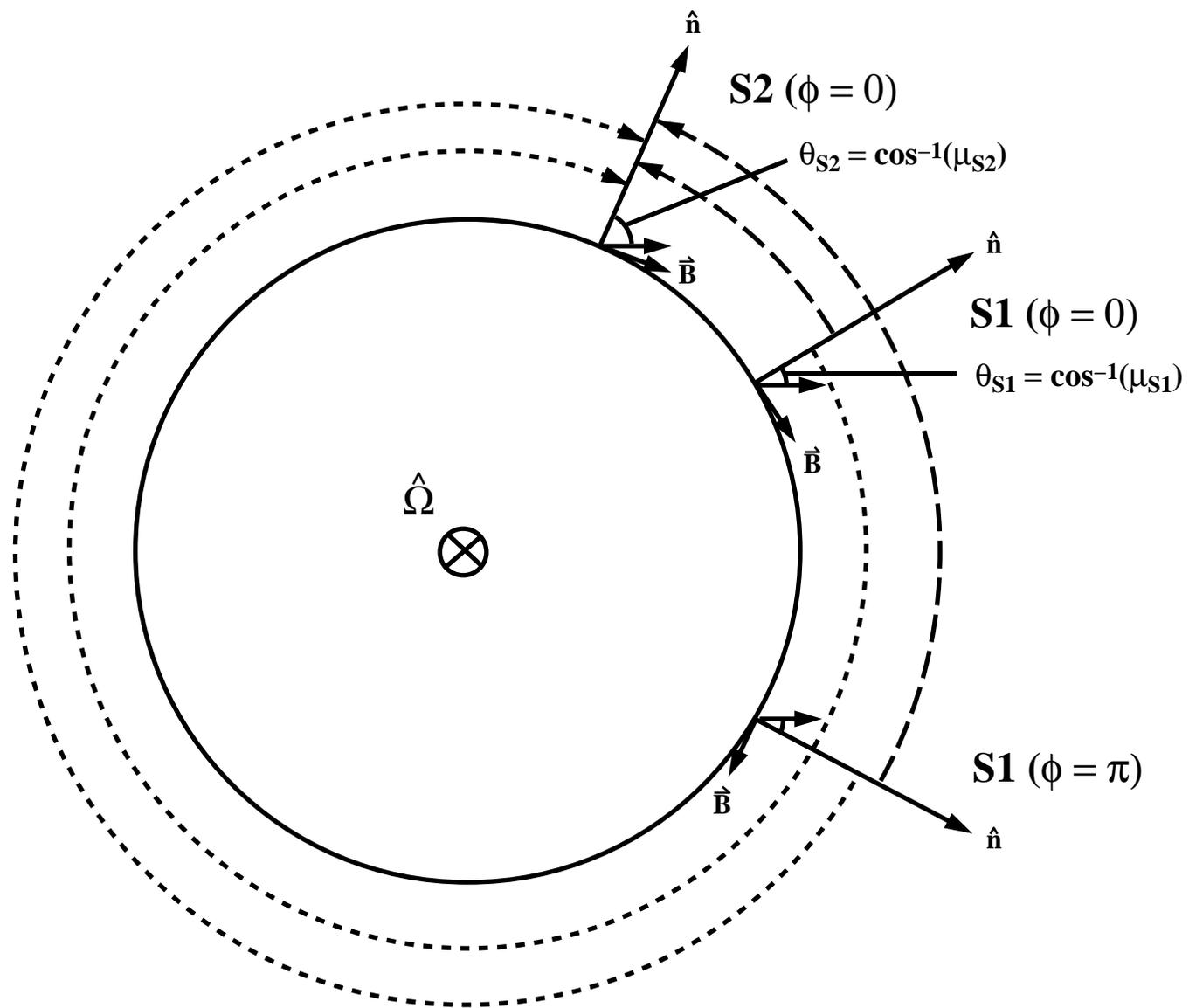
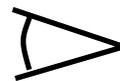
\end{document}